\documentclass[12pt,preprint]{aastex}

\newcommand{\rphk}{$R^\prime_{\rm HK}$}
\newcommand{\rhk}{$R^\prime_{\rm HK}$}

\newcommand{\logrphk}{$\log R^\prime_{\rm HK}$}
\newcommand{\logrphkbar}{$\overline{\log R^\prime_{\rm HK}}$}

\newcommand{\rx}{$R_{\rm X}$}
\newcommand{\loglx}{$\log L_{\rm X}$}
\newcommand{\logrx}{$\log R_{\rm X}$}
\newcommand{\loglxlbol}{log($L_{\rm X}$/$L_{bol}$)}

\newcommand{\smw}{$S_{MW}$}

\newcommand{\kms}{km\,s$^{-1}$}

\newcommand{\ro}{$R_o$}
\newcommand{\tauc}{$\tau_c$}

\newcommand{\bvo}{$(B - V)_0$}
\newcommand{\bvsun}{$(B-V)_{\odot}$}
\newcommand{\ebv}{E$(B-V)$}

\slugcomment{accepted for publication in ApJ - 10 July 2008}
\shorttitle{Activity Ages}
\shortauthors{Mamajek \& Hillenbrand}
\begin{document}
\title{Improved Age Estimation for Solar-Type Dwarfs Using
Activity-Rotation Diagnostics}

\author{Eric E. Mamajek\altaffilmark{1,2}}
\email{emamajek@cfa.harvard.edu}
\author{Lynne A. Hillenbrand\altaffilmark{3}}
\email{lah@astro.caltech.edu}
\altaffiltext{1}{Harvard-Smithsonian Center for Astrophysics, 60 Garden St., MS-42, Cambridge, MA 02138, USA}
\altaffiltext{2}{Clay Postdoctoral Fellow}
\altaffiltext{3}{Astronomy/Astrophysics, California Institute of Technology, Pasadena, CA 91125, USA}

\begin{abstract} While the strong anti-correlation between
chromospheric activity and age has led to the common use of the Ca II
H \& K emission index ($R^\prime_{\rm HK} = L_{HK}/L_{bol}$) as an
empirical age estimator for solar type dwarfs, existing activity-age
relations produce implausible ages at both high and low activity
levels. We have compiled $R^\prime_{\rm HK}$ data from the literature
for young stellar clusters, richly populating for the first time the
young end of the activity-age relation. Combining the cluster activity
data with modern cluster age estimates, and analyzing the
color-dependence of the chromospheric activity age index, we derive an
improved activity-age calibration for F7-K2 dwarfs (0.5 $<$ B--V $<$
0.9 mag).  We also present a more fundamentally motivated activity-age
calibration that relies on conversion of $R^\prime_{\rm HK}$ values
through the Rossby number to rotation periods, and then makes use of
improved gyrochronology relations. We demonstrate that our new
activity-age calibration has typical age precision of $\sim$0.2 dex
for normal solar-type dwarfs aged between the Hyades and the Sun
($\sim$0.6-4.5 Gyr). Inferring ages through activity-rotation-age
relations accounts for some color-dependent effects, and
systematically improves the age estimates (albeit only slightly).  We
demonstrate that coronal activity as measured through the fractional
X-ray luminosity ($R_X$= L$_{X}$/L$_{bol}$) has nearly the same age-
and rotation-inferring capability as chromospheric activity measured
through $R^\prime_{\rm HK}$. As a first application of our
calibrations, we provide new activity-derived age estimates for the
nearest 100 solar-type field dwarfs ($d$ $<$ 15 pc).
\end{abstract}


\section{Introduction}

Age is, arguably, the most difficult basic stellar quantity to
estimate for low-mass field dwarfs \citep[see e.g.][]{Mamajek07}.
Yet, the temporal evolution of phenomena such as stellar activity,
surface abundances, rotation, and circumstellar matter is of current
interest and within observational means for nearby stars.  Our
particular motivation for improving field star age estimates stems
from our interest in circumstellar disk evolution as executed via the
{\it Spitzer Space Telescope} Formation and Evolution of Planetary
Systems (FEPS)\footnote{http://feps.as.arizona.edu} Legacy Science
program which is surveying the dust surrounding solar-type stars
between $\sim$3 Myr and $\sim$3 Gyr \citep{Meyer04,
Kim05, Stauffer05, Hines06, Meyer06, Silverstone06, Hines07,
Moro-Martin07, Bouwman08, Meyer08, Hillenbrand08, Carpenter09}.

The most theoretically grounded stellar age estimator is the
Hertzsprung-Russell diagram, which predicts ages based on our
understanding of nuclear physics, stellar interior structure, and
stellar atmospheres.  It can be employed in stellar clusters for which
main sequence turn-off and/or turn-on ages are typically available and
to field stars of known distance that are in the pre-MS or post-MS
phases of stellar evolution.  Field stars, however, are generally main
sequence objects and, by definition, lack co-eval accompanying stellar
populations that might enable accurate age dating via standard H-R
diagram techniques. Thus proxy indicators of age are necessary.

\subsection{Chromospheric Activity as an Age Indicator}

Historically, a popular age estimator for field stars of roughly solar
mass has been the \rphk\, index which measures chromospheric emission
in the cores of the broad photospheric Ca II H \& K absorption lines,
normalized to the underlying photospheric spectrum.  Chromospheric
activity is generated through the stellar magnetic dynamo,
the strength of which appears to scale with rotation velocity
\citep{Kraft67, Noyes84, Montesinos01}.  Both chromospheric emission
and rotation are observationally constrained to decay with age
\citep{Wilson63, Skumanich72, Soderblom83, Soderblom91}. The angular
momentum loss is theoretically understood as due to mass loss in a
magnetized wind \citep{Schatzman62, Weber67, Mestel68}.

The chromospheric activity index \rphk\, is calculated from a
band-ratio measurement of the Ca H \& K emission line strength
\citep[the ``S-index'' or, when converted to the Mount Wilson system,
\smw;][]{Vaughan78,Vaughan80,Duncan91} from which the underlying
stellar photospheric contribution is then subtracted.  We refer the
reader to papers by \citet[][and references therein]{Noyes84,
Baliunas96, Baliunas95, Henry96, Wright04} for in-depth discussion of
how to measure \smw\ and \rphk, as well as the history of studies
using this index.  Our simple goal for this study is to provide an
\rphk\, vs. age relation applicable to sets of \rphk\, and \bvo\ data
(the latter derived from a spectral type or from a color) for
solar-type and near-solar metallicity dwarfs.

The activity-age data pair of highest quality is that for the Sun, and
our adopted values are listed in Table \ref{tab:sun}. The solar age is
presumed coincident with that of the oldest portions of meteorites
\citep[the Ca-Al-rich inclusions; 4.570 Gyr;][]{Baker05}. However, the
Sun and presumably most other stars exhibit activity cycles (with
period 11 years in the case of the Sun) as well as longer term
variations (e.g. the so-called Maunder minimum in the case of the
Sun).  Over the period 1966-1993, covering mostly solar cycles 20, 21,
and 22, \citet{Baliunas95} estimated the solar Mt. Wilson S-index to
be $\overline{S_{\odot}}$ = 0.179.  Over the period 1994-2006, mostly
solar cycle 23, \citet{Hall07} measured $\overline{S_{\odot}}$ =
0.170.  Using a mean solar $S$-value which is approximately weighted
by the span of measurements ($\overline{S_{\odot}}$ = 0.176; for
$\sim$1966-2006) and a mean solar color of \bv\, = 0.65 \citep{Cox00},
and using the equations from \citet{Noyes84}, we estimate the mean
solar activity to be \logrphk\, = --4.91.  We also give in Table
\ref{tab:sun} the 68\% and 95\% range of the observed solar \logrphk\,
due to variability.

\begin{deluxetable}{lcccl}
\setlength{\tabcolsep}{0.03in}
\tablewidth{0pt}
\tablecaption{Adopted Solar Data \label{tab:sun}}
\tablehead{
{(1)}        &{(2)}   & {(3)}   & {(4)}  \\
{Parameter}  &{Value} & {Units} & {Ref.}}
\startdata
\bvo\,                                    & 0.65   & mag & 1 \\
Age                                       & 4.570  & Gyr & 2 \\
$\overline{{\rm S}_{\odot}}$              & 0.176  & ... & 3 \\
$\overline{\log {\rm R}^\prime_{\rm HK}}$ & -4.906 & dex & 4 \\
\logrphk\, 68\%\, range      & -4.942 to -4.865 & dex & 5 \\
\logrphk\, 95\%\, range      & -4.955 to -4.832 & dex & 5 \\
\loglx                       & 27.35            & erg\,s$^{-1}$ & 6\\
\logrx (=\loglxlbol)         & -6.24            & dex & 6 \\
\enddata
\tablecomments{
References: (1) \citet{Cox00}, (2) minimum age from \citet{Baker05},
(3) time-weighted average of \citet{Baliunas96} and \citet{Hall07} for
1966-2006, (4) calculated using \bvo\, and mean S$_{\odot}$ via
\citet{Noyes84}, (5) calculating using solar 1$\AA$ K-index data from
\citet{Livingston07}, using relations from \citet{Radick98} and
\citet{Noyes84}, and adopting the solar \bvo\, color listed, (6)
soft X-ray (0.1-2.4 keV) luminosity and fraction luminosity 
estimated from \citet{Judge03}, with 50\% uncertainty. 
An uncertainty in the solar \bvo\, of $\pm$0.01 mag produces
a systematic uncertainty of the \logrphk\, values by $\mp$0.004 dex.
Note that the absolute calibration of the \logrphk\, values (as a physical
metric of chromospheric line losses) are probably only accurate to
$\sim$10\% \citep{Hartmann84,Noyes84}.}
\end{deluxetable}

\subsection{Shortcomings of Previous Activity-Age Calibrations}

Using the Sun as one anchor point, we can look to open clusters with
ages derived from other methods (e.g. the H-R diagram) in order to
populate an activity-age calibration.  There are four such \rphk\,
vs. age relations in the literature which have been used in age-dating
field stars: two from \citet{Soderblom91}, and one each from
\citet{Donahue93}, and \citet{Lachaume99}.  The activity-age relations
from \citet{Soderblom91} include a linear fit to age vs. activity for
members of clusters and binaries.  The second relation, often
overlooked, assumes a constant star-formation history and takes into
account kinematic disk heating. D. Soderblom (priv. comm.) has kindly
provided an analytic version of this alternative activity-age
relation.

That there are deficiencies with these existing calibrations can be
easily demonstrated.  For the \citet{Lachaume99} calibration, the
solar \rphk\, value adopted here (-4.91) would imply a solar age of
7.2 Gyr, which is clearly in error. The other two calibrations used
the Sun as one of their anchor points, but with slightly different
\rphk\, values (for the calibrations of \citet{Soderblom91} and
\citet{Donahue93}, one derives ages of 4.1 and 4.0 Gyr, respectively).
\citet{Soderblom91} do not advocate extrapolating either of their
activity-age relations to the young/active regime (\logrphk\, $>$
-4.4), however \citet{Donahue93} explicitly fit their activity-age
relation to age $\sim$10 Myr and \logrphk\, $\simeq$ -4.2 (anchoring
their fit to data for NGC 2264).  Given the observed activity levels
in the $\sim$5-15 Myr Sco-Cen OB association (\logrphk\, $\simeq$
--4.05; \S2), neither the fit from \citet{Donahue93} or extrapolating
the two fits from \citet{Soderblom91} estimates an age similar to the
isochronal value. Indeed, the commonly used fit of \citet{Donahue93}
would estimate an age of {\it one minute} for a star with \logrphk\,
$\simeq$ --4.05.  Given the paucity of young stars in the previous
calibrations, we should not be too surprised at the lack of agreement
with other age-dating methods at the high-activity end of the
relation.

\subsection{Potential for Improved Activity-Age Calibrations}

Clearly, an improved activity-age calibration is needed. Further, we
would like to understand and quantify the limitations of any such
relationship and hence its practical application.  We focus this paper
primarily on {\it refining the age-activity relation for solar-type
dwarfs}.  By ``solar-type'', we mean $\sim$F7-K2 or 0.5 $<$ \bvo\ $<$
0.9 mag, which is approximately the color range over which the
\citet{Noyes84} relation for the photospheric contribution to the
S-index is applicable, as well as the color range blanketed by recent
activity surveys. The F3V-F6V temperature region (0.42 $<$ \bvo\, $<$
0.5) appears to mark the transition where the rotation-activity
correlation breaks down, chromospheric activity diminishes, stellar
convective envelopes thin, and magnetic breaking becomes inefficient
\citep{Kraft67,Wolff85,GarciaLopez93}. By ``dwarfs'', we mean MS and
pre-MS stars, and explicitly exclude evolved stars more than one
magnitude above the MS.

There are three developments that make our investigation timely:

(1) Recently measured \rphk\, values for stars that belong to
age-dated young stellar aggregates (e.g. Sco-Cen, $\beta$ Pic, etc.).
These additions to the literature both broaden and strengthen modern
activity-age derivations relative to the data landscape of 1-2 decades
ago.

(2) The ages of well-studied nearby open clusters (e.g. $\alpha$~Per,
Pleiades) have been updated during the past decade. The most
noticeable difference relative to traditionally accepted age values is
the systematic shift towards older ages driven by results using the
Li-depletion boundary age estimation method
\citep[e.g.][]{Stauffer98,Barrado04}.

(3) Interest in circumstellar disk and planetary system evolution has
increased dramatically over the past five years.  The availability of
relevant infrared data, e.g. from {\it Spitzer} observations, begs for
a robust stellar age estimator in order to probe the collisional and
radiative evolution of debris disks, and the connection of such
phenomena to exo-solar planetary system dynamics. Similarly, exoplanet
discoveries over the past decade have motivated interest in the ages
of the parent field stars for comparison to the Sun and solar system.
In this paper we derive using samples drawn from cluster and moving
group populations (\S2) a new \rphk\ activity vs. age relation (\S3).
In \S4, we tie both chromospheric activity index (\rphk) and coronal
activity index (\rx\, = \loglxlbol) data to stellar rotation rates via
the Rossby number (i.e. secure an activity-rotation relation), and
attempt to derive independently an activity-age relation based on the
``gyrochronology'' rotation evolution formalism of \citet{Barnes07},
though with newly derived coefficients.  In an Appendix, we quantify
the correlation between fractional X-ray luminosity and Ca H\&K
activity for solar-type stars, and demonstrate that \rx, like \rphk,
can be used to derive quantitative age estimates.

\section{Data \label{data}}

\subsection{Ca II H \& K Data \label{ca_data}}

We have collected \rphk\, indices derived from S-values in the
tradition of the Mt. Wilson HK project.  Typical errors for single
observations due to measurement uncertainty and calibration to the
standard system combine to typically $\sim$0.1 dex
\citep[e.g.][]{Henry96,Paulson02,White07}.  Given the ubiquity of the
\rphk\, index in the literature, and the uniformity in its calculation
and calibration by previous authors, we make no attempt either to
improve upon the \rphk\, index, nor to correct for other effects
(i.e. metallicity\footnote{The near-solar metallicity
\citep[r.m.s. $\simeq$ 0.1 dex in Fe/H;][]{Twarog97} of many of the
nearest young open clusters and stellar aggregates which anchor the
activity-age relation is well established.  This finding extends to T
Tauri stars in the nearest star-forming regions \citep{Padgett96}.
However, recent analysis of the California-Carnegie Planet Search
Project sample by J. Wright (private communication; 2009, in prep.)
suggests that there are metallicity effects which can bias \rphk, most
severely for stars older than the Sun.}, gravity, etc.).

\rphk\, values were taken from many sources, including the large
multi-epoch surveys of \citet{Duncan91}, \citet{Baliunas96},
\citet{Wright04}, and \citet{Hall07}, the large single-epoch surveys
of \citet{Henry96}, \citet{Gray03}, and \citet{Gray06}, the smaller,
focused surveys of \citet{Soderblom93}, \citet{Soderblom98},
\citet{Paulson02}, \citet{Tinney02}, \citet{Jenkins06,Jenkins08}, and
\citet{White07}.  The S-values from \citet{Duncan91} were converted to
\rphk\, following \citet{Noyes84} using \bv\, colors from
\citet{Perryman97}.  Discussion of the calibration of the HK
observations onto the Mt. Wilson system are addressed in the
individual studies.  Single-epoch surveys typically give consistent
\logrphk\, values that agree at the $\sim$0.1 dex r.m.s. level
\citep[e.g.][]{Jenkins08}, likely due to observational errors in
evaluating the S-index plus intrinsic stellar variability.

As solar-type stars undergo major changes in their interior structure
at the end of their main sequence lifetime, and \citet{Wright04} has
demonstrated that evolved stars show systematically lower activity
levels, we restrict our sample to stars that are consistent with being
main sequence stars \citep[here defined as being within $\Delta$M$_V$
of 1 magnitude of the main sequence defined by][]{Wright05}. We
specifically retain pre-MS stars, however, as we are interested in
probing the activity-age relation towards the youngest ages.

Although stellar rotation varies slowly with time, rotation-driven
stellar activity varies on much shorter time scales e.g. years, weeks,
and days.  This variability is also taken -- in and of itself -- as an
age indicator with more rapid, stochastic, and high-amplitude
variability indicative of younger stars while regularly periodic, long
cycle, and low-amplitude variability characterizes older stars
\citep[e.g.][]{Radick95,Radick98,Hempelmann96,Baliunas98}.
\citet{Lockwood07} and \citet{Hall07} also provide recent synopses.

The physical mechanisms producing such variability include changes in
the filling factor of emitting regions, growth and decay of individual
emitting regions, and short and long-term activity cycles.  For
example, in the Sun as well as in other stars, there is considerable
variation in the observable $S$ through an 11 year cycle, by 10\%
\citep{White81}.  In M67 a substantial fraction of the stars exhibit
even larger variations \citep{Giampapa06}.  Evidence from the
California Planet Search \citep[][Fischer \& Isaacson 2008, private
communication]{Wright04} shows that the bulk of the sample exhibit
variations of a few to $\sim$10\% in S at activity levels $-4.9 < $
\logrphk\ $< -4.4$ with less variation at lower activity levels,
$<$2\% in S at \logrphk\ $< -5.1$.  Within samples of presumably
co-eval cluster stars, there is similar evidence of scatter in
\logrphk\, values for a given color (as we illustrate for our sample
in \S3.1) which can be interpreted as a mix of high and low activity
levels about the mean level characteristic of the cluster age.
Estimated variations on time scales up to a few percent of the solar
age correspond to $\sim$0.15 in \logrphk.

In Table \ref{tab:sun} we list the 68\% and 95\% ranges for the solar
\logrphk\, value from 1977-2008 as estimated from the data of
\citet{Livingston07}. During recent solar maxima \logrphk\, $\simeq$
-4.83, and during recent solar minima \logrphk\, $\simeq$ -4.96.
Through extrapolation of the chromospheric activity-cycle length
relation, \citet{Baliunas95b} extrapolate the solar activity during
the Maunder minimum period ($\sim$1645-1715) to be roughly \logrphk\,
$\simeq$ -5.10.

All of this implies errors in ages which we could quantify if we
understood the probability that an individual measurement reflects the
mean activity level for that star.  For our sample, the \logrphk\,
data is a mix of long-term multi-epoch averages along with some
single/few-epoch observations.  Most of the X-ray data (discussed
next) is single epoch observations of length a few hundred seconds.
The evidence on variability suggests caution in age derivation for
stars lacking activity index monitoring of sufficient duration such
that mean activity levels can be determined.  Hence, we expect some
uncertainty in ages derived from activity levels to be due to
variability.

\subsection{Rotation and X-ray Data \label{Rotation_X-ray}}

To augment our understanding of the activity-age relation, we also
compiled data that allowed us to explore the more fundamental
rotation-age relation. We created a database of solar-type stars
having \logrphk\ with complimentary estimates of color, rotation
period, and when available, fractional X-ray luminosity (\loglxlbol\,
= \logrx). We started with the compiled catalog of
\citet{Pizzolato03}, and added stars from the FEPS program that had
new rotation periods measured by G. Henry (private comm.). We removed
stars from the \citet{Pizzolato03} sample which had periods inferred
from chromospheric activity levels as in \citet{Saar97}, i.e. we
retain only those rotation periods measured from the observed
modulation of starspots or chromospheric activity.

X-ray luminosities for sample stars were calculated using the 0.1-2.4
keV X-ray count rates and HR1 hardness ratios from the {\it ROSAT}
All-Sky Survey \citep{Voges99,Voges00}
\footnote{One can convert count-rates and fluxes between ROSAT and
other X-ray bands can using the PIMMS tool
(http://cxc.harvard.edu/toolkit/pimms.jsp). For a brief discussion
regarding converting ROSAT and Chandra fluxes, see
\citet{Preibisch05}.}. X-ray count rate $f_X$ (ct s$^{-1}$) can be
converted to X-ray flux (ergs\,cm$^{-2}$\,sec$^{-1}$) in the low
column density regime via a conversion factor ($C_X$) formula from
\citet{Fleming95}:

\begin{equation}
C_X = (8.31 + 5.30~{\rm HR1})~\times 10^{-12}~{\rm ergs~cm^{-2}~ct^{-1}}
\end{equation}

\noindent Combining the X-ray flux $f_X$ and conversion
factor $C_X$ with distance $D$, one can estimate the stellar X-ray
luminosity L$_X$ (erg\,s$^{-1}$):

\begin{equation}
L_X = 4~ \pi~ D^2~ C_X~ f_X
\end{equation}

The final conversion to X-ray and bolometric luminosities used
parallaxes, V-band photometry, and B-V colors from {\it Hipparcos}
\citep{Perryman97} and bolometric corrections from \citet{Kenyon95}.

Our rotation-activity sample consists of 167 MS and pre-MS stars of
near-solar color (0.5 $<$ \bv\, $<$ 0.9 mag) with measured periods and
\logrphk.  Of these, 166 have X-ray luminosities and \logrx\, values
that can be estimated.  The three lacking X-ray data are 
unsurprisingly inactive (\logrphk\, $<$ -5.0). While the primary
focus on this paper is on using chromospheric activity to gauge
stellar ages, we recognize that X-ray luminosities are calculable for
many more stars than those with published \logrphk\,
measurements. Hence, in Appendix A we quantify the correlation between
chromospheric and X-ray activity for solar-type dwarfs.

\subsection{Field Binaries \label{Data_binaries}}

Solar-type dwarf binaries are a useful sample for two reasons in the
present investigation: examining whether there is a color-dependence
of \logrphk\, vs. age, and gauging the precision of the age estimates
derived from activity. The coevality of stellar binary components at
the $<$1 Myr level is well-motivated observationally
\citep[e.g.][]{Hartigan94,Hartigan03}.  We list three useful samples
for the purposes of exploring the age-activity relation.

First, for exploring the color-dependence of \logrphk\, for a given
age, we identify 21 ``color-separated'' binary systems in the
literature with \rphk\, measurements that have (1) photospheric \bv\,
colors differing between the two components by $>$0.05 mag, and (2)
\bv\, color for each component between 0.45 and $\approx$0.9
\citep[where the photospheric correction to \rphk\ is well
characterized;][]{Noyes84}.  These systems are listed in Table
\ref{tab:pairs}. As our primary focus is on systems of near-solar
metallicity, we exclude two very metal poor systems from the analysis
(HD 23439AB and HD 134439/40, both with [Fe/H] $\simeq$ -1.0
\citep{Thevenin99}), although inclusion of the pair would have
negligible impact on our findings.

Second, in Table \ref{tab:twins} we list solar-type binaries that met
the color range criterion (0.45 $<$ \bvo\, $<$ 0.9), but whose
components had near-identical colors (|$\Delta$\bvo|\, $<$ 0.05),
i.e. ``twin'' binaries. We include these systems in our analysis of
gauging the accuracy to which activity-derived ages can be
estimated. Lastly, following \citet{Barnes07}, we also identify five
field binaries from the literature having measured rotation periods,
and list their properties in Table \ref{tab:bin_per}. A few have
\bvo\, colors beyond the range where \logrphk\, is well-defined (i.e.
\bvo\, $>$ 0.9), however we include them in our sample for the
purposes of assessing the accuracy of the rotation vs. age vs. color
relation discussed in \S\ref{Gyro}.

\clearpage
\begin{deluxetable}{llccccl}
\tabletypesize{\scriptsize}
\setlength{\tabcolsep}{0.01in}
\tablewidth{0pt}
\tablecaption{\logrphk\, for Color-Separated Solar-type Dwarf Binaries \label{tab:pairs}}
\tablehead{
{(1)}    &{(2)}      &{(3)} &{(4)} &{(5)}      &{(6)}     &{(7)}\\
{A}      &{B}        &{A}   &{B}   &{A}        &{B}       &{}\\
{Name}   &{Name}     &{\bv} &{\bv} &{\logrphk} &{\logrphk}&{Refs.} 
}
\startdata
HD 531B    & HD 531A    & 0.67 & 0.75 & -4.28 & -4.39 & 1,2\\
HD 5190    & HD 5208    & 0.52 & 0.68 & -4.96 & -5.13 & 1,3\\
HD 6872A   & HD 6872B   & 0.47 & 0.54 & -4.86 & -4.96 & 1,2\\
HD 7439    & HD 7438    & 0.45 & 0.81 & -4.75 & -4.67 & 1,2,4\\
HD 13357A  & HD 13357B  & 0.67 & 0.72 & -4.74 & -4.61 & 1,2*\\
HD 14082A  & HD 14082B  & 0.52 & 0.62 & -4.41 & -4.37 & 1,2\\
HD 26923   & HD 26913   & 0.57 & 0.68 & -4.50 & -4.39 & 1,5\\
HD 28255A  & HD 28255B  & 0.62 & 0.69 & -4.89 & -4.65 & 1,6\\
HD 53705   & HD 53706   & 0.62 & 0.78 & -4.93 & -5.01 & 1,3\\
HD 59099   & HD 59100   & 0.49 & 0.63 & -4.72 & -4.98 & 1,3*\\
HD 73668A  & HD 73668B  & 0.61 & 0.81 & -4.88 & -4.66 & 1,2\\
HD 103432  & HD 103431  & 0.71 & 0.76 & -4.82 & -4.73 & 1,2,7*\\
HD 116442  & HD 116443  & 0.78 & 0.87 & -4.94 & -4.94 & 1,2\\
HD 118576  & GJ 9455B   & 0.64 & 0.85 & -4.92 & -4.73 & 1,7\\
HD 128620  & HD 128621  & 0.63 & 0.84 & -5.00 & -4.92 & 3,8\\
HD 134331  & HD 134330  & 0.62 & 0.72 & -4.82 & -4.82 & 1,3\\
HD 135101A & HD 135101B & 0.68 & 0.74 & -5.11 & -5.01 & 1,2,4\\
HD 137763  & HD 137778  & 0.79 & 0.87 & -4.97 & -4.37 & 1,2\\
HD 142661  & HD 142661B & 0.55 & 0.81 & -4.94 & -4.58 & 1,4*\\
HD 144087  & HD 144088  & 0.75 & 0.85 & -4.66 & -4.60 & 1,2\\
HD 219175A & HD 219175B & 0.54 & 0.65 & -4.99 & -4.89 & 1,7*,2\\
\enddata
\tablecomments{References:
(1) \citet{Perryman97},
(2) \citet{Wright04},
(3) \citet{Henry96},
(4) \citet{Gray03},
(5) \citet{Baliunas96},
(6) \citet{Tinney02},
(7) \citet{Duncan91},
(8) \citet{Bessell81}
"*" implies that the published $S_{MW}$ value from the cited
survey was converted to \logrphk\, by the author following \citet{Noyes84}.
}
\end{deluxetable}

\begin{deluxetable}{llccccl}
\tabletypesize{\scriptsize}
\setlength{\tabcolsep}{0.03in}
\tablewidth{0pt}
\tablecaption{\logrphk\, for Near-Identical Solar-type Dwarf Binaries\label{tab:twins}}
\tablehead{
{(1)}    &{(2)}      &{(3)} &{(4)} &{(5)}      &{(6)}     &{(7)}\\
{A}      &{B}        &{A}   &{B}   &{A}        &{B}       &{}\\
{Name}   &{Name}     &{\bv} &{\bv} &{\logrphk} &{\logrphk}&{Refs.}}
\startdata
HD 9518A   & HD 9518B   & 0.53 & 0.54 & -5.12 & -5.00 & 1,2\\
HD 10361   & HD 10360   & 0.85 & 0.88 & -4.88 & -4.75 & 3,4\\
HD 20807   & HD 20766   & 0.60 & 0.64 & -4.79 & -4.65 & 1,4\\
HD 84612   & HD 84627   & 0.52 & 0.53 & -4.83 & -4.81 & 1,4\\
HD 92222A  & HD 92222B  & 0.59 & 0.59 & -4.44 & -4.51 & 2,5\\
HD 98745   & HD 98744   & 0.54 & 0.54 & -5.04 & -5.21 & 1,2\\
HD 103743  & HD 103742  & 0.64 & 0.67 & -4.81 & -4.83 & 1,4\\
HD 111484A & HD 111484B & 0.56 & 0.56 & -4.71 & -4.81 & 1,2\\
HD 145958A & HD 145958B & 0.76 & 0.80 & -4.94 & -4.94 & 1,2\\
HD 154195A & HD 154195B & 0.61 & 0.61 & -4.87 & -4.88 & 1,4*\\
HD 155886  & HD 155885  & 0.85 & 0.86 & -4.57 & -4.56 & 6,7,8*\\
HD 167216  & HD 167215  & 0.53 & 0.58 & -5.05 & -5.12 & 1,2\\
HD 179957  & HD 179958  & 0.64 & 0.64 & -5.05 & -5.08 & 2,3\\
HD 186408  & HD 186427  & 0.64 & 0.66 & -5.10 & -5.08 & 1,2\\
\enddata
\tablecomments{References:
(1) \citet{Perryman97},
(2) \citet{Wright04},
(3) \citet{Mermilliod91},
(4) \citet{Henry96},
(5) \bvo\, inferred from spectral type,
(6) \citet{Gliese91},
(7) \citet{Baliunas96},
(8) \citet{Baliunas95}.
"*" implies that the published $S_{MW}$ value from the cited survey
was recalculated to \logrphk\, by the author using the color listed
and following \citet{Noyes84}.} 
\end{deluxetable}

\clearpage
\begin{deluxetable}{llccccl}
\setlength{\tabcolsep}{0.03in}
\tablewidth{0pt}
\tablecaption{Field Binaries With Rotation Periods\label{tab:bin_per}}
\tablehead{
{(1)}    &{(2)} &{(3)} &{(4)} &{(5)}   &{(6)}   &{(7)}\\
{A}      &{B}   &{A}   &{B}   &{A}     &{B}     &{}\\
{Name}   &{Name}&{\bv} &{\bv} &{Per(d)}&{Per(d)}&{Refs.}}
\startdata	   			   
HD 131156A & HD 131156B & 0.73 & 1.16 &  6.31  & 11.94  & 1,2\\
HD 128620  & HD 128621  & 0.63 & 0.84 & 25.6   & 36.9   & 3,4,5,6\\
HD 155886  & HD 155885  & 0.85 & 0.86 & 20.69  & 21.11  & 2,6\\
HD 201091  & HD 201092  & 1.07 & 1.31 & 35.37  & 37.84  & 1,2\\ 
HD 219834A & HD 219834B & 0.79 & 0.90 & 42     & 43     & 7,8
\enddata
\tablecomments{References:
(1) \citet{Perryman97},
(2) \citet{Donahue96},
(3) \citet{Bessell81},
(4) E. Guinan (priv. comm.),
(5) \citet{Jay97},
(6) \citet{Hallam91},
(7) \citet{Hoffleit91},
(8) \citet{Mermilliod91},
(9) \citet{Baliunas96}. The period for HD 128620 ($\alpha$ Cen A) is
a mean (25.6 days) from values given by E. Guinan (priv. comm.;
22\,$\pm$\,3 day) and Hallam et al. (1991; 28.8\,$\pm$\,2.5 days), and
is consistent within the constraints from $v$sin$i$ and $p$-mode
rotational splitting \citep{Fletcher06,Bazot07}.
}
\end{deluxetable}

\clearpage
\begin{deluxetable}{lllcrcrcrcrrl}
\tabletypesize{\scriptsize}
\setlength{\tabcolsep}{0.03in}
\tablewidth{0pt}
\tablecaption{Members of Stellar Aggregates with \logrphk\, Measurements \label{tab:cluster_ca}}
\tablehead{
{(1)} &{(2)}  &{(3)}  &{(4)}&{(5)} &{(6)} &{(7)} &{(8)} &{(9)} &{(10)}    &{(11)}     &{(12)}&{(13)}\\
{Name}&{Alias}&{Alias}&{\bv}&{Ref.}&{\ebv}&{Ref.}&{\bvo}&{Ref.}&{\logrphk}&{N$_{obs}$}&{Ref.}&{Group}\\
{}    &{}     &{}     &{mag}&{}    &{mag} &{}    &{mag} &{}    &{dex}     &{}         &{}    &{}
}
\startdata
TYC 6779-1372-1 & ScoPMS 5 & HD 142361  & 0.71 & 1 & 0.10 & 4,2 & 0.62 & 4,2 & -4.01 & 2 & 6 & US  \\
TYC 6793-501-1 & ScoPMS 60 & HD 146516  & 0.79 & 4 & 0.20 & 4,5 & 0.59 & 2,4 & -4.09 & 1 & 6 & US  \\
TYC 6215-184-1 & ScoPMS 214 &    ...     & 1.24 & 4 & 0.30 & 4,5 & 0.82 & 4,2 & -4.17 & 1 & 6 & US  \\
TYC 6785-476-1 & PZ99 J154106.7-265626 &    ...     & 0.92 & 8 & 0.50 & 7 & 0.74 & 2,7 & -3.88 & 1 & 6 & US  \\
TYC 6208-1543-1 & PZ99 J160158.2-200811 &    ...     & 1.10 & 1 & 0.30 & 7 & 0.68 & 2,7 & -3.92 & 1 & 6 & US  \\
2UCAC 22492947 & PZ99 J161329.3-231106 &    ...     & ... & ... & 0.60 & 5,7 & 0.86 & 2,7 & -4.28 & 1 & 6 & US  \\
TYC 6793-1406-1 & PZ99 J161618.0-233947 &    ...     & 0.64 & 1 & 0.40 & 5,7 & 0.74 & 2,7 & -4.07 & 1 & 6 & US  \\
TYC 6779-305-1 & V1149 Sco & HD 143006  & 0.75 & 1 & 0.07 & 1,2 & 0.68 & 2 & -4.05 & 1 & 6 & US  \\
TYC 6779-305-1 & V1149 Sco & HD 143006  & 0.75 & 1 & 0.07 & 1,2 & 0.68 & 2 & -4.03 & 4 & 3 & US  \\
HIP 84586  & V824 Ara & HD 155555  & 0.80 & 1 & 0.00 & 9 & 0.80 & 1,9 & -3.97 & ... & 10 & $\beta$ Pic  \\
HIP 92680  & PZ Tel & HD 174429  & 0.78 & 1 & 0.00 & 9 & 0.78 & 1,9 & -3.78 & 1 & 11 & $\beta$ Pic  \\
HIP 92680  & PZ Tel & HD 174429  & 0.78 & 1 & 0.00 & 9 & 0.78 & 1,9 & -3.84 & ... & 12 & $\beta$ Pic  \\
HIP 25486  & HR 1817 & HD 35850  & 0.55 & 1 & 0.00 & 9 & 0.55 & 1,9 & -4.08 & ... & 10 & $\beta$ Pic  \\
HIP 25486  & HR 1817 & HD 35850  & 0.55 & 1 & 0.00 & 9 & 0.55 & 1,9 & -4.22 & 5 & 3 & $\beta$ Pic  \\
HIP 25486  & HR 1817 & HD 35850  & 0.55 & 1 & 0.00 & 9 & 0.55 & 1,9 & -4.29 & 1 & 6 & $\beta$ Pic  \\
TYC 7310-2431 1 & MML 52 &    ...     & 0.97 & 14 & 0.05 & 14 & 0.62 & 2,14 & -4.12 & 2 & 6 & UCL  \\
TYC 7319-749 1 & MML 58 &    ...     & 0.88 & 13 & 0.14 & 14 & 0.81 & 2,14 & -4.20 & 2 & 6 & UCL  \\
TYC 7822-158 1 & MML 63 &    ...     & 0.87 & 13 & 0.23 & 14 & 0.80 & 2,14 & -4.02 & 1 & 6 & UCL  \\
HIP 76673 & MML 69 & HD 139498  & 0.75 & 1 & 0.09 & 14 & 0.68 & 2,14 & -4.04 & 1 & 6 & UCL  \\
TYC 7331-782 1 & MML 70 &    ...     & 0.95 & 14 & 0.15 & 14 & 0.82 & 2,14 & -4.06 & 1 & 6 & UCL  \\
TYC 7333-1260 1 & MML 74 & HD 143358  & 0.73 & 14 & 0.05 & 14 & 0.59 & 2,14 & -4.04 & 2 & 6 & UCL  \\
HIP 59764  & SAO 251810 & HD 106506  & 0.60 & 1 & 0.06 & 15 & 0.55 & 1,15 & -3.95 & 1 & 11 & LCC  \\
HIP 59764  & SAO 251810 & HD 106506  & 0.60 & 1 & 0.06 & 15 & 0.55 & 1,15 & -3.97 & ... & 12 & LCC  \\
HIP 66941  & SAO 252423 & HD 119022  & 0.74 & 1 & 0.12 & 14 & 0.62 & 1,14 & -4.03 & ... & 11 & LCC  \\
HIP 66941  & SAO 252423 & HD 119022  & 0.74 & 1 & 0.12 & 14 & 0.62 & 1,14 & -4.06 & ... & 12 & LCC  \\
HIP 490  & SAO 214961 & HD 105  & 0.59 & 1 & 0.00 & 9 & 0.59 & 1,9 & -4.36 & 1 & 11 & Tuc  \\
HIP 490  & SAO 214961 & HD 105  & 0.59 & 1 & 0.00 & 9 & 0.59 & 1,9 & -4.41 & 7 & 3 & Tuc  \\
HIP 1481  & SAO 248159 & HD 1466  & 0.54 & 1 & 0.00 & 9 & 0.67 & 1,9 & -4.36 & 1 & 11 & Tuc  \\
HIP 105388  & SAO 246975 & HD 202917  & 0.69 & 1 & 0.00 & 9 & 0.69 & 1,9 & -4.06 & 1 & 11 & Tuc  \\
HIP 105388  & SAO 246975 & HD 202917  & 0.69 & 1 & 0.00 & 9 & 0.69 & 1,9 & -4.09 & ... & 12 & Tuc  \\
HIP 105388  & SAO 246975 & HD 202917  & 0.69 & 1 & 0.00 & 9 & 0.69 & 1,9 & -4.22 & 4 & 3 & Tuc  \\
HIP 116748A  & DS Tuc A & HD 222259A  & 0.68 & 1 & 0.00 & 9 & 0.68 & 1,9 & -4.00 & ... & 12 & Tuc  \\
HIP 116748A  & DS Tuc A & HD 222259A  & 0.68 & 1 & 0.00 & 9 & 0.68 & 1,9 & -4.09 & 1 & 11 & Tuc  \\
TYC 3319-306-1 & Cl Melotte 20 350 &    ...     & 0.69 & 19 & 0.10 & 17 & 0.60 & 17,19 & -4.04 & 1 & 6 & $\alpha$ Per  \\
TYC 3319-306-1 & Cl Melotte 20 350 &    ...     & 0.69 & 19 & 0.10 & 17 & 0.60 & 17,19 & -4.21 & 1 & 6 & $\alpha$ Per  \\
TYC 3315-1080-1 & Cl Melotte 20 373 &    ...     & 0.77 & 20 & 0.10 & 17 & 0.67 & 20,17 & -4.04 & 2 & 6 & $\alpha$ Per  \\
TYC 3319-589-1 & Cl Melotte 20 389 &    ...     & 0.67 & 16 & 0.10 & 17 & 0.57 & 16,17 & -4.53 & 1 & 6 & $\alpha$ Per  \\
TYC 3320-1283-1 & Cl Melotte 20 622 &    ...     & 0.82 & 19 & 0.10 & 17 & 0.72 & 17,19 & -3.78 & 1 & 6 & $\alpha$ Per  \\
2UCAC 47964793 & Cl Melotte 20 696 &    ...     & 0.74 & 19 & 0.10 & 17 & 0.64 & 17,19 & -4.21 & 1 & 6 & $\alpha$ Per  \\
TYC 3320-545-1 & Cl Melotte 20 699 &    ...     & 0.70 & 16 & 0.10 & 17 & 0.60 & 16,17 & -4.05 & 2 & 6 & $\alpha$ Per  \\
TYC 3320-423-1 & Cl Melotte 20 750 &    ...     & 0.59 & 19 & 0.10 & 17 & 0.49 & 17,19 & -4.80 & 1 & 6 & $\alpha$ Per  \\
TYC 3320-2239-1 & Cl Melotte 20 767 &    ...     & 0.61 & 19 & 0.10 & 17 & 0.52 & 17,19 & -4.62 & 2 & 6 & $\alpha$ Per  \\
TYC 3320-583-1 & Cl Melotte 20 935 &    ...     & 0.63 & 19 & 0.10 & 17 & 0.53 & 17,19 & -4.16 & 1 & 6 & $\alpha$ Per  \\
TYC 3321-1655-1 & Cl Melotte 20 1101 &    ...     & 0.69 & 20 & 0.10 & 17 & 0.59 & 17,19 & -4.00 & 1 & 6 & $\alpha$ Per  \\
TYC 3325-753-1 & Cl Melotte 20 1234 &    ...     & 0.72 & 16 & 0.10 & 17 & 0.62 & 16,17 & -4.53 & 1 & 6 & $\alpha$ Per  \\
2UCAC 47800056 & Cl* Melotte 20 AP 93 &    ...     & 0.94 & 18 & 0.10 & 17 & 0.84 & 17,18 & -4.05 & 1 & 6 & $\alpha$ Per  \\
TYC 1799-118-1 & Cl Melotte 22 102 &    ...     & 0.72 & 19 & 0.04 & 19,21 & 0.68 & 21 & -4.45 & 1 & 6 & Pleiades  \\
TYC 1799-118-1 & Cl Melotte 22 102 &    ...     & 0.72 & 19 & 0.04 & 19,21 & 0.68 & 21 & -4.48 & 1 & 6 & Pleiades  \\
TYC 1799-102-1 & Cl Melotte 22 120 &    ...     & 0.71 & 19 & 0.04 & 19,21 & 0.67 & 21 & -4.35 & 1 & 6 & Pleiades  \\
TYC 1799-1268-1 & Cl Melotte 22 129 &    ...    & 0.88 & 19 & 0.05 & 19,21 & 0.83 & 21 & -4.27 & 1 & 23 & Pleiades  \\
TYC 1799-1037-1 & Cl Melotte 22 164 & HD 23158  & 0.49 & 19 & 0.03 & 19,21 & 0.46 & 21 & -4.33 & 2 & 23 & Pleiades  \\
TYC 1803-1351-1 & Cl Melotte 22 173 &    ...     & 0.85 & 19 & 0.04 & 19,21 & 0.81 & 21 & -4.20 & 1 & 6 & Pleiades  \\
TYC 1803-8-1 & Cl Melotte 22 174 &    ...     & 0.85 & 19 & 0.04 & 19,21 & 0.81 & 21 & -3.48 & 1 & 6 & Pleiades  \\
TYC 1799-1224-1 & Cl Melotte 22 233 & HD 23195  & 0.53 & 19 & 0.03 & 19,21 & 0.49 & 21 & -4.72 & 2 & 23 & Pleiades  \\
TYC 1803-818-1 & Cl Melotte 22 250 &    ...     & 0.69 & 19 & 0.05 & 19,21 & 0.64 & 21 & -4.49 & 1 & 6 & Pleiades  \\
TYC 1799-963-1 & Cl Melotte 22 296 &    ...    & 0.84 & 19 & 0.04 & 19,21 & 0.80 & 21 & -3.90 & 1 & 21 & Pleiades  \\
TYC 1803-574-1 & Cl Melotte 22 314 &    ...     & 0.66 & 19 & 0.04 & 19,21 & 0.61 & 21 & -4.21 & 1 & 6 & Pleiades  \\
TYC 1803-542-1 & Cl Melotte 22 405 &    ...    & 0.54 & 19 & 0.04 & 19,21 & 0.49 & 21 & -4.42 & 3 & 23 & Pleiades  \\
TYC 1803-808-1 & Cl Melotte 22 489 &    ...    & 0.63 & 19 & 0.10 & 19,21 & 0.53 & 21 & -3.94 & 2 & 23 & Pleiades  \\
TYC 1803-1061-1 & Cl Melotte 22 514 &    ...     & 0.70 & 19 & 0.04 & 19,21 & 0.66 & 21 & -4.34 & 1 & 6 & Pleiades  \\
TYC 1803-1156-1 & Cl Melotte 22 571 &    ...     & 0.78 & 19 & 0.03 & 19,21 & 0.75 & 21 & -4.40 & 1 & 6 & Pleiades  \\
GSC 1799-960 & Cl Melotte 22 625 &    ...    & 1.17 & 19 & 0.36 & 19,21 & 0.82 & 21 & -3.85 & 1 & 21 & Pleiades  \\
TYC 1799-974-1 & Cl Melotte 22 708 &    ...    & 0.61 & 19 & 0.03 & 19,21 & 0.58 & 21 & -3.88 & 2 & 23 & Pleiades  \\
TYC 1803-156-1 & Cl Melotte 22 727 &    ...    & 0.55 & 19 & 0.03 & 19,21 & 0.52 & 21 & -3.78 & 4 & 23 & Pleiades  \\
TYC 1803-944-1 & Cl Melotte 22 739 &    ...    & 0.62 & 19 & 0.04 & 19,21 & 0.59 & 21 & -3.97 & 1 & 21 & Pleiades  \\
TYC 1799-978-1 & Cl Melotte 22 745 & HD 282969  & 0.52 & 19 & 0.03 & 19,21 & 0.50 & 21 & -4.43 & 1 & 23 & Pleiades  \\
TYC 1800-1917-1 & Cl Melotte 22 923 &    ...    & 0.62 & 19 & 0.04 & 19,21 & 0.58 & 21 & -4.23 & 2 & 23 & Pleiades  \\
TYC 1804-2129-1 & Cl Melotte 22 996 &    ...    & 0.65 & 19 & 0.04 & 19,21 & 0.60 & 21 & -4.25 & 3 & 23 & Pleiades  \\
TYC 1804-2366-1 & Cl Melotte 22 1015 &    ...     & 0.65 & 19 & 0.04 & 19,21 & 0.61 & 21 & -4.55 & 1 & 6 & Pleiades  \\
TYC 1800-1620-1 & Cl Melotte 22 1117 &    ...    & 0.72 & 19 & 0.04 & 19,21 & 0.68 & 21 & -4.57 & 2 & 23 & Pleiades  \\
TYC 1800-1774-1 & Cl Melotte 22 1182 &    ...     & 0.64 & 19 & 0.04 & 19,21 & 0.60 & 21 & -4.44 & 1 & 6 & Pleiades  \\
TYC 1800-1627-1 & Cl Melotte 22 1200 &    ...     & 0.54 & 19 & 0.03 & 19,21 & 0.51 & 21 & -4.68 & 1 & 6 & Pleiades  \\
TYC 1804-2205-1 & Cl Melotte 22 1207 &    ...    & 0.63 & 19 & 0.04 & 19,21 & 0.59 & 21 & -4.29 & 1 & 23 & Pleiades  \\
TYC 1800-1616-1 & Cl Melotte 22 1215 &    ...    & 0.64 & 19 & 0.04 & 19,21 & 0.60 & 21 & -4.26 & 1 & 23 & Pleiades  \\
TYC 1800-1683-1 & Cl Melotte 22 1613 &    ...    & 0.54 & 19 & 0.05 & 19,21 & 0.49 & 21 & -4.42 & 1 & 23 & Pleiades  \\
TYC 1800-1632-1 & Cl Melotte 22 1726 & HD 23713  & 0.54 & 19 & 0.04 & 19,21 & 0.51 & 21 & -4.44 & 2 & 23 & Pleiades  \\
TYC 1804-2140-1 & Cl Melotte 22 1776 & HD 282958  & 0.72 & 19 & 0.04 & 19,21 & 0.68 & 21 & -4.07 & 1 & 21 & Pleiades  \\
TYC 1804-2140-1 & Cl Melotte 22 1776 &    ...     & 0.72 & 19 & 0.04 & 19,21 & 0.68 & 21 & -4.30 & 1 & 6 & Pleiades  \\
TYC 1800-1852-1 & Cl Melotte 22 1797 &    ...    & 0.56 & 19 & 0.04 & 19,21 & 0.52 & 21 & -4.36 & 1 & 23 & Pleiades  \\
TYC 1800-1716-1 & Cl Melotte 22 1856 &    ...    & 0.56 & 19 & 0.04 & 19,21 & 0.51 & 21 & -4.39 & 1 & 23 & Pleiades  \\
2UCAC 40300217 & Cl Melotte 22 2027 &    ...    & 0.86 & 19 & 0.04 & 19,21 & 0.82 & 21 & -4.71 & 1 & 23 & Pleiades  \\
2UCAC 39967447 & Cl Melotte 22 2106 &    ...    & 0.86 & 19 & 0.04 & 19,21 & 0.82 & 21 & -4.19 & 2 & 23 & Pleiades  \\
2UCAC 39967447 & Cl Melotte 22 2106 &    ...     & 0.86 & 19 & 0.04 & 19,21 & 0.82 & 21 & -3.94 & 1 & 6 & Pleiades  \\
2UCAC 39967452 & Cl Melotte 22 2126 &    ...    & 0.85 & 19 & 0.04 & 19,21 & 0.81 & 21 & -4.14 & 2 & 23 & Pleiades  \\
2UCAC 39967452 & Cl Melotte 22 2126 &    ...    & 0.85 & 19 & 0.04 & 19,21 & 0.81 & 21 & -4.16 & 1 & 21 & Pleiades  \\
TYC 1800-1091-1 & Cl Melotte 22 2147 &    ...    & 0.81 & 19 & 0.03 & 19,21 & 0.78 & 21 & -4.11 & 2 & 23 & Pleiades  \\
TYC 1800-1091-1 & Cl Melotte 22 2147 &    ...     & 0.81 & 19 & 0.03 & 19,21 & 0.78 & 21 & -3.94 & 1 & 6 & Pleiades  \\
TYC 1804-1179-1 & Cl Melotte 22 2278 &    ...     & 0.87 & 19 & 0.04 & 19,21 & 0.83 & 21 & -4.19 & 1 & 6 & Pleiades  \\
TYC 1800-471-1 & Cl Melotte 22 2506 &    ...     & 0.60 & 19 & 0.05 & 19,21 & 0.55 & 21 & -4.43 & 1 & 6 & Pleiades  \\
TYC 1804-305-1 & Cl Melotte 22 2644 &    ...     & 0.74 & 19 & 0.04 & 22 & 0.70 & 19,22 & -4.42 & 1 & 6 & Pleiades  \\
TYC 1800-1526-1 & Cl Melotte 22 2786 &    ...     & 0.61 & 19 & 0.04 & 19,21 & 0.56 & 21 & -4.38 & 1 & 6 & Pleiades  \\
TYC 1804-1400-1 & Cl Melotte 22 3097 &    ...     & 0.74 & 19 & 0.04 & 19,21 & 0.70 & 21 & -4.23 & 1 & 6 & Pleiades  \\
TYC 1804-1400-1 & Cl Melotte 22 3097 &    ...     & 0.74 & 19 & 0.04 & 19,21 & 0.70 & 21 & -4.29 & 1 & 6 & Pleiades  \\
TYC 1800-1415-1 & Cl Melotte 22 3179 &    ...     & 0.57 & 19 & 0.03 & 19,21 & 0.53 & 21 & -4.55 & 1 & 6 & Pleiades  \\
TYC 1813-126-1 & Cl* Melotte 22 PELS 191 &    ...     & 0.71 & 1 & 0.04 & 1,21 & 0.67 & 21 & -4.38 & 1 & 6 & Pleiades  \\
HIP 13806 & Cl Melotte 25 153 &    ...    & 0.85 & 1 & 0.00 & 24 & 0.85 & 1,24 & -4.38 & 18 & 25 & Hyades  \\
HIP 14976 & SAO 56256 & HD 19902  & 0.73 & 1 & 0.00 & 24 & 0.73 & 1,24 & -4.57 & 10 & 25 & Hyades  \\
HIP 14976 & SAO 56256 & HD 19902  & 0.73 & 1 & 0.00 & 24 & 0.73 & 1,24 & -4.60 & 1 & 3 & Hyades  \\
HIP 15310 & Cl Melotte 25 2 & HD 20439  & 0.62 & 1 & 0.00 & 24 & 0.62 & 1,24 & -4.49 & 3 & 23 & Hyades  \\
HIP 15310 & Cl Melotte 25 2 & HD 20439  & 0.62 & 1 & 0.00 & 24 & 0.62 & 1,24 & -4.54 & 13 & 25 & Hyades  \\
HIP 16529 & Cl Melotte 25 4 &    ...    & 0.84 & 1 & 0.00 & 24 & 0.84 & 1,24 & -4.37 & 9 & 25 & Hyades  \\
HIP 18327 & Cl Melotte 25 7 & HD 258252  & 0.90 & 1 & 0.00 & 24 & 0.90 & 1,24 & -4.36 & 8 & 25 & Hyades  \\
HIP 19098 & Cl Melotte 25 228 & HD 285367  & 0.89 & 1 & 0.00 & 24 & 0.89 & 1,24 & -4.39 & 8 & 25 & Hyades  \\
HIP 19148 & Cl Melotte 25 10 & HD 25825  & 0.59 & 1 & 0.00 & 24 & 0.59 & 1,24 & -4.47 & 8 & 25 & Hyades  \\
HIP 19148 & Cl Melotte 25 10 & HD 25825  & 0.59 & 1 & 0.00 & 24 & 0.59 & 1,24 & -4.48 & 13 & 3 & Hyades  \\
HIP 19148 & Cl Melotte 25 10 & HD 25825  & 0.59 & 1 & 0.00 & 24 & 0.59 & 1,24 & -4.57 & 3 & 23 & Hyades  \\
HIP 19261 B & Cl Melotte 25 12 & HD 26015B  & 0.65 & 1 & 0.00 & 24 & 0.65 & 1,24 & -4.28 & 8 & 25 & Hyades  \\
HIP 19781 & Cl Melotte 25 17 & HD 26756  & 0.69 & 1 & 0.00 & 24 & 0.69 & 1,24 & -4.42 & 21 & 25 & Hyades  \\
HIP 19781 & Cl Melotte 25 17 & HD 26756  & 0.69 & 1 & 0.00 & 24 & 0.69 & 1,24 & -4.47 & 27 & 23 & Hyades  \\
HIP 19786 & Cl Melotte 25 18 & HD 26767  & 0.64 & 1 & 0.00 & 24 & 0.64 & 1,24 & -4.39 & 2 & 23 & Hyades  \\
HIP 19786 & Cl Melotte 25 18 & HD 26767  & 0.64 & 1 & 0.00 & 24 & 0.64 & 1,24 & -4.44 & 15 & 25 & Hyades  \\
HIP 19786 & Cl Melotte 25 18 & HD 26767  & 0.64 & 1 & 0.00 & 24 & 0.64 & 1,24 & -4.48 & 9 & 3 & Hyades  \\
HIP 19793 & Cl Melotte 25 15 & HD 26736  & 0.66 & 1 & 0.00 & 24 & 0.66 & 1,24 & -4.42 & 24 & 23 & Hyades  \\
HIP 19793 & Cl Melotte 25 15 & HD 26736  & 0.66 & 1 & 0.00 & 24 & 0.66 & 1,24 & -4.42 & 17 & 25 & Hyades  \\
HIP 19796 & Cl Melotte 25 19 & HD 26784  & 0.51 & 1 & 0.00 & 24 & 0.51 & 1,24 & -4.49 & 1 & 23 & Hyades  \\
HIP 19796 & Cl Melotte 25 19 & HD 26784  & 0.51 & 1 & 0.00 & 24 & 0.51 & 1,24 & -4.54 & 11 & 25 & Hyades  \\
HIP 20130 & Cl Melotte 25 26 & HD 27250  & 0.74 & 1 & 0.00 & 24 & 0.74 & 1,24 & -4.45 & 8 & 25 & Hyades  \\
HIP 20130 & Cl Melotte 25 26 & HD 27250  & 0.74 & 1 & 0.00 & 24 & 0.74 & 1,24 & -4.47 & 12 & 23 & Hyades  \\
HIP 20146 & Cl Melotte 25 27 & HD 27282  & 0.72 & 1 & 0.00 & 24 & 0.72 & 1,24 & -4.45 & 45 & 23 & Hyades  \\
HIP 20146 & Cl Melotte 25 27 & HD 27282  & 0.72 & 1 & 0.00 & 24 & 0.72 & 1,24 & -4.46 & 9 & 25 & Hyades  \\
HIP 20237 & Cl Melotte 25 31 & HD 27406  & 0.56 & 1 & 0.00 & 24 & 0.56 & 1,24 & -4.45 & 8 & 25 & Hyades  \\
HIP 20237 & Cl Melotte 25 31 & HD 27406  & 0.56 & 1 & 0.00 & 24 & 0.56 & 1,24 & -4.48 & 161 & 23 & Hyades  \\
HIP 20480 & Cl Melotte 25 42 & HD 27732  & 0.76 & 1 & 0.00 & 24 & 0.76 & 1,24 & -4.46 & 8 & 25 & Hyades  \\
HIP 20480 & Cl Melotte 25 42 & HD 27732  & 0.76 & 1 & 0.00 & 24 & 0.76 & 1,24 & -4.48 & 10 & 23 & Hyades  \\
HIP 20492 & Cl Melotte 25 46 & HD 27771  & 0.85 & 1 & 0.00 & 24 & 0.85 & 1,24 & -4.39 & 8 & 25 & Hyades  \\
HIP 20492 & Cl Melotte 25 46 & HD 27771  & 0.85 & 1 & 0.00 & 24 & 0.85 & 1,24 & -4.81 & 1 & 23 & Hyades  \\
HIP 20557 & Cl Melotte 25 48 & HD 27808  & 0.52 & 1 & 0.00 & 24 & 0.52 & 1,24 & -4.50 & 8 & 25 & Hyades  \\
HIP 20557 & Cl Melotte 25 48 & HD 27808  & 0.52 & 1 & 0.00 & 24 & 0.52 & 1,24 & -4.52 & 183 & 23 & Hyades  \\
HIP 20577 & Cl Melotte 25 52 & HD 27859  & 0.60 & 1 & 0.00 & 24 & 0.60 & 1,24 & -4.45 & 102 & 23 & Hyades  \\
HIP 20577 & Cl Melotte 25 52 & HD 27859  & 0.60 & 1 & 0.00 & 24 & 0.60 & 1,24 & -4.47 & 1 & 6 & Hyades  \\
HIP 20577 & Cl Melotte 25 52 & HD 27859  & 0.60 & 1 & 0.00 & 24 & 0.60 & 1,24 & -4.47 & 9 & 25 & Hyades  \\
HIP 20741 & Cl Melotte 25 64 & HD 20899  & 0.66 & 1 & 0.00 & 24 & 0.66 & 1,24 & -4.47 & 9 & 25 & Hyades  \\
HIP 20741 & Cl Melotte 25 64 & HD 28099  & 0.66 & 1 & 0.00 & 24 & 0.66 & 1,24 & -4.50 & 81 & 23 & Hyades  \\
HIP 20741 & Cl Melotte 25 64 & HD 28099  & 0.66 & 1 & 0.00 & 24 & 0.66 & 1,24 & -4.62 & 1 & 6 & Hyades  \\
HIP 20815 & Cl Melotte 25 65 & HD 28205  & 0.54 & 1 & 0.00 & 24 & 0.54 & 1,24 & -4.58 & 8 & 25 & Hyades  \\
HIP 20815 & Cl Melotte 25 65 & HD 28205  & 0.54 & 1 & 0.00 & 24 & 0.54 & 1,24 & -4.60 & 144 & 23 & Hyades  \\
HIP 20826 & Cl Melotte 25 66 & HD 28237  & 0.56 & 1 & 0.00 & 24 & 0.56 & 1,24 & -4.46 & 8 & 25 & Hyades  \\
HIP 20826 & Cl Melotte 25 66 & HD 28237  & 0.56 & 1 & 0.00 & 24 & 0.56 & 1,24 & -4.46 & 2 & 23 & Hyades  \\
HIP 20826 & Cl Melotte 25 66 & HD 28237  & 0.56 & 1 & 0.00 & 24 & 0.56 & 1,24 & -4.48 & 5 & 3 & Hyades  \\
HIP 20826 & Cl Melotte 25 66 & HD 28237  & 0.56 & 1 & 0.00 & 24 & 0.56 & 1,24 & -4.55 & 1 & 6 & Hyades  \\
HIP 20850 & Cl Melotte 25 178 & HD 28258  & 0.84 & 1 & 0.00 & 24 & 0.84 & 1,24 & -4.43 & 9 & 25 & Hyades  \\
HIP 20899 & Cl Melotte 25 73 & HD 28344  & 0.61 & 1 & 0.00 & 24 & 0.61 & 1,24 & -4.44 & 33 & 23 & Hyades  \\
HIP 20899 & Cl Melotte 25 73 & HD 28344  & 0.61 & 1 & 0.00 & 24 & 0.61 & 1,24 & -4.46 & 7 & 3 & Hyades  \\
HIP 20899 & Cl Melotte 25 73 & HD 28344  & 0.61 & 1 & 0.00 & 24 & 0.61 & 1,24 & -4.50 & 10 & 25 & Hyades  \\
HIP 20899 & Cl Melotte 25 73 & HD 28344  & 0.61 & 1 & 0.00 & 24 & 0.61 & 1,24 & -4.59 & 1 & 6 & Hyades  \\
HIP 20951 & Cl Melotte 25 79 & HD 285733  & 0.83 & 1 & 0.00 & 24 & 0.83 & 1,24 & -4.44 & 12 & 23 & Hyades  \\
HIP 20951 & Cl Melotte 25 79 & HD 285733  & 0.83 & 1 & 0.00 & 24 & 0.83 & 1,24 & -4.52 & 1 & 6 & Hyades  \\
HIP 20951 & Cl Melotte 25 79 & HD 285773  & 0.83 & 1 & 0.00 & 24 & 0.83 & 1,24 & -4.44 & 9 & 25 & Hyades  \\
HIP 20978 & Cl Melotte 25 180 & HD 28462  & 0.86 & 1 & 0.00 & 24 & 0.86 & 1,24 & -4.27 & 9 & 23 & Hyades  \\
HIP 20978 & Cl Melotte 25 180 & HD 28462  & 0.86 & 1 & 0.00 & 24 & 0.86 & 1,24 & -4.29 & 1 & 6 & Hyades  \\
HIP 20978 & Cl Melotte 25 180 & HD 28462  & 0.86 & 1 & 0.00 & 24 & 0.86 & 1,24 & -4.41 & 7 & 25 & Hyades  \\
HIP 21099 & Cl Melotte 25 87 & HD 28593  & 0.73 & 1 & 0.00 & 24 & 0.73 & 1,24 & -4.46 & 21 & 23 & Hyades  \\
HIP 21099 & Cl Melotte 25 87 & HD 28593  & 0.73 & 1 & 0.00 & 24 & 0.73 & 1,24 & -4.48 & 9 & 25 & Hyades  \\
HIP 21112 & Cl Melotte 25 88 & HD 28635  & 0.54 & 1 & 0.00 & 24 & 0.54 & 1,24 & -4.39 & 1 & 6 & Hyades  \\
HIP 21112 & Cl Melotte 25 88 & HD 28635  & 0.54 & 1 & 0.00 & 24 & 0.54 & 1,24 & -4.56 & 10 & 25 & Hyades  \\
HIP 21112 & Cl Melotte 25 88 & HD 28635  & 0.54 & 1 & 0.00 & 24 & 0.54 & 1,24 & -4.56 & 33 & 23 & Hyades  \\
HIP 21317 & Cl Melotte 25 97 & HD 28892  & 0.63 & 1 & 0.00 & 24 & 0.63 & 1,24 & -4.45 & 8 & 25 & Hyades  \\
HIP 21317 & Cl Melotte 25 97 & HD 28992  & 0.63 & 1 & 0.00 & 24 & 0.63 & 1,24 & -4.48 & 60 & 23 & Hyades  \\
HIP 21317 & Cl Melotte 25 97 & HD 28992  & 0.63 & 1 & 0.00 & 24 & 0.63 & 1,24 & -4.52 & 1 & 6 & Hyades  \\
HIP 21637 & Cl Melotte 25 105 & HD 29419  & 0.58 & 1 & 0.00 & 24 & 0.58 & 1,24 & -4.52 & 8 & 25 & Hyades  \\
HIP 21654 & Cl Melotte 25 106 & HD 29461  & 0.66 & 1 & 0.00 & 24 & 0.66 & 1,24 & -4.55 & 15 & 3 & Hyades  \\
HIP 21654 & Cl Melotte 25 106 & HD 29461  & 0.66 & 1 & 0.00 & 24 & 0.66 & 1,24 & -4.58 & 1 & 6 & Hyades  \\
HIP 22203 & Cl Melotte 25 142 & HD 30246  & 0.67 & 1 & 0.00 & 24 & 0.66 & 1,24 & -4.63 & 1 & 6 & Hyades  \\
HIP 22422 & Cl Melotte 25 118 & HD 30589  & 0.58 & 1 & 0.00 & 24 & 0.58 & 1,24 & -4.75 & 1 & 23 & Hyades  \\
HIP 22422 & Cl Melotte 25 118 & HD 30589  & 0.58 & 1 & 0.00 & 24 & 0.58 & 1,24 & -4.82 & 10 & 25 & Hyades  \\
HIP 23069 & Cl Melotte 25 127 & HD 31609  & 0.74 & 1 & 0.00 & 24 & 0.74 & 1,24 & -4.45 & 7 & 25 & Hyades  \\
HIP 23498 & Cl Melotte 25 187 & HD 32347  & 0.77 & 1 & 0.00 & 24 & 0.77 & 1,24 & -4.44 & 7 & 25 & Hyades  \\
HIP 23750 & Cl* Melotte 25 S 140 & HD 240648  & 0.73 & 1 & 0.00 & 24 & 0.73 & 1,24 & -4.43 & 7 & 25 & Hyades  \\
TYC 1265-569-1 & Cl Melotte 25 49 & HD 27835  & 0.59 & 19 & 0.00 & 24 & 0.59 & 19,24 & -4.62 & 1 & 6 & Hyades  \\
TYC 1265-569-1 & Cl Melotte 25 49 & HD 27835  & 0.60 & 1 & 0.00 & 24 & 0.60 & 1,24 & -4.52 & 8 & 25 & Hyades  \\
TYC 1265-569-1 & Cl Melotte 25 49 & HD 27835  & 0.60 & 1 & 0.00 & 24 & 0.60 & 1,24 & -4.53 & 12 & 23 & Hyades  \\
TYC 1266-1012-1 & Cl Melotte 25 91 & HD 28783  & 0.88 & 19 & 0.00 & 24 & 0.88 & 19,24 & -4.47 & 3 & 23 & Hyades  \\
TYC 1266-1012-1 & Cl Melotte 25 91 & HD 28783  & 0.88 & 19 & 0.00 & 24 & 0.88 & 19,24 & -4.80 & 1 & 6 & Hyades  \\
TYC 1266-1175-1 & Cl Melotte 25 99 & HD 29159  & 0.87 & 19 & 0.00 & 24 & 0.87 & 19,24 & -4.38 & 8 & 25 & Hyades  \\
TYC 1266-1175-1 & Cl Melotte 25 99 & HD 29159  & 0.87 & 19 & 0.00 & 24 & 0.87 & 19,24 & -4.40 & 9 & 23 & Hyades  \\
TYC 1266-1175-1 & Cl Melotte 25 99 & HD 29159  & 0.87 & 19 & 0.00 & 24 & 0.87 & 19,24 & -4.69 & 1 & 6 & Hyades  \\
TYC 1266-1286-1 & Cl Melotte 25 92 & HD 28805  & 0.73 & 1 & 0.00 & 24 & 0.73 & 1,24 & -4.44 & 7 & 25 & Hyades  \\
TYC 1266-1286-1 & Cl Melotte 25 92 & HD 28805  & 0.74 & 19 & 0.00 & 24 & 0.74 & 19,24 & -4.45 & 18 & 23 & Hyades  \\
TYC 1266-1286-1 & Cl Melotte 25 92 & HD 28805  & 0.74 & 19 & 0.00 & 24 & 0.74 & 19,24 & -4.61 & 1 & 6 & Hyades  \\
TYC 1266-149-1 & Cl Melotte 25 93 & HD 28878  & 0.89 & 19 & 0.00 & 24 & 0.89 & 19,24 & -4.40 & 8 & 25 & Hyades  \\
TYC 1266-149-1 & Cl Melotte 25 93 & HD 28878  & 0.89 & 19 & 0.00 & 24 & 0.89 & 19,24 & -4.63 & 1 & 6 & Hyades  \\
HIP 8486 & GJ 9061B & HD 11131  & 0.65 & 1 & 0.00 & 9 & 0.65 & 1,9 & -4.47 & 4 & 11 & UMa  \\
HIP 8486 & GJ 9061B & HD 11131  & 0.65 & 1 & 0.00 & 9 & 0.65 & 1,9 & -4.52 & ... & 31 & UMa  \\
HIP 19859 & HR 1322 & HD 26923  & 0.57 & 1 & 0.00 & 9 & 0.57 & 1,9 & -4.55 & ... & 31 & UMa  \\
HIP 19859 & HR 1322 & HD 26923  & 0.57 & 1 & 0.00 & 9 & 0.57 & 1,9 & -4.52 & ... & 31 & UMa  \\
HIP 21276 & GJ 3295 & HD 28495  & 0.76 & 1 & 0.00 & 9 & 0.76 & 1,9 & -4.34 & 6 & 3 & UMa  \\
HIP 27072 & HR 1983 & HD 38393  & 0.48 & 1 & 0.00 & 9 & 0.48 & 1,9 & -4.77 & 3 & 11 & UMa  \\
HIP 27072 & HR 1983 & HD 38393  & 0.48 & 1 & 0.00 & 9 & 0.48 & 1,9 & -4.82 & ... & 31 & UMa  \\
HIP 27913 & HR 2047 & HD 39587  & 0.59 & 1 & 0.00 & 9 & 0.59 & 1,9 & -4.46 & ... & 31 & UMa  \\
HIP 27913 & HR 2047 & HD 39587  & 0.59 & 1 & 0.00 & 9 & 0.59 & 1,9 & -4.43 & ... & 31 & UMa  \\
HIP 36704 & HR 8883 & HD 59747  & 0.86 & 1 & 0.00 & 9 & 0.86 & 1,9 & -4.37 & 1 & 3 & UMa  \\
HIP 36704 & HR 8883 & HD 59747  & 0.86 & 1 & 0.00 & 9 & 0.86 & 1,9 & -4.46 & ... & 31 & UMa  \\
HIP 42438 & HR 3391 & HD 72905  & 0.62 & 1 & 0.00 & 9 & 0.62 & 1,9 & -4.40 & 3 & 3 & UMa  \\
HIP 42438 & HR 3391 & HD 72905  & 0.62 & 1 & 0.00 & 9 & 0.62 & 1,9 & -4.48 & 1 & 6 & UMa  \\
HIP 80686 & HR 6098 & HD 147584  & 0.56 & 1 & 0.00 & 9 & 0.56 & 1,9 & -4.56 & 1 & 11 & UMa  \\
HIP 80686 & HR 6098 & HD 147584  & 0.56 & 1 & 0.00 & 9 & 0.56 & 1,9 & -4.58 & ... & 31 & UMa  \\
HIP 88694 & HR 6748 & HD 165185  & 0.61 & 1 & 0.00 & 9 & 0.61 & 1,9 & -4.54 & ... & 31 & UMa  \\
HIP 115312 & HR 8883 & HD 220096  & 0.82 & 1 & 0.00 & 9 & 0.82 & 1,9 & -4.39 & 1 & 3 & UMa  \\
2UCAC 35931542 & Cl* NGC 2682 SAND 0603 &    ...     & 0.59 & 32 & 0.04 & 33 & 0.55 & 32,33 & -4.74 & ... & 34 & M67  \\
2UCAC 35931521 & Cl* NGC 2682 SAND 0621 &    ...     & 0.66 & 32 & 0.04 & 33 & 0.62 & 32,33 & -4.83 & ... & 34 & M67  \\
2UCAC 35931673 & Cl* NGC 2682 SAND 0724 &    ...     & 0.00 & ... & 0.04 & 33 & 0.63 & 34 & -4.86 & ... & 34 & M67  \\
2UCAC 35931593 & Cl* NGC 2682 SAND 0746 &    ...     & 0.71 & 32 & 0.04 & 33 & 0.67 & 32,33 & -4.85 & ... & 34 & M67  \\
2UCAC 35931670 & Cl* NGC 2682 SAND 0747 &    ...     & 0.70 & 32 & 0.04 & 33 & 0.67 & 32,33 & -4.47 & ... & 34 & M67  \\
2UCAC 35931634 & Cl* NGC 2682 SAND 0748 &    ...     & 0.83 & 32 & 0.04 & 33 & 0.79 & 32,33 & -4.75 & ... & 34 & M67  \\
2UCAC 35931585 & Cl* NGC 2682 SAND 0753 &    ...     & 0.63 & 32 & 0.04 & 33 & 0.59 & 32,33 & -4.77 & ... & 34 & M67  \\
2UCAC 35931642 & Cl* NGC 2682 SAND 0770 &    ...     & 0.68 & 32 & 0.04 & 33 & 0.64 & 32,33 & -4.88 & ... & 34 & M67  \\
2UCAC 35931570 & Cl* NGC 2682 SAND 0777 &    ...     & 0.67 & 32 & 0.04 & 33 & 0.64 & 32,33 & -4.82 & ... & 34 & M67  \\
2UCAC 35931615 & Cl* NGC 2682 SAND 0779 &    ...     & 0.69 & 32 & 0.04 & 33 & 0.65 & 32,33 & -4.94 & ... & 34 & M67  \\
2UCAC 35931637 & Cl* NGC 2682 SAND 0785 &    ...     & 0.70 & 32 & 0.04 & 33 & 0.66 & 32,33 & -4.79 & ... & 34 & M67  \\
2UCAC 35931665 & Cl* NGC 2682 SAND 0789 &    ...     & 0.66 & 32 & 0.04 & 33 & 0.62 & 32,33 & -4.82 & ... & 34 & M67  \\
2UCAC 35931671 & Cl* NGC 2682 SAND 0801 &    ...     & 0.72 & 32 & 0.04 & 33 & 0.68 & 32,33 & -4.98 & ... & 34 & M67  \\
2UCAC 35931641 & Cl* NGC 2682 SAND 0802 &    ...     & 0.72 & 32 & 0.04 & 33 & 0.68 & 32,33 & -4.92 & ... & 34 & M67  \\
2UCAC 35931626 & Cl* NGC 2682 SAND 0829 &    ...     & 0.63 & 32 & 0.04 & 33 & 0.59 & 32,33 & -4.88 & ... & 34 & M67  \\
2UCAC 35931675 & Cl* NGC 2682 SAND 0937 &    ...     & 0.59 & 32 & 0.04 & 33 & 0.55 & 32,33 & -4.84 & ... & 34 & M67  \\
2UCAC 35931686 & Cl* NGC 2682 SAND 0942 &    ...     & 0.63 & 32 & 0.04 & 33 & 0.59 & 32,33 & -4.78 & ... & 34 & M67  \\
2UCAC 35931848 & Cl* NGC 2682 SAND 0943 &    ...     & 0.76 & 32 & 0.04 & 33 & 0.72 & 32,33 & -5.08 & ... & 34 & M67  \\
2UCAC 35931810 & Cl* NGC 2682 SAND 0945 &    ...     & 0.67 & 32 & 0.04 & 33 & 0.63 & 32,33 & -4.83 & ... & 34 & M67  \\
2UCAC 35931701 & Cl* NGC 2682 SAND 0951 &    ...     & 0.72 & 32 & 0.04 & 33 & 0.68 & 32,33 & -4.94 & ... & 34 & M67  \\
2UCAC 35931726 & Cl* NGC 2682 SAND 0958 &    ...     & 0.00 & ... & 0.04 & 33 & 0.62 & 34 & -4.82 & ... & 34 & M67  \\
2UCAC 35931749 & Cl* NGC 2682 SAND 0963 &    ...     & 0.71 & 32 & 0.04 & 33 & 0.67 & 32,33 & -5.05 & ... & 34 & M67  \\
2UCAC 35931815 & Cl* NGC 2682 SAND 0965 &    ...     & 0.76 & 32 & 0.04 & 33 & 0.72 & 32,33 & -4.83 & ... & 34 & M67  \\
2UCAC 35931793 & Cl* NGC 2682 SAND 0969 &    ...     & 0.67 & 32 & 0.04 & 33 & 0.63 & 32,33 & -4.83 & ... & 34 & M67  \\
GSC 814-1735   & Cl* NGC 2682 SAND 0981 &    ...     & 0.71 & 32 & 0.04 & 33 & 0.67 & 32,33 & -5.00 & ... & 34 & M67  \\
2UCAC 35931819 & Cl* NGC 2682 SAND 0982 &    ...     & 0.61 & 32 & 0.04 & 33 & 0.57 & 32,33 & -4.66 & ... & 34 & M67  \\
2UCAC 35931700 & Cl* NGC 2682 SAND 0991 &    ...     & 0.68 & 32 & 0.04 & 33 & 0.65 & 32,33 & -4.96 & ... & 34 & M67  \\
2UCAC 35931814 & Cl* NGC 2682 SAND 1004 &    ...     & 0.76 & 32 & 0.04 & 33 & 0.72 & 32,33 & -4.86 & ... & 34 & M67  \\
2UCAC 35931816 & Cl* NGC 2682 SAND 1012 &    ...     & 0.74 & 32 & 0.04 & 33 & 0.70 & 32,33 & -4.80 & ... & 34 & M67  \\
2UCAC 35931821 & Cl* NGC 2682 SAND 1014 &    ...     & 0.71 & 32 & 0.04 & 33 & 0.67 & 32,33 & -4.72 & ... & 34 & M67  \\
2UCAC 35931731 & Cl* NGC 2682 SAND 1033 &    ...     & 0.61 & 32 & 0.04 & 33 & 0.57 & 32,33 & -4.74 & ... & 34 & M67  \\
2UCAC 35931828 & Cl* NGC 2682 SAND 1041 &    ...     & 0.73 & 32 & 0.04 & 33 & 0.69 & 32,33 & -4.93 & ... & 34 & M67  \\
2UCAC 35931843 & Cl* NGC 2682 SAND 1048 &    ...     & 0.69 & 32 & 0.04 & 33 & 0.65 & 32,33 & -4.92 & ... & 34 & M67  \\
GSC 814-1295   & Cl* NGC 2682 SAND 1050 &    ...     & 0.66 & 32 & 0.04 & 33 & 0.62 & 32,33 & -4.38 & ... & 34 & M67  \\
2UCAC 35931775 & Cl* NGC 2682 SAND 1057 &    ...     & 0.68 & 32 & 0.04 & 33 & 0.64 & 32,33 & -4.82 & ... & 34 & M67  \\
GSC 814-1233   & Cl* NGC 2682 SAND 1064 &    ...     & 0.66 & 32 & 0.04 & 33 & 0.62 & 32,33 & -4.94 & ... & 34 & M67  \\
GSC 814-1221   & Cl* NGC 2682 SAND 1065 &    ...     & 0.80 & 32 & 0.04 & 33 & 0.76 & 32,33 & -4.85 & ... & 34 & M67  \\
2UCAC 35931734 & Cl* NGC 2682 SAND 1068 &    ...     & 0.75 & 32 & 0.04 & 33 & 0.71 & 32,33 & -4.87 & ... & 34 & M67  \\
2UCAC 35931840 & Cl* NGC 2682 SAND 1078 &    ...     & 0.66 & 32 & 0.04 & 33 & 0.62 & 32,33 & -4.88 & ... & 34 & M67  \\
2UCAC 35931804 & Cl* NGC 2682 SAND 1087 &    ...     & 0.64 & 32 & 0.04 & 33 & 0.60 & 32,33 & -4.82 & ... & 34 & M67  \\
2UCAC 35931713 & Cl* NGC 2682 SAND 1089 &    ...     & 0.67 & 32 & 0.04 & 33 & 0.63 & 32,33 & -4.98 & ... & 34 & M67  \\
2UCAC 35931762 & Cl* NGC 2682 SAND 1093 &    ...     & 0.64 & 32 & 0.04 & 33 & 0.60 & 32,33 & -4.78 & ... & 34 & M67  \\
2UCAC 35931696 & Cl* NGC 2682 SAND 1095 &    ...     & 0.65 & 32 & 0.04 & 33 & 0.61 & 32,33 & -4.92 & ... & 34 & M67  \\
2UCAC 35931717 & Cl* NGC 2682 SAND 1096 &    ...     & 0.66 & 32 & 0.04 & 33 & 0.62 & 32,33 & -4.88 & ... & 34 & M67  \\
2UCAC 35931684 & Cl* NGC 2682 SAND 1106 &    ...     & 0.69 & 32 & 0.04 & 33 & 0.65 & 32,33 & -5.06 & ... & 34 & M67  \\
GSC 814-1789   & Cl* NGC 2682 SAND 1107 &    ...     & 0.64 & 32 & 0.04 & 33 & 0.60 & 32,33 & -4.62 & ... & 34 & M67  \\
2UCAC 35931906 & Cl* NGC 2682 SAND 1203 &    ...     & 0.71 & 32 & 0.04 & 33 & 0.68 & 32,33 & -4.82 & ... & 34 & M67  \\
2UCAC 35931884 & Cl* NGC 2682 SAND 1208 &    ...     & 0.00 & ... & 0.04 & 33 & 0.79 & 34 & -4.84 & ... & 34 & M67  \\
2UCAC 35931970 & Cl* NGC 2682 SAND 1212 &    ...     & 0.78 & 32 & 0.04 & 33 & 0.74 & 32,33 & -4.86 & ... & 34 & M67  \\
2UCAC 35931925 & Cl* NGC 2682 SAND 1213 &    ...     & 0.60 & 32 & 0.04 & 33 & 0.56 & 32,33 & -4.81 & ... & 34 & M67  \\
2UCAC 35931900 & Cl* NGC 2682 SAND 1218 &    ...     & 0.68 & 32 & 0.04 & 33 & 0.65 & 32,33 & -4.88 & ... & 34 & M67  \\
2UCAC 35931880 & Cl* NGC 2682 SAND 1246 &    ...     & 0.69 & 32 & 0.04 & 33 & 0.65 & 32,33 & -4.93 & ... & 34 & M67  \\
2UCAC 35931918 & Cl* NGC 2682 SAND 1247 &    ...     & 0.62 & 32 & 0.04 & 33 & 0.58 & 32,33 & -4.70 & ... & 34 & M67  \\
2UCAC 35931894 & Cl* NGC 2682 SAND 1248 &    ...     & 0.62 & 32 & 0.04 & 33 & 0.58 & 32,33 & -4.78 & ... & 34 & M67  \\
2UCAC 35931973 & Cl* NGC 2682 SAND 1249 &    ...     & 0.78 & 32 & 0.04 & 33 & 0.74 & 32,33 & -4.91 & ... & 34 & M67  \\
2UCAC 35931980 & Cl* NGC 2682 SAND 1251 &    ...     & 0.75 & 32 & 0.04 & 33 & 0.71 & 32,33 & -4.80 & ... & 34 & M67  \\
2UCAC 35931939 & Cl* NGC 2682 SAND 1252 &    ...     & 0.64 & 32 & 0.04 & 33 & 0.60 & 32,33 & -4.81 & ... & 34 & M67  \\
2UCAC 35931911 & Cl* NGC 2682 SAND 1255 &    ...     & 0.67 & 32 & 0.04 & 33 & 0.63 & 32,33 & -4.80 & ... & 34 & M67  \\
GSC 814-1973   & Cl* NGC 2682 SAND 1258 &    ...     & 0.00 & ... & 0.04 & 33 & 0.61 & 34 & -4.92 & ... & 34 & M67  \\
GSC 814-1679   & Cl* NGC 2682 SAND 1260 &    ...     & 0.62 & 32 & 0.04 & 33 & 0.59 & 32,33 & -4.79 & ... & 34 & M67  \\
2UCAC 35931940 & Cl* NGC 2682 SAND 1269 &    ...     & 0.76 & 32 & 0.04 & 33 & 0.72 & 32,33 & -4.99 & ... & 34 & M67  \\
2UCAC 35931858 & Cl* NGC 2682 SAND 1278 &    ...     & 0.78 & 32 & 0.04 & 33 & 0.74 & 32,33 & -5.00 & ... & 34 & M67  \\
2UCAC 35931865 & Cl* NGC 2682 SAND 1289 &    ...     & 0.76 & 32 & 0.04 & 33 & 0.72 & 32,33 & -5.03 & ... & 34 & M67  \\
2UCAC 35931886 & Cl* NGC 2682 SAND 1307 &    ...     & 0.81 & 32 & 0.04 & 33 & 0.77 & 32,33 & -5.05 & ... & 34 & M67  \\
2UCAC 35931913 & Cl* NGC 2682 SAND 1318 &    ...     & 0.62 & 32 & 0.04 & 33 & 0.58 & 32,33 & -4.64 & ... & 34 & M67  \\
2UCAC 35931949 & Cl* NGC 2682 SAND 1330 &    ...     & 0.66 & 32 & 0.04 & 33 & 0.62 & 32,33 & -4.62 & ... & 34 & M67  \\
2UCAC 36114630 & Cl* NGC 2682 SAND 1341 &    ...     & 0.74 & 32 & 0.04 & 33 & 0.71 & 32,33 & -4.72 & ... & 34 & M67  \\
2UCAC 35932025 & Cl* NGC 2682 SAND 1406 &    ...     & 0.55 & 32 & 0.04 & 33 & 0.51 & 32,33 & -4.75 & ... & 34 & M67  \\
GSC 814-2433   & Cl* NGC 2682 SAND 1420 &    ...     & 0.63 & 32 & 0.04 & 33 & 0.59 & 32,33 & -4.75 & ... & 34 & M67  \\
2UCAC 35932087 & Cl* NGC 2682 SAND 1426 &    ...     & 0.62 & 32 & 0.04 & 33 & 0.58 & 32,33 & -4.80 & ... & 34 & M67  \\
2UCAC 35932080 & Cl* NGC 2682 SAND 1446 &    ...     & 0.61 & 32 & 0.04 & 33 & 0.58 & 32,33 & -4.76 & ... & 34 & M67  \\
2UCAC 35932033 & Cl* NGC 2682 SAND 1449 &    ...     & 0.66 & 32 & 0.04 & 33 & 0.62 & 32,33 & -4.85 & ... & 34 & M67  \\
2UCAC 35932057 & Cl* NGC 2682 SAND 1452 &    ...     & 0.67 & 32 & 0.04 & 33 & 0.63 & 32,33 & -4.35 & ... & 34 & M67  \\
2UCAC 35932039 & Cl* NGC 2682 SAND 1462 &    ...     & 0.67 & 32 & 0.04 & 33 & 0.64 & 32,33 & -4.89 & ... & 34 & M67  \\
2UCAC 35932031 & Cl* NGC 2682 SAND 1473 &    ...     & 0.78 & 32 & 0.04 & 33 & 0.74 & 32,33 & -5.13 & ... & 34 & M67  \\
2UCAC 35932013 & Cl* NGC 2682 SAND 1477 &    ...     & 0.72 & 32 & 0.04 & 33 & 0.68 & 32,33 & -4.98 & ... & 34 & M67  \\
\enddata
\tablecomments{References and notes:
(1) \citet{Perryman97},
(2) unreddened \bv\ color appropriate for spectral type given by other reference,
(3) \citet{Wright04},
(4) \citet{Walter94},
(5) I have assumed $A_V$/\ebv\, = 3.1 in converting a $A_V$ value to E(B-V),
(6) \citet{White07},
(7) \citet{Preibisch99},
(8) \citet{Hog00} (converted to Johnson using \citet{Mamajek06},
(9) star is within 75\,pc and presumed to have no reddening,
(10) \citet{Gray06},
(11) \citet{Henry96},
(12) \citet{Soderblom98},
(13) \citet{Wichmann97},
(14) \citet{Mamajek02},
(15) \citet{Nordstrom04},
(16) \citet{Prosser92},
(17) \citet{Crawford74},
(18) \citet{Messina01},
(19) \citet{Mermilliod91},
(20) \citet{Stauffer89},
(21) \citet{Soderblom93},
(22) \citet{Stauffer87},
(23) \citet{Duncan91}, converted to to \logrphk\, following \citet{Noyes84},
(24) \citet{Taylor06},
(25) \citet{Paulson02},
(26) \citet{Upgren85},
(27) \citet{Reid92},
(28) \citet{Weis82},
(29) \citet{vanAltena69},
(30) \citet{Weis88},
(31) \citet{Gray03},
(32) \citet{Montgomery93},
(33) \citet{VandenBerg04},
(34) \citet{Giampapa06}.
}
\end{deluxetable}

\clearpage
\begin{deluxetable}{lrcccccc}
\setlength{\tabcolsep}{0.03in}
\tablewidth{0pt}
\tablecaption{Cluster \logrphk\ Values \label{tab:cluster}}
\tablehead{
{(1)}        &{(2)} & {(3)}   & {(4)}              &{(5)}  &{(6)}&{(7)}               &{(8)}\\
{Group}      &{Age} & {Refs.} & {\logrphk}         & {68\%}&{N}  &{activity-color}    &{\logrphkbar}\\
{Name}       &{Myr} & {}      & {median  }         & {CL}  &{}   &{slope $m$}         &{\bvsun}}
\startdata					 				       
USco         &   5  & 1,2,3   & -4.05\,$\pm$\,0.03 & 0.13  &  9  & -0.73\,$\pm$\,0.62 & -4.01\\
$\beta$ Pic  &  12  & 4,5     & -4.03\,$\pm$\,0.13 & 0.23  &  6  &  1.40\,$\pm$\,0.30 & -4.06\\
UCL+LCC      &  16  & 6,7     & -4.04\,$\pm$\,0.01 & 0.07  & 10  & -0.37\,$\pm$\,0.27 & -4.04\\
Tuc-Hor      &  30  & 6,7     & -4.16\,$\pm$\,0.13 & 0.16  &  8  &  3.02\,$\pm$\,0.45 & -4.23\\
$\alpha$ Per &  85  & 9,10,11 & -4.16\,$\pm$\,0.08 & 0.27  & 13  &  2.04\,$\pm$\,1.52 & -4.16\\
Pleiades     & 130  & 9,11,12 & -4.33\,$\pm$\,0.04 & 0.24  & 56  &  0.75\,$\pm$\,0.24 & -4.27\\
UMa          & 500  & 13      & -4.48\,$\pm$\,0.03 & 0.09  & 17  &  0.80\,$\pm$\,0.27 & -4.50\\
Hyades       & 625  & 11,14   & -4.47\,$\pm$\,0.01 & 0.09  & 87  &  0.14\,$\pm$\,0.13 & -4.50\\ 
M67          & 4000 & 15,16   & -4.84\,$\pm$\,0.01 & 0.11  & 76  & -1.03\,$\pm$\,0.23 & -4.85
\enddata
\tablecomments{Columns:
(1) name of group,
(2) age,
(3) age and membership references,
(4) \logrphk\, median and uncertainty \citep{Gott01},
(5) 68\% confidence intervals on \logrphk,
(6) number of data points per bin,
(7) OLS bisector slope $m$ = $\Delta$\logrphk/$\Delta$\bv\, and uncertainty,
(8) mean \logrphk\, interpolated at solar \bvo. 
OLS (Y|X) slopes and uncertainties were calculated using 10$^4$
jackknife sampling simulations, except for $\beta$ Pic and Tuc-Hor
where the slope was analytic calculated, due to their small sample
size. Estimation of the solar \logrphk\, value is discussed in \S1.
References: (1) \citet{Preibisch02},
(2) \citet{Preibisch99},
(3) \citet{Walter94},
(4) \citet{Ortega02},
(5) \citet{Zuckerman04},
(6) \citet{Mamajek02},
(7) \citet{deZeeuw99},
(8) \citet{Mamajek04},
(9) \citet{Barrado04},
(10) \citet{Makarov06},
(11) this work (\S2.2),
(12) \citet{Duncan91},
(13) \citet{King03}
(14) \citet{Perryman98},
(15) \citet{VandenBerg04},
(16) \citet{Giampapa06}, selected from \citet{Girard89}.
}
\end{deluxetable}
\clearpage

\subsection{Cluster Ages, Membership, and Activity \label{cluster_membership}}

We turn now to a detailed discussion of our cluster samples.
Kinematic membership of individual stars to their assigned groups was
scrutinized with modern astrometric data (i.e. {\it Hipparcos},
Tycho-2, and UCAC2 catalogs) either by the authors or through
examination of recently published kinematic studies, or
both. Assessment of whether the stars' proper motions were consistent
with membership follows the methodology in \citet{Mamajek05}. Table
\ref{tab:cluster_ca} lists the members of the stellar groups along
with their relevant color and activity, and Table
\ref{tab:cluster} summarizes the cluster ages, the number of
published \logrphk\, values for cluster members, and a summary of
activity statistics.  In total there are 274 published \logrphk\,
measurements for 206 stars in our cluster database.  In the following
subsections we briefly review the stellar groups and references for
their membership and ages.

\subsubsection{Young Associations}

Members of Upper Sco were taken from \citet{Preibisch99} and
\citet{Walter94}; we adopt the mean group age (5 Myr) from
\citet{Preibisch02}.  Memberships and mean ages for the $\beta$ Pic
and Tuc-Hor moving groups (12 and 30 Myr ages, respectively) were
taken from \citet{Zuckerman04}, and HD 105 was added as a Tuc-Hor
member following \citet{Mamajek04}.  In Tuc-Hor, only stars
demonstrated by \citet{Mamajek04} to be near and co-moving with the
$\beta$ Tuc nucleus were retained for our activity-age calibration.
Members of Lower Cen-Cru (LCC) and Upper Cen-Lup (UCL) were taken from
\citet{deZeeuw99} and \citet{Mamajek02}, and mean group ages were
adopted from \citet{Mamajek02}. \citet{Preibisch08} suggest that LCC
shows evidence for substructure and a probable age gradient (the more
populous northern part appears to be $\sim$17 Myr, while the less
populous southern part appears to be $\sim$12 Myr), however 16 Myr is
a reasonable mean age for the group, and given the lack of evolution
in \logrphk\, between $\sim$10$^{6}$ and 10$^{8}$ yr, the choice of
adopted age has negligible impact on our analysis.  Furthermore, to
improve the statistics we combined the UCL and LCC groups, which are
approximately coeval and whose individual \rphk\, measurements were
similar.

We have decided to not include members of a few nearby stellar groups
in our calibration of the activity vs. age relation: AB Dor, Her-Lyr,
and Castor.  Although there are solar-type members of the nearby AB
Dor moving group, we do not include its members for the following
reasons: (1) its age is controversial
\citep{Close05,Luhman05,Ortega07}, (2) it is not clear that a clean
separation between membership within a supposedly ``coeval'' AB Dor
group \citep{Zuckerman04} and the ``non-coeval'' Pleiades B4 moving
group \citep{Asiain99,Famaey07} has been demonstrated, and (3) the
range of acceptable velocities for membership in the AB Dor group
seems rather large for a coeval group \citep{Zuckerman04} compared to
OB associations and clusters \citep[e.g.][]{Briceno07}. The coevality
and evidence for a common origin for members of the Her-Lyr and Castor
groups has also not been sufficiently demonstrated for inclusion in a
sample of calibration stars.

\subsubsection{$\alpha$ Per, Pleiades, \& UMa}

The $\alpha$ Per members have been confirmed kinematically by
\citet{Makarov06} for all of the cluster candidates except Cl Melotte
20 696 and AP 93. We find that the UCAC2 proper motions for both of
these stars are statistically consistent with membership, and include
them in our $\alpha$ Per sample. For the age of $\alpha$ Per, we adopt
the most recent Li-depletion boundary value from \citet[][85
Myr]{Barrado04}.

For the Pleiades, all of the \rphk\, measurements of candidate members
from \citet{Duncan91}, \citet{Soderblom93}, and \citet{White07}, were
considered. We independently tested the kinematic membership of each
of these stars to the Pleiades using Tycho-2 or UCAC2 proper motions
and the group proper motion from \citet{Robichon99}.  All of the
objects have motions within 2$\sigma$ of the Pleiades mean motion
(although \#571 is a marginal case, but supporting evidence suggests
that this is probably a bona fide member).  \citet{Deacon04}
independently assign high membership probability to the Pleiades for
stars \#102, 129, 173, 296, 314, 514, 923, 1776, 1015, 1207, 3097.
For the age of the Pleiades, we adopt the recent Li-depletion boundary
estimate from \citet[][130 Myr]{Barrado04}.

An extensive study of the age, membership, and activity of the Ursa
Major cluster was undertaken by \citet{King03}, and we include their
``Y'' or ``Y?'' candidate members in our census for that
cluster. Recently, \citet{King05} reevaluated the age of UMa, and
claimed that the system appears to be approximately coeval with the
Hyades and Coma Ber clusters (all $\sim$0.6 Gyr) but ``with the Hyades
perhaps being only 100 Myr older''. This assessment is apparent in
visual inspection of Fig. 2 of \citet{King05} of the main sequence
turn-offs with overlaid evolutionary tracks appropriate for the
metallicities of UMa and the Hyades. Based on this, we adopt the age
of UMa from \citet{King03}, 500 Myr.

\subsubsection{Hyades}

The Hyades is the best studied cluster in terms of its chromospheric
activity.  Our primary source of membership assignment and age (625
Myr) for the Hyades is \citet{Perryman98}, 
adopting their members constrained both by proper motions
and RVs. Additional non-{\it Hipparcos} Hyades candidates were gleaned
from the \logrphk\, surveys of \citet{Duncan91}, \citet{Paulson02},
and \citet{White07}, including Cl Melotte 25 \#s 49, 91, 92, 93, 99,
183, and Cl Melotte 25 VA \#s 115, 146, 354, 383, 502, and
637. Tycho-2 and UCAC2 proper motions for these stars were tested for
Hyades membership using the \citet{deBruijne01} convergent point, and
all of these candidates are kinematically consistent with Hyades
membership with moving cluster distances of $\sim$44-52 pc.

\begin{figure}
\epsscale{1}
\plotone{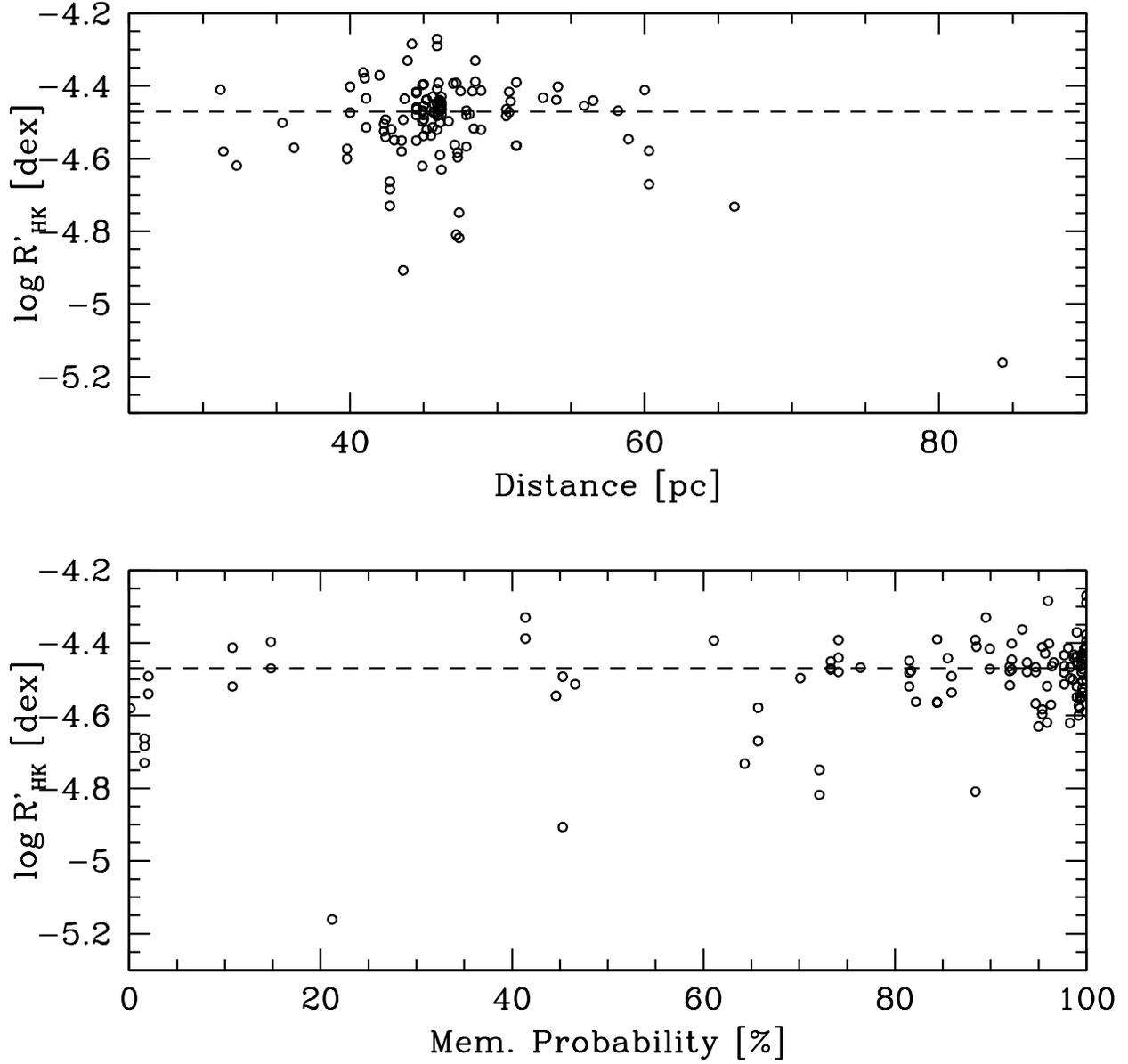}
\caption{(top) Moving cluster distance vs. \logrphk\, for candidate
Hyades members. (bottom) Membership probability vs. \logrphk\, for
candidate Hyades members.
\label{fig:rhk_dist_Hyad}}
\end{figure}

Among the \logrphk\, data for Hyades members were a handful of
remarkably active and inactive stars. Further investigation of these
objects was warranted to see whether we should include them in our
sample statistics (critical for establishing what the spread in
plausible activity levels is for stars of a given age).  To see if the
extreme outliers might be dominated by ``supercluster'' members or
interlopers that might be unrelated to the Hyades nucleus, we plotted
moving cluster distance vs. \logrphk\ and membership probability vs.
\logrphk\, in Figure \ref{fig:rhk_dist_Hyad}. The moving cluster
distances and probabilities were calculated following
\citet{Mamajek05} using the \citet{deBruijne01} convergent point
solution with {\it Hipparcos}, Tycho-2, or UCAC2 proper motions (in
order of preference). An intrinsic velocity dispersion of 1 \kms\, was
assumed in the membership probability estimation, with relative
ranking seen as more important than absolute values.

A few things are apparent from Fig. \ref{fig:rhk_dist_Hyad}. The
\logrphk\, values for the high membership probability Hyads (P $>$
75\%) are consistent with a median value of \logrphk\, = -4.47 and
remarkably small r.m.s. of 0.08 dex\footnote{Our literature search for
Hyades activity measurements yielded three extremely active outliers
which are excluded in our analysis: Cl Melotte 25 \#s 76, 105, and
127. Coincidentally, the \logrphk\, values for all three stars were
estimated from single observations by the Mt. Wilson survey that all
took place 22 July 1977. All three were also observed by
\citet{Paulson02}, and their \logrphk\, values for these stars are
more in line with other Hyades (\logrphk\, = -4.47, -4.52, and -4.45,
for \# 76, 105, and 127, respectively).  The idea that three Hyads
could be flaring simultaneously on the same night at unprecedentedly
high levels is extremely unlikely, so we exclude these Mt. Wilson
observations from our statistics.}. The lower membership probability
objects (P $<$ 75\%) have a lower median \logrphk\, (-4.51) and higher
r.m.s. (0.14 dex).  We attribute this to the likely inclusion in the
current list of Hyades candidates of older field interlopers.  It is
apparent that the stars at $d$ $<$ 40 pc and $d$ $>$ 60 pc tend to be
less frequently active, probably due to inclusion of interlopers.

In summary, for our activity study, we conservatively include only
those Hyades stars with membership probabilities $>$50\% and cluster
parallax distances within 1 tidal radius ($\pm$10 pc) of the mean
distance (46.3 pc); \citet{Perryman98}.
\footnote{ Due to the distance constraint, we reject from our sample:
HIP 10672, 13600, 13976, 15563, 15720, 17766, 19386, 19441, 20949,
21741, 22566, 24923, 25639. Due to low membership probability, we
reject from our sample: HIP 15304, 17609, 19082, 19834, 20082, 20719,
21280, 22380.  Some stars failed both the distance and the membership
criteria: HIP 19386 \& 20441.  {\it Some}, and possibly even {\it
most} of the rejected stars in the first two lists may be bona fide
Hyades members, although the stars in the last list are almost
certainly non-members.  Our goal is to create as clean a sample of
Hyades members as possible for the study of their activity -- hence
our stringent membership criteria.  We do not necessarily recommend
rejecting these stars from future studies of the Hyades.  Our
selection criterion clips the two most inactive Hyades candidates
studied by \citet{Paulson02}: HIP 25639 (\logrphk\, = -5.38; $d$ = 86
pc, HIP 19386 (\logrphk\, = -5.16; $d$ = 84 pc).  The least active
Hyad that satisfies our membership and color criteria is HIP 22422
\citep[\logrphk\, = -4.82;][]{Paulson02}, while the most active is HIP
20978 \citep[\logrphk\, = -4.27;][]{Duncan91}.  }

\subsubsection{M67 \label{M67}}

We adopt an age of 4.0 Gyr for the M67 cluster from
\citet{Sarajedini99} and \citet{VandenBerg04}, and include the M67
membership and HK observations of \citet{Giampapa06} in our analysis.
The \logrphk\, values listed in Table \ref{tab:cluster_ca} were
converted from the HK emission equivalent widths by M. Giampapa
(priv. comm.). The candidate RS CVn Sanders 1112 is listed in Table
\ref{tab:cluster_ca}, but was excluded from the analysis (with
\logrphk\, = -4.11).

\subsubsection{Ancillary Cluster Data \label{Ancillary}}

We believe that the cluster membership assignments in Table
\ref{tab:cluster_ca} are quite reliable.  Any interlopers among the
samples that we may not have caught are small in number, and will have
negligible impact on our findings.  The ages reflect modern
astrophysical understanding and are systematically older than those
used in previous age-\logrphk\ calibrations.

Notably the current sample is sparsely populated at ages of $>$1 Gyr.
The historical lack of $>$1 Gyr-old clusters in the age-activity
calibration is due to the deficiency of nearby ($<$100-200 pc) older
clusters with solar-type members bright enough for observations with
the Mt. Wilson photometer.

To overcome this shortcoming, Barry and collaborators determined
Mt. Wilson S-values with a lower resolution system
\citep{Barry87,Barry88}. \citet{Soderblom91} argued that the Barry et
al. S-values were not on the Mt. Wilson system, but that a linear
correction could remedy this. While Soderblom's correction is not
well-constrained at the high or low activity regimes, we none-the-less
use it to correct cluster mean \logrphk\, values from \citet{Barry87}
to \logrphk\, values on the Mt. Wilson system. These ancillary cluster
age-activity data are compiled in Table \ref{tab:cluster_other}.  We
omit a datum for the $\sim$3 Myr-old cluster NGC 2264 for two reasons:
(1) Soderblom's (1991) correction for the \citet{Barry87} data does
not extend to activity levels this high, and (2) the extrapolated mean
\logrphk\, value for NGC 2264 (-4.26) is $\sim$0.2 dex lower than the
mean values for the similarly aged Upper Sco, UCL, LCC, and $\beta$
Pic groups \footnote{Notably, the form of the
\citet{Donahue93} relation at high activity levels is driven largely
by the NGC 2264 datum.}.  The \citet{Barry87} data are nominally
corrected to a standard color of \bvo\, = 0.60; however, for our
purposes the differences are negligible. As a check on the Soderblom
et al. conversion, we find a nearly identical median \logrphk\, value
at solar color for the M67 sample (-4.86) as found in the
high-resolution HK study of \citet{Giampapa06} (-4.85).

There is a clear need for more modern derivation of \logrphk\,
activity diagnostics in fiducial older clusters such as M~34, Coma
Ber, NGC~752, and NGC~188.  Recent studies of H\&K emission in
such members of older clusters \citep[e.g.][]{Pace04} did not provide
\logrphk\, values, only emission line fluxes.  Attempts by the authors
and D. Soderblom (priv. comm.) to tie these observations to the
Mt. Wilson system have thus far failed.

\begin{deluxetable}{lrcccc}
\setlength{\tabcolsep}{0.03in}
\tablewidth{0pt}
\tablecaption{\logrphk\, Data For Ancillary Samples \label{tab:cluster_other}}
\tablehead{
{(1)}     &{(2)} & {(3)}  &   {(4)}    & {(5)}      &{(6)}     \\
{Cluster} &{Age} & {Age}  & {original} & {corrected}&{\logrphk}\\
{Name}    &{Myr} & {Ref.} & {\logrphk} & {\logrphk} &{Ref.}}
\startdata
M 34      &  200 & 1      & -4.4:      &  ...       & 2 \\
Coma Ber  &  600 & 3      & -4.51      & -4.43      & 4 \\
NGC 752   & 2000 & 5      & -4.70      & -4.70      & 4 \\
M 67      & 4000 & 6,7    & -4.82      & -4.86      & 4 \\
NGC 188   & 6900 & 6,7    & -4.98      & -5.08      & 4 \\
old field & 8000 & 8      & -4.99      &  ...       & 8 
\enddata
\tablecomments{
Columns:
(1) name of group,
(2) age,
(3) age reference,
(4) originally quoted mean \logrphk\, value,
(5) corrected mean \logrphk\, value (only relevant for ref. 4),
(6) activity references.
References:
(1) \citet{Jones97},
(2) visual inspection of Fig. 1 of \citet{King03}
(3) \citet{King05},
(4) data from \citet{Barry88} corrected following \citet{Soderblom91},
(5) \citet{Dinescu95}
(6) \citet{Sarajedini99},
(7) \citet{VandenBerg04},
(8) this study (\S\ref{Old_Field}).
}
\end{deluxetable}

\subsubsection{Field Stars with Precise Isochronal Ages \label{Old_Field}}

To further augment the activity data for old stellar samples, we
consider an additional sample of solar-type field dwarfs with
well-constrained isochronal ages.  \citet[][hereafter VF05]{Valenti05}
report spectroscopic properties and isochronal age estimates for 1040
solar-type field dwarfs in the Keck, Lick, and AAT planet search
samples (the ``SPOCs'' sample). After estimating accurate
temperatures, luminosities, metallicities, and $\alpha$-element
enhancements, VF05 interpolate isochronal ages for each star on the
Yonsei-Yale evolutionary tracks \citep{Yi03}.  From their sample of
1040 solar-type stars (which includes some evolved stars), VF05 were
able to constrain isochronal ages for 57 stars (5.5\%) to better than
20\% in both their positive and negative age uncertainties.  As our
activity-relation is currently poorly constrained at the old ages
(given the lack of suitable cluster samples), we include VF05
solar-type dwarfs within 1 mag of the MS and isochronal ages of 5-15
Gyr.  The stars in this sample that have published \logrphk\, data are
listed in Table \ref{tab:VF05_old}. As the sample is sparse (N = 23),
to put it on equal footing with the cluster samples we simply treat it
as a single ``cluster'' with median age 8.0 \,$\pm$\,0.7 ($\pm$3.9;
68\%CL) Gyr or log\,$\tau$ = 9.90\,$\pm$\,0.04 ($\pm$0.19 ; 68\%CL)
dex. The mean activity for the sample is \logrphkbar\, =
-4.99\,$\pm$\,0.02 dex ($\pm$0.07; 68\%CL). The mean color for the
sample is similar to that of the Sun: $\overline{\bv}$ $\simeq$
0.62 mag.

\begin{deluxetable}{lccccc}
\tabletypesize{\scriptsize}
\setlength{\tabcolsep}{0.03in}
\tablewidth{0pt}
\tablecaption{Old Solar-Type Dwarfs From VF05 With Age Uncertainties of $<$20\% \label{tab:VF05_old}}
\tablehead{
{(1)}    &{(2)}  &{(3)}   &{(4)}     &{(5)}   &{(6)} \\
{HD}     &{\bv}  &{$\tau$}&{\logrphk}&{Ref.}  &{$\Delta$M$_V$}\\
{}       &{mag}  &{Gyr}   &{dex}     &{}      &{mag}
}
\startdata
3823     &  0.564  &  6.7   &  -4.97 &  1  &  -0.37  \\
20794    &  0.711  &  13.5  &  -4.98 &  2  &  +0.18  \\
22879    &  0.554  &  13.9  &  -4.92 &  3  &  +0.58  \\
32923    &  0.657  &  9.0   &  -5.15 &  4  &  -0.93  \\
34297    &  0.652  &  13.4  &  -4.93 &  2  &  -0.30  \\
36108    &  0.590  &  7.1   &  -5.01 &  1  &  -0.37  \\
38283    &  0.584  &  5.7   &  -4.97 &  2  &  -0.56  \\
45289    &  0.673  &  7.6   &  -5.01 &  2  &  -0.49  \\
51929    &  0.585  &  12.4  &  -4.86 &  2  &  +0.14  \\
95128    &  0.624  &  5.0   &  -5.02 &  4  &  -0.34  \\
122862   &  0.581  &  5.9   &  -4.99 &  1  &  -0.61  \\
142373   &  0.563  &  7.7   &  -5.11 &  3  &  -0.63  \\
143761   &  0.612  &  8.7   &  -5.04 &  5  &  -0.37  \\
153075   &  0.581  &  11.2  &  -4.88 &  2  &  +0.15  \\
157214   &  0.619  &  11.6  &  -5.00 &  4  &  -0.01  \\
186408   &  0.643  &  5.8   &  -5.05 &  4  &  -0.43  \\
186427   &  0.661  &  8.0   &  -5.04 &  4  &  -0.26  \\
190248   &  0.751  &  6.2   &  -5.00 &  2  &  -0.78  \\
191408   &  0.868  &  15.0  &  -4.99 &  2  &  +0.39  \\
193307   &  0.549  &  5.7   &  -4.90 &  2  &  -0.43  \\
196378   &  0.544  &  5.3   &  -4.95 &  1  &  -0.91  \\
201891   &  0.525  &  14.5  &  -4.86 &  3  &  +0.65  \\
210918   &  0.648  &  8.5   &  -4.95 &  2  &  -0.27  \\
\enddata
\tablecomments{Columns: 
(1) HD name,
(2) \bv\, color from \citet{Perryman97},
(3) isochronal age in Gyr (VF05; uncertainties $<$20\%),
(4) chromospheric activity \logrphk,
(5) activity reference,
(6) difference between stellar absolute magnitude
and that for MS star of same \bv\, color.
References:
(1) \citet{Jenkins06},
(2) \citet{Henry96},
(3) \citet{Wright04},
(4) \citet{Hall07},
(5) \citet{Baliunas96}.
}
\end{deluxetable}

\section{Ca II H\&K Analysis \label{analysis}}

With established membership lists and assembled \rphk\, values
deriving from a few large, homogeneous spectroscopic surveys, we
proceed in this section to derive a modern activity-age relationship.
We first consider various second parameter effects,
e.g. color/temperature/mass, surface gravity, and composition.  We
investigate color dependencies by examining the \rphk\, diagnostic for
binary pairs having the same age/composition but substantial
temperature differences (\S3.1.1) and then for kinematic groups
sampling a range of masses at different ages (\S3.1.2). We proceed in
\S3.2 to derive a preliminary empirical activity-age relation based on
cluster and solar \logrphk\, data.

\subsection{Systematics in \rphk\, \label{color_dep}}

There is some evidence that \rphk\, varies systematically not only as
a function of age, but at a given age with stellar color (i.e. mass).
Specifically, while \citet{Soderblom91} found a flat \logrphk\,
vs. \bvo\, relation for halo stars, they found a significant positive
slope for members of the Hyades cluster ($m$ =
$\Delta$\logrphk/$\Delta$(\bv) = 0.391). Elsewhere in the literature,
it appears that the color-dependence of \logrphk\, is largely ignored.

Spectral dependencies of \rphk\, could systematically impact our
calibration of \rphk\ as an age estimator, if the distribution of
colors differs amongst the different associations and clusters in our
sample.  To test whether the activity-age relation may be mass
dependent, we study both binary pairs and kinematic groups, presuming
in the respective samples that the components have the same age but
different masses, and look for trends in \rphk\, with color.

\subsubsection{Trends Among Binary Pairs \label{color_binaries}}

\begin{figure}
\epsscale{1}
\plotone{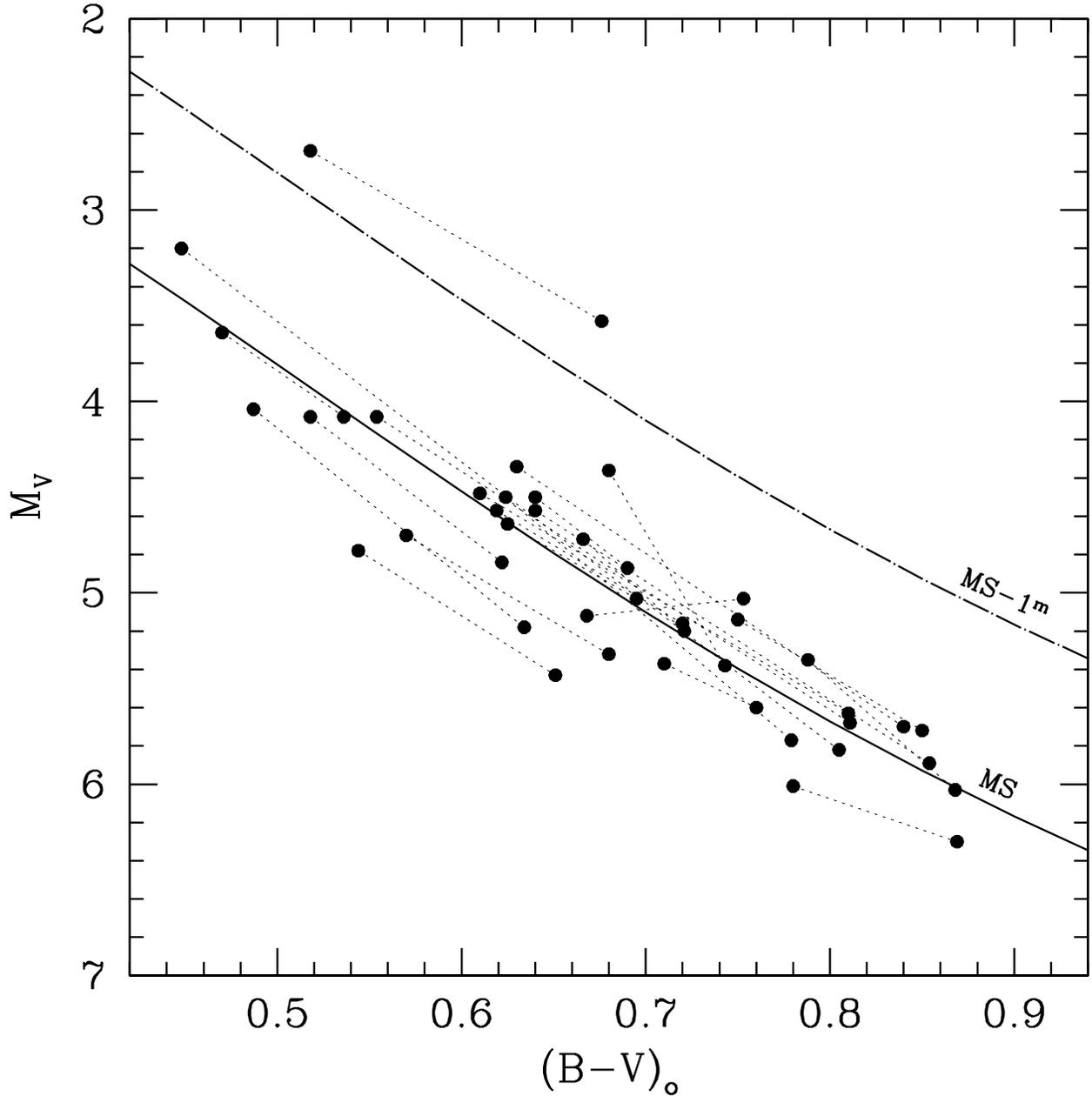}
\caption{Color vs. absolute magnitude for 23 non-identical
($\Delta$(B-V) $\ge$ 0.05) stellar binaries (see
\S\ref{color_binaries}). {\it Thin shorted-dashed lines} connect the
stellar binary components ({\it filled circles}). The {\it solid line}
is the main sequence from \citet{Wright05}, and the {\it dash-dotted
line} is 1 mag brighter than the main sequence (approximately
segregating post-MS stars from MS stars).  The system above the ``MS
minus 1 mag'' line (HD 5208) was retained as its color-magnitude slope
was consistent with being a system of two MS stars. As the system has
roughly solar metallicity \citep{Marsakov95}, it is possible that its
{\it Hipparcos} parallax is significantly in
error.\label{fig:pair_mv}.}
\end{figure}

We plot in Fig. \ref{fig:pair_mv} color vs. absolute magnitude for the
field binaries from Table \ref{tab:pairs} with significant color
difference ($\Delta$\bv $>$ 0.05 mag). The reddening towards these
stars is small, according to their spectral types and \bv\, colors, as
well as their proximity to the Sun (most are within $<$75 pc, and
likely have negligible reddening). As can be seen, the pairs are
generally aligned with the main sequence, although it is apparent that
the systems have a modest range in metallicities which slide their
individual main sequences above and below the mean field MS.

\begin{figure}
\epsscale{1}
\plotone{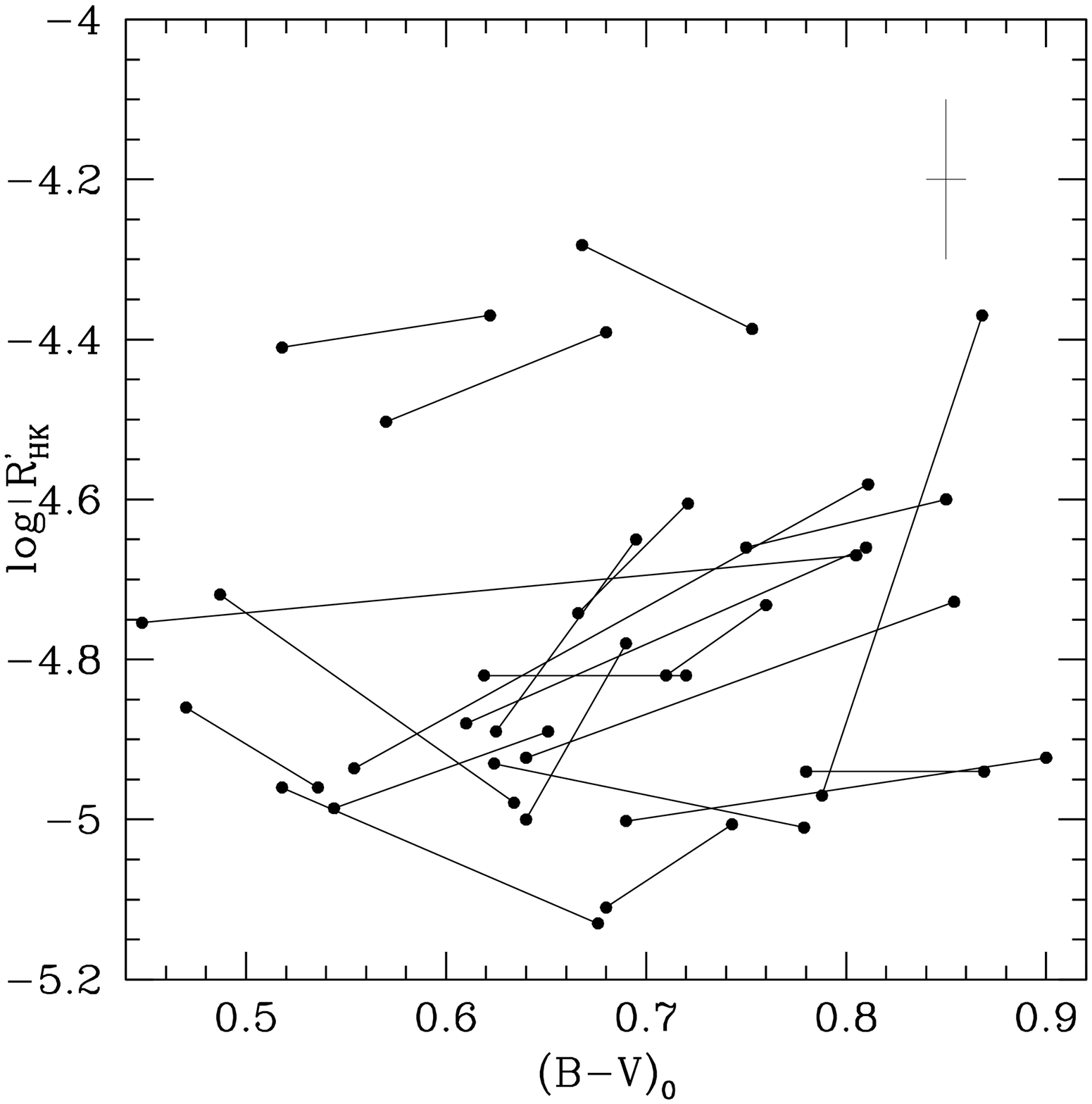}
\caption{Color versus activity for 23 non-identical ($\Delta$(B-V)
$\ge$ 0.05) stellar binaries (see \S\ref{color_binaries}). A typical
errorbar for (B-V) colors ($\pm$0.01 mag) and for a single \logrphk\,
observation ($\pm$0.1 dex) is illustrated by the cross. The pair on
the right side with the large slope is the pathological binary HD
137763.
\label{fig:pairs}}
\end{figure}


In Fig. \ref{fig:pairs} we show \rphk\ as a function of color for the
24 pairs.  Interstellar reddening, which should be negligible, should
affect both components equally and therefore should not influence
measurements of the activity-color mean slope.  There is a range of
slopes ($m$ = $\Delta$\logrphk/$\Delta$(\bv)) characterizing the
sample, with some negative and some positive.  A statistical analysis
of the individual slopes shows that one system is statistically
deviant \citep[HD 137763\footnote{HD 137763 appears to be a true
pathology. While the B component HD 137778 is clearly an active K2V
dwarf, the A component is an inactive spectroscopic binary with the
highest measured eccentricity ever reported \citep[$e$ =
0.975;][]{Pourbaix04}. The spectroscopic companion Ab is likely
applying torques to the primary \citep{Duquennoy92}, altering its
rotational evolution. \label{HD_137763}} ; rejected by Chauvenet's
criterion;][]{Bevington92}, and that the mean slope is $\overline{m}$
= 0.51\,$\pm$\,0.29.  The true median of the slope is
$\widetilde{m}$ = 0.60$^{+0.34}_{-0.27}$ \citep{Gott01}.

While the binary data alone are within $\sim$2$\sigma$ of zero slope
in $\Delta$\logrphk/$\Delta$(\bv), there is some hint that the slope
is indeed slightly positive. \citet{Donahue98} made a plot similar to
Fig. \ref{fig:pairs} (indeed, using many of the same systems), but did
not explicitly state any conclusions regarding the existence of a
color trend.  As noted above, there is likely a range of ages
represented by these binary pairs; we investigate now whether the
observed variation in slope of \rphk\ with color can be correlated
with stellar age.

\subsubsection{Trends Among Stellar Kinematic Groups \label{color_groups}}

\begin{figure}
\epsscale{1}
\plotone{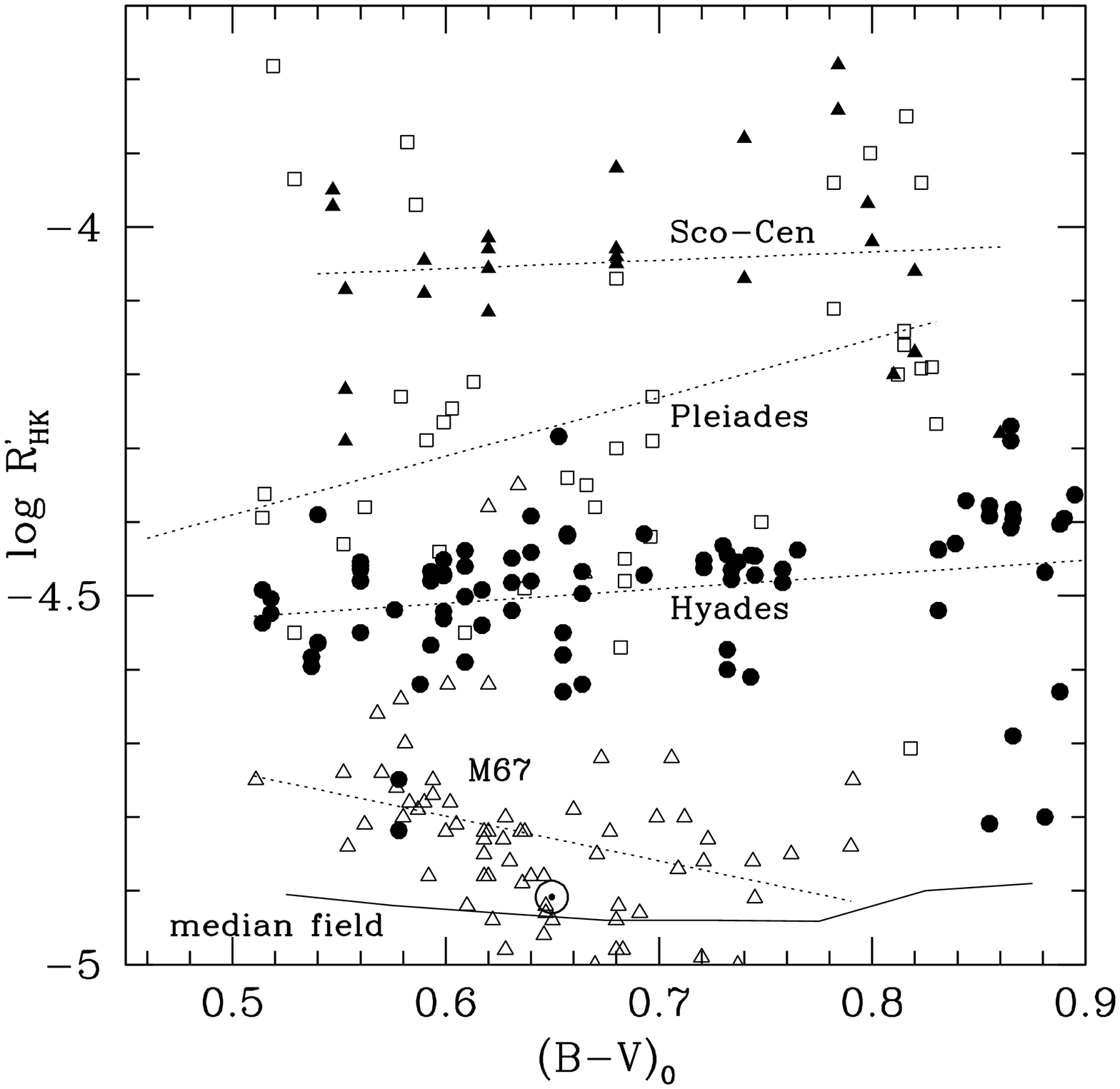}
\caption{\bvo\, vs. \logrphk\, for members of several stellar clusters
in Table \ref{tab:cluster_ca}.  {\it Filled triangles} are $\sim$5-16
Myr Sco-Cen members (incl. Upper Sco, $\beta$ Pic, UCL, LCC), {\it
open squares} are $\sim$130 Myr-old Pleiads, {\it filled circles} are
$\sim$625 Myr-old Hyads, and {\it open triangles} are $\sim$4 Gyr-old
M67 members.  Linear fits to the cluster data are {\it dashed lines}.
The {\it circle-dot} is the Sun. The {\it solid line} represents the
median \logrphk\, for solar-type field stars (median \logrphk\, values
for 8 color bins from a sample of 1572 unique stars in the activity
surveys of \citet{Henry96} and \citet{Wright04}).
\label{fig:cluster_bv_rhk}}
\end{figure}

In Fig. \ref{fig:cluster_bv_rhk}, we plot \logrphk\, vs \bv\,
color for the separate kinematically defined groups in our study.
From 10$^4$ jackknife sampling simulations, the slopes ($m$ =
$\Delta$\logrphk/$\Delta$(\bv)) for each group were evaluated using
ordinary least squares linear regression with \logrphk\, as the
dependent variable and \bvo\, as the independent variable \citep[OLS
(Y|X);][]{Isobe90}. These slopes, along with the median \logrphk\,
values, are provided in Table \ref{tab:cluster_ca}.

Examination of Table \ref{tab:cluster_ca} shows that divining a unique
slope applicable to all solar-type stars at all activity levels is not
be feasible.  The $<$100 Myr-old groups show a wide range of slopes
(-1 $<$ $m$ $<$ 3) with typically large uncertainties, but a mean
slope for the ensemble of $m$ = 0.91\,$\pm$\,0.40.  The $\sim$0.1-0.5
Gyr Pleiades and UMa clusters show similarly steep slopes of
0.75\,$\pm$\,0.24 and 0.80\,$\pm$\,0.27, respectively. These values
are $\sim$2$\sigma$ steeper than the slope for the $\sim$0.6 Gyr
Hyades (0.14\,$\pm$\,0.13). The oldest cluster (M67) also has the most
negative slope (-1.0\,$\pm$\,0.2). Together, the data suggest that the
slope $\Delta$\logrphk/$\Delta$(\bv) may flatten as a function of age.
The mean slope for all of the clusters combined is $m$ =
0.37\,$\pm$\,0.14, essentially identical to the Hyades slope ($m$ =
0.39) found by \citet{Soderblom85}. However, our Hyades slope appears
to be flatter than that derived by \citet{Soderblom85} due to
inclusion of additional lower activity stars at the blue and red edges
of our color range.  A sample of $\sim$1500 unique solar-type field
stars from the combined surveys of \citet{Wright04} and
\citet{Henry96} is statistically consistent with having zero slope
(see Fig. \ref{fig:cluster_bv_rhk}). Similarly, \citet{Soderblom91}
report a negligible slope for a sample of solar-type halo field stars.

For either the cluster (plus older field) sample alone or the binary
sample alone, the significance of the activity-color slope is
$<3\sigma$.  However, based on the fact that the measured slopes are
consistent between these populations in the mean, and systematic with
stellar age, we conclude that there is indeed an activity-color
correlation that needs to be taken into account.

\subsection{\rphk\,--Age Calibration Using Cluster Stars}

\subsubsection{Assembled Cluster Data}

\begin{figure}
\epsscale{.6}
\plotone{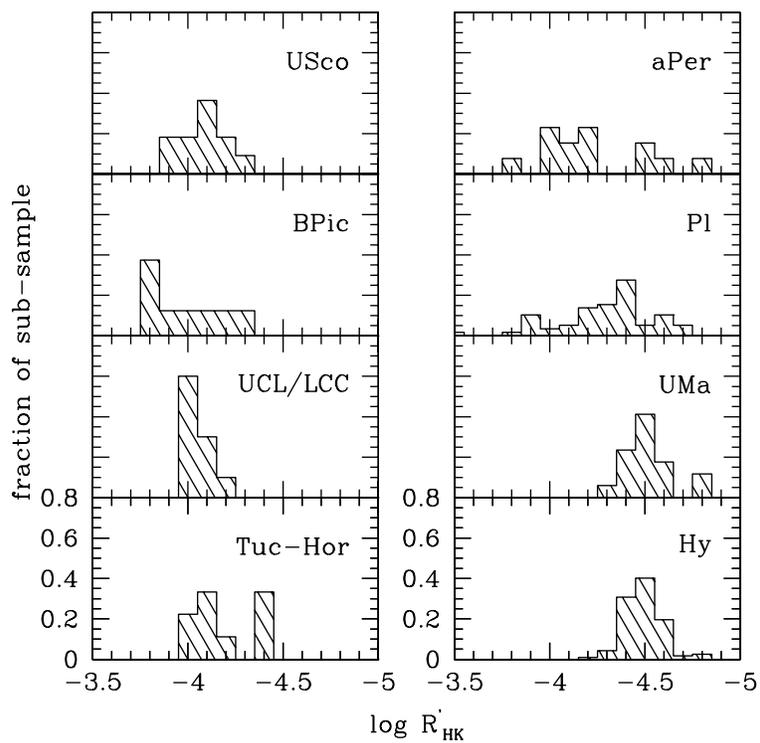}
\caption{Normalized histograms showing the distribution of \logrphk\,
values within each stellar cluster or association, as compiled in
Table \ref{tab:cluster_ca}.  Individual kinematic groups show a
dispersion in activity that is driven by both measurement error and
astrophysical variation; the latter appears to be at a maximum at
$\alpha$ Per and Pleiades ages. \label{fig:cluster_rhk_hist}}
\end{figure}

With estimates of the mean \logrphk\, values and color trends for
stellar samples of known age, we can proceed towards an improved
activity-age relation.  In Fig. \ref{fig:cluster_rhk_hist}, we plot
histograms of the distribution of \logrphk\, values for the stellar
groups in our study (Table \ref{tab:cluster}).  For each cluster, we
use the individual $\Delta$\logrphk/$\Delta$(\bv) slopes calculated
above to interpolate a mean \logrphk\, value for a hypothetical
cluster member of solar color ((B-V)$_{\odot}$ = 0.65 mag).  These are
quoted in the last column of Table \ref{tab:cluster_ca} and adopted in
the analysis that follows.

\subsubsection{A New \rphk\,--Age Relation \label{Relation}}

\begin{figure}
\epsscale{1}
\plotone{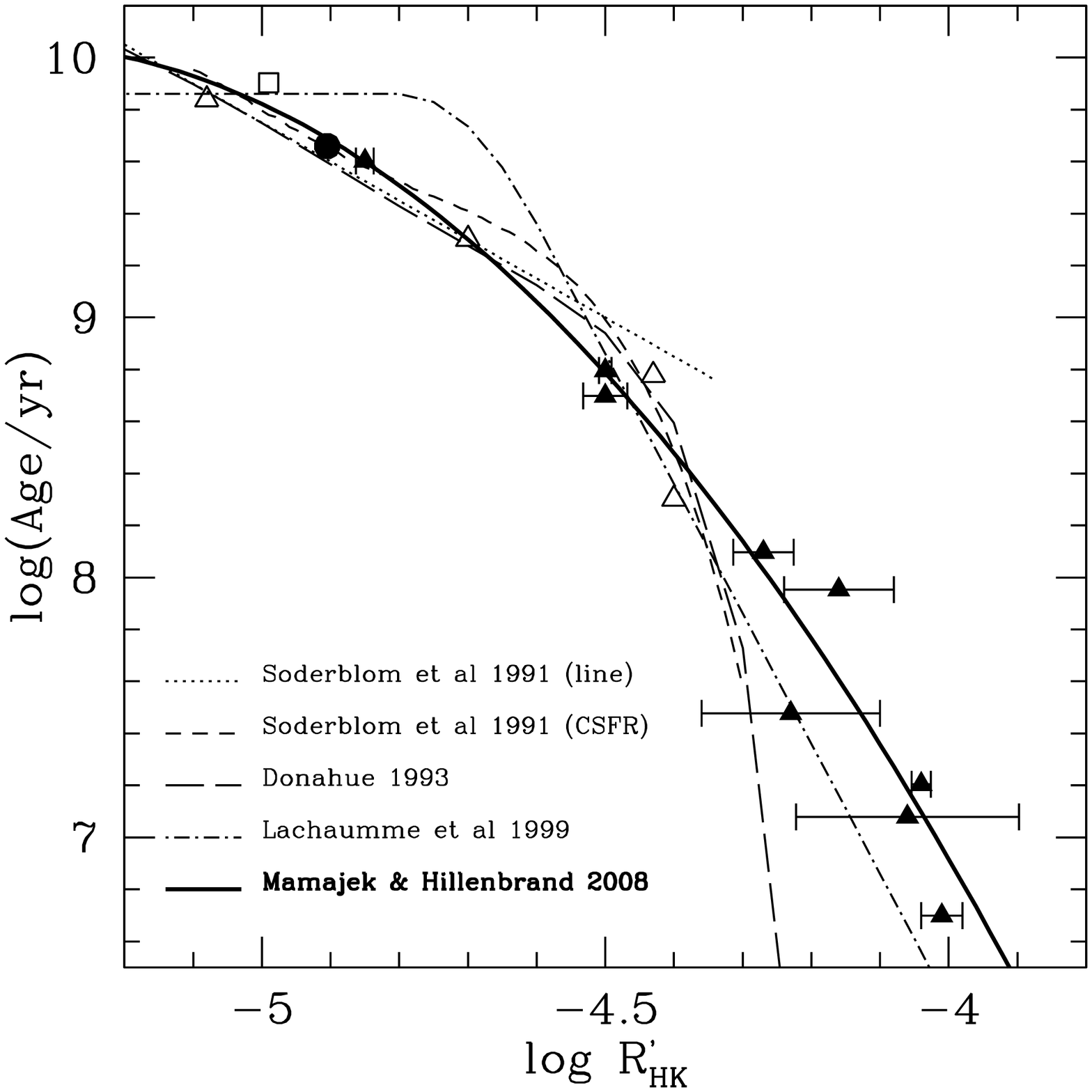}
\caption{Mean \logrphk\, cluster values (interpolated to solar \bv)
vs. cluster age. {\it Filled triangles} are cluster mean \logrphk\,
values. {\it Open triangles} are ancillary cluster mean \logrphk\,
values listed in Table \ref{tab:cluster_other}.  {\it Open square} is
the mean datum for the 5-15 Gyr-old solar-type dwarfs from
\citet{Valenti05} with isochronal age uncertainties of $<$20\%. The
{\it filled circle} is the Sun. Previously published activity-age
relations are plotted as dotted and/or dashed
curves. \citet{Soderblom91} attempted two fits: ({\it dotted}) a
linear fit to his cluster data, and ({\it long dashed}) a fit that
assumes a constant star-formation rate (CSFR) taking into account disk
heating. Our best fit polynomial to the data in Tables
\ref{tab:cluster} and \ref{tab:cluster_other} is the {\it dark solid
line} (Eqn. \ref{rhk_new_calib}).
\label{fig:new_old_clusters}}
\end{figure}

In Fig. \ref{fig:new_old_clusters}, we plot the mean \logrphk\, values
vs. cluster age.  The data are the combined set of: individually
assessed \logrphk\, measurements from Table \ref{tab:cluster_ca} along
with their 1-$\sigma$ confidence levels, and adopted mean \logrphk\,
values from Table \ref{tab:cluster_other}. In both cases the ordinate
values have been corrected to a nominally solar-color population.  The
best unweighted quadratic fit to the cluster data\footnote{If the
``classical'' ages for the $\alpha$ Per and Pleiades clusters are
adopted \citep[51 Myr and 77 Myr, respectively;][]{Mermilliod81}
instead of the Li-depletion ages, there is negligible impact on this
fit: \rm{log}($\tau$) = -36.331 - 17.213\, \rm{log}(R$^{'}_{HK}$) -
1.5977\, \rm{log}(R$^{'}_{HK}$)$^2$. The general effect is that the
very active stars become roughly $\sim$5\% younger.} is:

\begin{equation}
\log\,\tau = -38.053 - 17.912\, \log\,R^{'}_{HK} - 1.6675\, \log\,(R^{'}_{HK})^2
\label{rhk_new_calib}
\end{equation}

and its inverse (better fit as a trinomial)

\begin{equation}
\log\,R^{'}_{HK} = 8.94 - 4.849\, \log\,\tau\, + 0.624\, (\log\,\tau)^2 - 0.028\, (\log\,\tau)^3
\label{rhk_new_calib_inverse}
\end{equation}

where $\tau$ is the age in years, and where the fit is only
appropriate approximately between \logrphk\, values of -4.0 and -5.1
and log($\tau$) of 6.7 and 9.9 (the approximate range covered by our
cluster samples).  Our new function is plotted with the cluster mean
activity values and the previously published activity-age relations in
Fig. \ref{fig:new_old_clusters}.  Along the active sequence (-5.0 $<$
\logrphk\, $<$ -4.3) corresponding to ages older than the Pleiades,
the observed r.m.s. in the fit is only log($\tau$/yr) = 0.11 dex
($\sim$29\%). When the lower-accuracy ancillary cluster data
(\S\ref{Ancillary}) are removed, the r.m.s. for \logrphk\, $<$ -4.3 is
only $\sim$0.07 dex in log($\tau$/yr).  We believe the latter value is
more representative of the fidelity of our activity-age relation
(Eqn. \ref{rhk_new_calib}).  For the very active stars having
\logrphk\, $>$ -4.3, the r.m.s. in the fit is log($\tau$/yr) = 0.23
dex ($\sim$60\%).  While the age calibration has an unquantified
systematic uncertainty due to the uncertainty in the cluster age
scale, these r.m.s. values represent lower limits on the calibration
uncertainty assigned to ages from \logrphk\, measurements.

What is the typical uncertainty due to observational uncertainties or
variability? To quantify this we apply equation \ref{rhk_new_calib} to
our binary and cluster samples. For the binary samples, the mean age
inferred for the binary from the two \logrphk\, values is assumed to
be the correct system age.  Among the 20 color-separated solar-type
dwarf binaries in Table \ref{tab:pairs}, the mean dispersion in the
ages for the 40 components is $\pm$0.15 dex (1$\sigma$). Among the 14
near-identical solar-type dwarf binaries in Table \ref{tab:twins}, the
mean dispersion in the ages for the 28 components is $\pm$0.07 dex
(1$\sigma$).  The age dispersions observed among the various stellar
samples are summarized in column 2 of Table \ref{tab:rhk_test}.
Applying the relation to the well-populated Hyades and M67 activity
samples yields dispersions in the predicted ages of 0.25 dex and 0.20
dex, respectively. Hence we see slightly larger dispersions in
inferred age from among the cluster samples than among the binary
samples -- the reasons for which are not entirely clear.  Taking into
account observational uncertainties, calibration uncertainties, and
astrophysical scatter, we conclude that for solar-type dwarfs older
than a few hundred Myr the revised activity-age yields age estimates
with total accuracy $\sim$60\% (0.25 dex). For younger stars, the
uncertainty is approximately 1 dex.  In \S\ref{Implications}, we will
compare these results to those of an alternative technique -- tying
together age-rotation and rotation-activity relations to quantify the
activity-age relation as a function of color, which somewhat reduces
the scatter.

Equation \ref{rhk_new_calib} is clearly an improvement on the
previously published activity-age relations given the copious amount
of new activity data that we have incorporated into our fit,
especially for young clusters.  However, some caveats to general
applicability remain. For example, our analysis was unable to
constrain quantitatively how the color-activity slope evolves with
age. It is apparent from our cluster data that were we to adopt
equation (3) for all solar type stars, we would introduce systematic
age effects as a function of stellar color (mass).  We are thus
motivated to see if we can find an empirical means of taking into
account the color(mass)-dependent evolution of activity as a function
of age.

\begin{deluxetable}{lcc}
\tablewidth{0pt}
\tablecaption{Dispersions in Age Estimates \label{tab:rhk_test}}
\tablehead{
{(1)}             &{(2)} & {(3)}\\
{Sample}          &{$\sigma$(A)} & {$\sigma$(B)}\\
{}                &{(dex)}       & {(dex)}}
\startdata
Upper Sco         & 0.60 & ... \\
$\beta$ Pic       & 1.06 & ... \\
UCL+LCC           & 0.31 & ... \\
Tuc-Hor           & 0.66 & ... \\
$\alpha$ Per      & 1.01 & ... \\
Pleiades          & 1.12 & 1.06\\
Ursa Major        & 0.25 & 0.23\\
Hyades            & 0.25 & 0.22\\
M67               & 0.20 & 0.24\\
Color-Sep. Pairs  & 0.15 & 0.07\\
Near-Ident. Pairs & 0.07 & 0.05\\
Sun               & 0.06 & 0.05\\
\enddata
\tablecomments{
A: 68\%CL range in ages derived from \logrphk-age formula
(Eqn. \ref{rhk_new_calib}).  B: 68\%CL range in ages derived from
\logrphk\, $\rightarrow$ R$_o$ $\rightarrow$ Period $\rightarrow$
$\tau$ (\S\ref{Rossby} and \S\ref{Gyro}).
}
\end{deluxetable}

\section{Activity Ages Via the Rossby Number and Gyrochronology \label{rossby_gyro}}

Thus far we have focused on calibrating the \logrphk\, vs. age
relation empirically using cluster and young association stars of
``known'' age. In this section, we demonstrate that an age
vs. activity calibration can also be derived by combining the observed
correlation between Rossby number and \logrphk\, demonstrated by
\citet{Noyes84} with a rendition of the empirical ``gyrochronology''
rotational evolution formalism of \citet{Barnes07}.  In this section
we update both the activity vs. Rossby number relation of
\citet{Noyes84}, and the rotation vs. age relation of
\citet{Barnes07}, and then combine these into an activity-age relation
to be compared to the activity-age relation in \S3
(Equation \ref{rhk_new_calib}).

\subsection{Rossby Number vs. Activity \label{Rossby}}

\subsubsection{Rossby Number Correlated with \rhk\, Measuring Chromospheric Activity \label{rossby_rphk}}

In their classic chromospheric activity study, \citet{Noyes84} attempt
to understand the evolution of \logrphk\, in terms of the stellar
dynamo \citep[e.g.][]{Parker79}. Chromospheric activity is a
manifestation of heating by surface magnetic fields, which for the Sun
are presumed to be generated near the base of the convective
envelope. Chromospheric activity should, theoretically, scale with
magnetic dynamo number; however dynamo models are parameterized by
variables whose functional forms remain poorly constrained both
observationally and theoretically \citep[e.g.][]{Noyes84, Donahue96,
Montesinos01, Charbonneau01}. \citet{Noyes84} demonstrated that the
mean levels of stellar chromospheric activity for solar-type dwarfs
decay as Rossby number increases. The Rossby number \ro\, is
parameterized as the stellar rotation period $P$ divided by the
convective turnover time \tauc\, or \ro\ = $P$/\tauc.  Some
assumptions are necessary in arriving at values for \ro\,.

First, stars are not rigid rotators, so any estimate of the rotation
rate of an unresolved stellar disk via either chromospheric activity
or starspot modulation will be a latitudinal mean that may vary with
time during the course of stellar activity cycles
\citep{Donahue96}. Second, the Rossby number is dependent on a
convective turnover time that is {\it an estimate}, based directly on
stellar interior models \citep[e.g.][]{Kim96} or informed by the
models but empirically calibrated \citep[e.g.][]{Noyes84}. Multiple
studies have attempted to quantify the convective turnover time for
solar-type main sequence stars \citep{Noyes84,Stepien94,Kim96} and
pre-main sequence stars \citep{Jung07}. \citet{Montesinos01} show that
the \citet{Noyes84} color vs. convective turnover time relation
produces the tightest correlation between activity and Rossby number
when compared to modern stellar models using Mixing Length Theory
(MLT) and Full Turbulence Spectrum (FTS) treatments of convection.

In light of the \citet{Montesinos01} results, we adopt the
\citet{Noyes84} convective turnover time relation as a function of
\bv\, color.  Indeed, from our own data set, the need for a
color-dependent normalization of the rotation periods, i.e. the use of
Rossby number \ro, is readily apparent from examination of period vs.
activity in which the color stratification is obvious.  One caveat is
that while the color-dependent convective turnover time should be
adequate for main sequence stars, it will be systematically in error
for pre-main sequence stars. As it is often unclear whether a given
field star is pre-MS or MS (in most cases due to inadequate or lacking
distance information) {\it we adopt the MS convective turnover times
for our calculations, independent of other age-constraining
considerations}. 

\begin{figure}
\epsscale{1}
\plotone{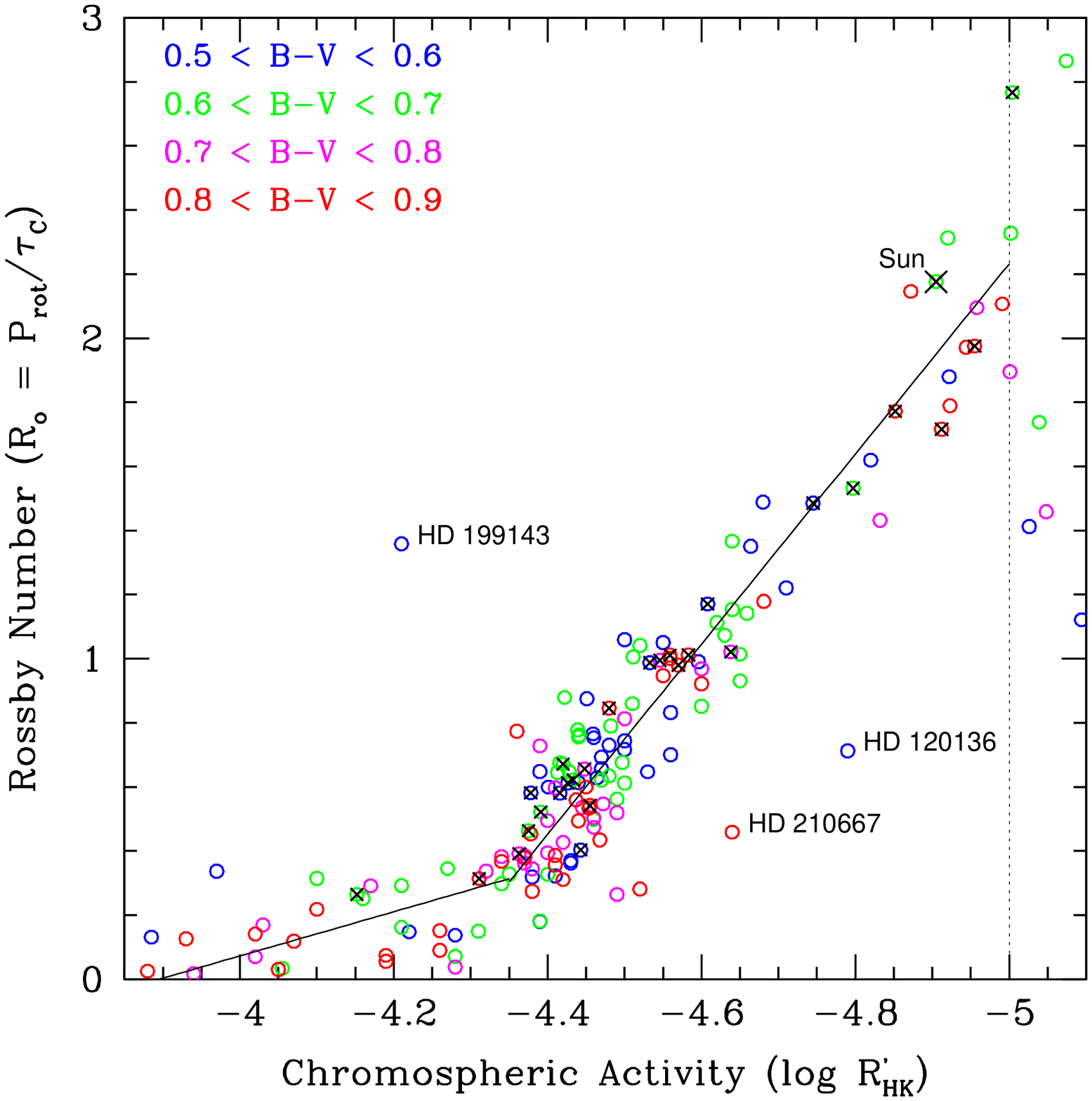}
\caption{Rossby number (\ro) versus \logrphk\, for 169 solar-type MS
or pre-MS stars with 0.5 $<$ \bvo\, $<$ 0.9 (the sample described in
\S\ref{Rotation_X-ray}).  Stars are color-coded according to the
legend.  Mt. Wilson HK survey stars with multi-seasonal mean periods
from \citet{Donahue96} and multi-decadal mean \logrphk\, from
\citet{Baliunas96} are flagged with crosses.  The best linear fits in
the very active and active regimes are plotted (equations
\ref{eqn:ro_inactive} and \ref{eqn:ro_active}). Stars with
\logrphk\, $<$ -5.0 appear to have a poor correlation between
\logrphk\, and \ro, possibly due to the increasingly important towards
low activity levels of gravity and metallicity on the photospheric
subtraction (J. Wright, priv. comm.). The Sun is marked with a
large circle with X.
\label{fig:rorphk}}
\end{figure}

In Fig. \ref{fig:rorphk}, we plot chromospheric activity
\logrphk\, vs.  Rossby number \ro.  The colored circles represent 169
solar-type MS and pre-MS (rejecting evolved stars more than 1
magnitude above the MS) stars having 0.5 $<$ \bvo\, $<$ 0.9 mag and
both measured periods and \logrphk. A subsample of 28 of these stars
have multi-seasonal mean rotation periods and \logrphk\, from
\citet{Donahue96} and \citet{Baliunas96}. These stars have the best
determined rotation periods and mean \logrphk\, values and are flagged
with black crosses in the figure. With few exceptions, the
Donahue-Baliunas stars all have published metallicity values within
$\pm$0.5 dex of solar, and the majority are within $\pm$0.2 dex of
solar \citep{Cayrel97,Cayrel01,Nordstrom04,Valenti05}.

Fig. \ref{fig:rorphk} suggests that the rotation vs. activity relation
should be clarified in three activity regimes.  In the ``very active''
regime\footnote{Note that the monikers ``very active'', ``active'', and
``inactive'' have been defined somewhat differently in other
papers \citep[e.g.][]{Henry96,Saar99,Wright04}. We delimit them based on the
appearance of Fig. \ref{fig:rorphk}.}  (\logrphk\, $>$ -4.3) there
appears to be little correlation between \logrphk\, and \ro\, (Pearson
$r$ = -0.24).  In the ``active'' regime (-5.0 $<$ \logrphk\, $<$
-4.3) there is a very strong anti-correlation between activity and
Rossby number (Pearson $r$ = -0.94). Curiously, in the ``active" and
``very active" regimes the vertical scatter at a given activity level is
roughly constant with \logrphk.  In the ``inactive'' regime
(\logrphk\, $<$ -5.0), the correlation between activity and Rossby
number is again very weak (Pearson $r$ = +0.33).  The inactive
regime (\logrphk\, $<$ -5.0) is exactly where \citet{Wright04b}
suggest that the age-activity correlation fails based on correlation
of inferred \logrphk\, with height above the main sequence.  Wright
(2009, in prep.) suggests that the definition of \logrphk\, may
require inclusion of a gravity-sensitive correction.  For the purposes
of our study, we omit the inactive stars (\logrphk\, $<$ -5.0)
from further rotation-activity analysis.


In Fig. \ref{fig:rorphk} we fit a OLS bisector line to the
``active'' (-5.0 $<$ \logrphk\, $<$ -4.3) sequence of solar-type
dwarfs, finding:

\begin{equation}
R_o = (0.808\pm0.014) - (2.966\pm0.098)(\log R^\prime_{\rm HK} + 4.52)
\label{eqn:ro_inactive}
\end{equation} 
and
\begin{equation}
\log R^\prime_{\rm HK} = (-4.522\pm0.005) - (0.337\pm0.011)(R_o - 0.814).
\label{eqn:rhk_inactive}
\end{equation} 

In this activity-rotation regime, the r.m.s. of the fits is $\sim$0.16
in \ro\, and $\sim$0.05 in \logrphk. Two obvious outliers were omitted
in the analysis (HD 210667 and HD 120136)\footnote{Multiple
independent estimates of \logrphk\, have been reported for HD 210667
\citep{Duncan91,Henry96,Gray03,Wright04,White07} and for HD 120136
\citep{Duncan91,Baliunas96,Wright04,Hall07}, so their activity levels
are well-constrained.  HD 210667 would appear to be normal inactive
star in Fig. \ref{fig:rorphk} if its period were 2$\times$ that
reported by \citet[][9.1 days]{Strassmeier00}, so it is possible that
this is a case of period aliasing (i.e. its true period is $\sim$18
days?). The other outlier is the famous star HD 120136 ($\tau$ Boo),
one of the first stars discovered to have a Hot Jupiter
\citep{Butler97}. Mean rotation periods have been reported by
\citet[][3.2\,$\pm$\,0.5 day]{Henry00} and \citet[][3.5\,$\pm$\,0.7
day]{Walker08}, and the rotation rate is suspiciously close to the
orbital period of 3.3 days for the planet \citep{Butler97}.
\citet{Walker08} concludes that the planetary companion is
magnetically inducing long-lived active regions on the star.
\citet{Henry00} similarly noted that the measured rotation period for
$\tau$ Boo is significantly shorter than what one would infer from its
activity level. Figure \ref{fig:rorphk} suggests that $\tau$ Boo's
unusual Rossby number vs. activity behavior is mimicked by $<$few\% of
solar-type field dwarfs.}.


In the ``very active'' regime in Fig. \ref{fig:rorphk} (\logrphk\, $>$
-4.3) the correlation between rotation and activity is very
weak. However we can still assess the empirical relation between
rotation and activity in this regime, even if the predictability of
the dependent variable on the independent variable is weak. Omitting
the outlier HD 199143 (a pre-MS late-F binary), for the stars with
\ro\, $<$ 0.4 in Fig. \ref{fig:rorphk}, we fit:

\begin{equation}
R_o = (0.233\,\pm\,0.015) - (0.689\,\pm\,0.063)(\log\,R^{\prime}_{HK} + 4.23)
\label{eqn:ro_active}
\end{equation} 

\begin{equation}
\log\,R^{\prime}_{HK} = (-4.23\,\pm\,0.02) - (1.451\,\pm\,0.131)(R_o - 0.233)
\label{eqn:rhk_active}
\end{equation} 

The r.m.s. of the fits is $\sim$0.10 in \ro\, and $\sim$0.16 in
\logrphk. While the r.m.s. in \ro\, is less for the very active than for
the active sequence, the fractional uncertainty in \ro\,
($\sim$50\%) is larger.  As the low Pearson $r$ for the data in the
active regime of Fig. \ref{fig:rorphk} reflects, the power to predict
activity given \ro, or vice versa, is limited with our current
toolkit.  The enhanced scatter for the very active stars is likely due
to: (1) increased variability, (2) only one or few \logrphk\,
measurements, and (3) inclusion of likely MS as well as pre-MS stars,
implying a spread in convective turnover times that is not being taken
into account.  For our purposes, we match the very active and active
sequence fits (Eqns. \ref{eqn:rhk_inactive}, \ref{eqn:ro_inactive},
\ref{eqn:rhk_active}, and \ref{eqn:ro_active}) at \logrphk\, = -4.35
and \ro\, = 0.32.

\subsubsection{Rossby Number Correlated with \rx\, Measuring Coronal Activity \label{rossby_rx}}

In addition to their chromospheric activity quantified via fractional
Ca II H\&K luminosity, \logrphk, young stars are often noted for
copious coronal activity and X-ray emission.  There appear to be at
least two rotation-activity regimes inferred from X-ray surveys
\citep[e.g.][]{Pizzolato03}: a ``saturated'' regime for very active,
fast-rotating stars where there is little correlation between rotation
and activity (\logrx\, $\simeq$ -3.2), and a ``non-saturated'' regime
of slower-rotating, lower activity stars where rotation and X-ray
emission are correlated (-7 $<$ \logrx\, $<$ -4). As rotation slows
with stellar age, one would surmise that X-ray emission (especially
``non-saturated'') can be a useful tool for estimating the ages of
solar-type star.

\begin{figure}
\epsscale{1}
\plotone{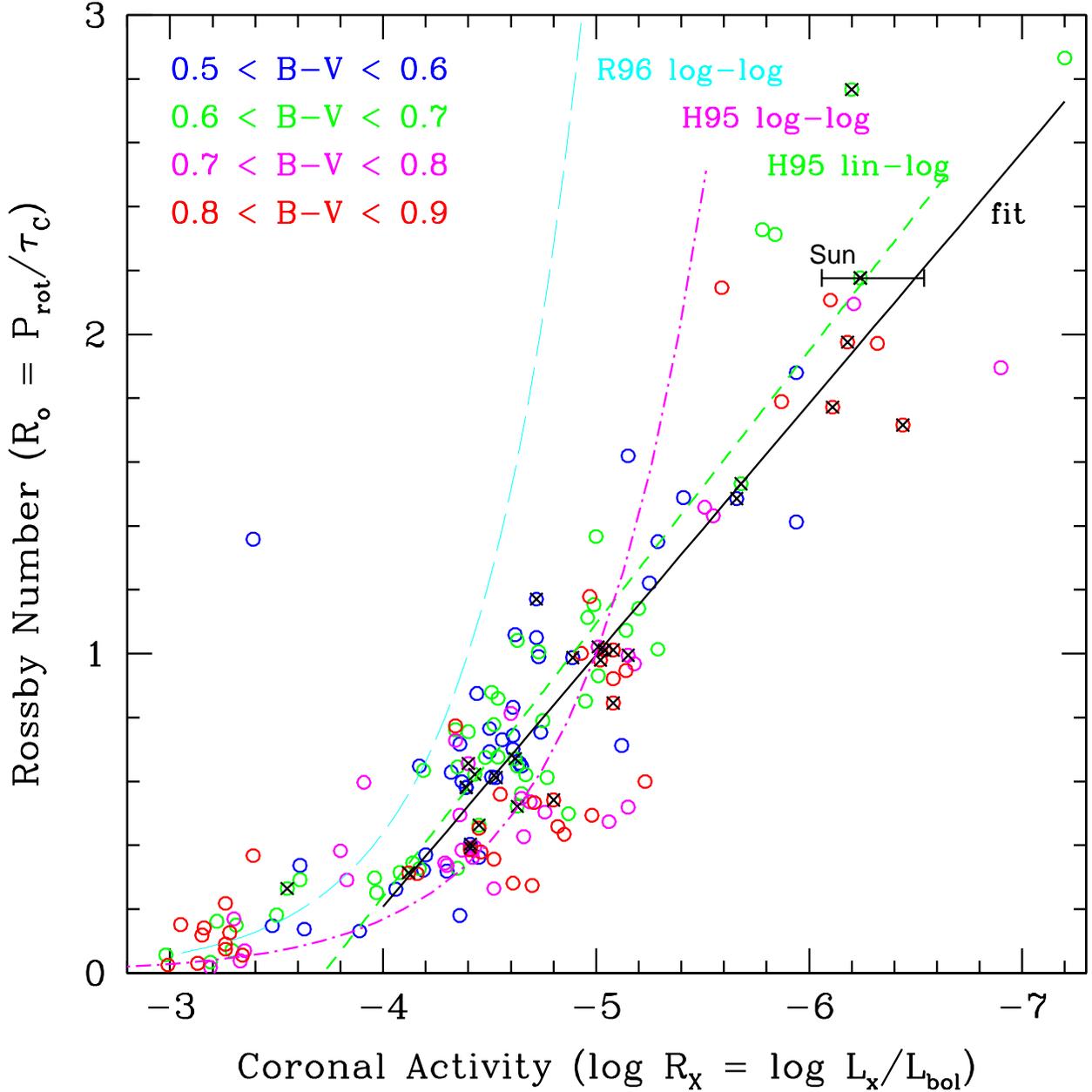}
\caption{\logrx\, vs. Rossby number \ro\, for stars in our sample of
solar-type stars with known rotation periods and chromospheric and
X-ray activity levels. Donahue-Baliunas stars with well-determined
periods also have dark Xs. Previously published \rx\, vs. \ro\, fits
are drawn: {\it cyan long-dashed line} is a log-log fit from
\citet{Randich96}, {\it magenta dot-dashed line} is a log-log fit from
\citet{Hempelmann95}, and the {\it green dashed line} is a linear-log
fit from \citet{Hempelmann95}. Our new log-linear fit for stars in the
range -7 $<$ \logrx\, $<$ -4 is the {\it solid dark line}, consistent
with the Hempelmann linear-log relation. Saturated X-ray emission
(\logrx\, $>$ -4) is consistent with \ro\, $<$ 0.5.
\label{fig:logrx_logro}}
\end{figure}

In Figure \ref{fig:logrx_logro} we show that coronal activity can be
related to Rossby number in a manner similar to that displayed in
Figure \ref{fig:rorphk} for the relation of chromospheric activity and
Rossby number \citep[see also e.g.][and references
therein]{Hempelmann95,Randich96,Pizzolato03}. As previous authors have
noted, finding a simple function form that adequately describes the
relationship between \logrx\, and \ro\, over the full range of
activity data available is difficult \citep[e.g.][]{Hempelmann95}.
Fig. \ref{fig:logrx_logro} shows three previous fits to the \logrx\,
vs. \ro\, data, one from \citet{Randich96} and two from
\citet{Hempelmann95}, plotted over the full range of activity sampled
by the respective authors. The \citet{Randich96} fit in
Fig. \ref{fig:logrx_logro}) comes from a very small sample of stars in
the young $\alpha$ Per cluster, and appears to miss the majority of
the data. The \citet{Hempelmann95} log-log fit in
Fig. \ref{fig:logrx_logro}) passes through the majority of
intermediate activity stars, but is a poor fit for the low-activity
stars (overestimating the Sun's X-ray emission by an order of
magnitude). The \logrx\, vs. \ro\, (log-linear) fit of
\citet{Hempelmann95} is satisfactory for the intermediate and
low-activity stars, but extrapolation above \logrx\, $>$ -4 (i.e.  the
saturated X-ray regime) is not recommended.

Following \citet{Hempelmann95}, we fit a log-linear regression to the
rotation-activity data. The range of fractional X-ray luminosities
over which there is a good correlation between \logrx\, and \ro\, (-7
$<$ \logrx\, $<$ -4) approximately overlaps the ``active'' regime in
Fig. \ref{fig:rorphk} (-5 $<$ \logrphk\, $<$ -4.3; see Appendix).
Over the ``active'' sequence, the fit

\begin{equation}
{\rm R_o} = (0.86 \pm 0.02) - (0.79 \pm 0.05)\,(\log R_X + 4.83)
\label{eqn:ro_logrx}
\end{equation} 

\noindent produces an r.m.s. scatter of 0.25 in Rossby number \ro. The
inverse relation is:

\begin{equation}
\log R_X = (-4.83 \pm 0.03) - (1.27 \pm 0.08)\,({\rm R_o} - 0.86)
\label{eqn:logrx_ro}
\end{equation} 

\noindent with an r.m.s. of 0.29 dex in \logrx. The correlation
between \logrx\, and Rossby number is very strong (Pearson $r$ =
-0.89). These fits are not applicable for stars with \logrx\, $>$ -4
that are nearing the saturated X-ray emission regime.  Saturated X-ray
emission appears to imply Rossby numbers \ro\, $<$ 0.5 (rotation
period $<$ 6 days for a G2 dwarf), and hence can be used to estimate
an upper limit to the rotation period. This transition region is
similar to that seen for \logrphk\, near \logrphk\, $\simeq$ -4.3
(Fig. \ref{fig:rorphk}). In the Appendix, we further quantify the
relationship between these chromospheric and coronal activity
indicators.

The r.m.s. scatter in \ro\, values inferred from \logrx\, is
comparable to that inferred from \logrphk\, values in
single-measurement or multi-year surveys (\S\ref{Rossby}; 0.25
vs. 0.16 in $\sigma$(\ro)), although the scatter for averaged data
from multi-decade Mt. Wilson HK observations is smaller (0.10 in
$\sigma$(\ro)). This suggests that soft X-ray luminosities can be used
to infer the rotation rate of old solar-type dwarfs almost as
accurately as most \logrphk\, values in the literature.

\subsubsection{Considerations for a Rotation-Activity-Age Relation \label{considerations}}

In the next section (\S\ref{Gyro}), we will attempt to derive a
rotation vs. age relation for solar-type dwarfs of a given color. Our
end goal is to combine an activity-rotation relation with a
rotation-age relation (next section; \S\ref{Gyro}) to produce an
activity-age relation to compare to equation \ref{rhk_new_calib}. As
we intend to infer rotation rates from activity levels, we would like
to know how accurately the uncertainty in \ro\, reflects the
uncertainty in rotation period from equations \ref{eqn:ro_active} and
\ref{eqn:ro_inactive}. From the definition of the Rossby number (\ro\,
= P/\tauc), the uncertainty in period is $\sigma_{P}$\, $\approx$\,
$\tau_{c}$\, $\sigma_{R_o}$. While \tauc\, varies from star-to-star as
a function of color, its mean value in our color range of interest is
$\sim$15 days, and hence a typical uncertainty in the predicted period
$\sigma_P$ is $\sim$1.5 days (ranging from $\sim$0.8 days for the
late-Fs to $\sim$2.2 days for the late-Ks). A good approximation for
the uncertainty in the period (in days) inferred from \logrphk\, for
late-F through early-K stars is:

\begin{equation}
\sigma_{P}\, \simeq\, 4.4 (B-V) (\sigma_{R_o}/0.1) - 1.7
\end{equation}

where $\sigma_{R_o}$ $\simeq$ 0.1 for stars with multi-decadal
\logrphk\, means \citep[i.e. Mt. Wilson HK survey stars,
e.g.]{Baliunas96}. For stars from \citet{Wright04}, with typically
dozens of \logrphk\, measurements over a span of a few years, the
scatter in $R_o$ as a function of \logrphk\, is $\sigma_{R_o}$
$\simeq$ 0.17. For stars with measured rotation periods, but with a
few to tens of \logrphk\, measurements
\citep[e.g.][]{Duncan91,Henry96,White07}, the scatter in $R_o$ as a
function of \logrphk\, is $\sigma_{R_o}$ $\simeq$ 0.2. In the limit of
a single \logrphk\, measurement, it appears that one should be able to
estimate $R_o$ to $\sim$0.2-0.3 1$\sigma$ accuracy for solar-type
dwarfs. This is comparable to the accuracy in \ro\, that single X-ray
observations can produce ($\sigma_{R_o}$ $\simeq$ 0.25;
\S\ref{rossby_rx}). Surveying the suite of coronal and X-ray activity
indicators published for thousands of stars, it appears that we can
predict rotation for the majority to better than $\pm$0.25 in \ro.

For the Sun's observed mean rotation period as measured through the
Mt. Wilson S-index \citep[26.09 day;][]{Donahue96}, one would predict
the Sun's mean chromospheric activity to be \logrphk\, = -4.98.  This
can be compared to the observed value, time-averaged over several
solar cycles, of -4.91 (\S 1.1).  As the observed r.m.s. in \logrphk\,
vs. \ro\, along the inactive sequence is only $\sim$0.05 dex, the
Sun's past 40 years of activity appears to be only $\sim$1.3$\sigma$
higher than predicted for its period. This corroborates previous
findings that Sun appears to have more or less normal activity for its
rotation period \citep[e.g.][]{Noyes84}. 


From the results of large chromospheric activity surveys
\citep[e.g.][]{Henry96,Wright04} for solar-type stars within 1 mag of
the MS, it appears that $\sim$76\% of solar-type field stars fall
within the active sequence (-5.0 $<$ \logrphk\, $<$ -4.35), $\sim$3\%
fall within the very active sequence (\logrphk\, $>$ -4.35), and
$\sim$21\% are inactive (\logrphk\, $<$ -5.0). The coronal activity
surveys show a similar distribution.  Hence, for roughly
three-quarters of the solar-type dwarfs, we have a well-determined
empirical rotation-activity relation where we can reliably use
activity to predict rotation period, or vice versa.  This corroborates
the results of \cite{Noyes84}. More importantly, we provide a modern,
well-established activity-rotation relationship using the best
available data. Our next step is to revisit the rotation-age
relationship, with the eventual goal of producing an
activity-rotation-age relationship with more predictive power than an
activity-age relation.

\subsection{Gyrochronology \label{Gyro}}

In the course of their evolution, solar-type stars lose angular
momentum via magnetic breaking due to their mass loss \citep{Weber67}.
This inexorably leads to a steady slowdown in rotation rates, first
quantified by \citet{Skumanich72} as projected rotation speed
$v$\,sin\,$i$ $\propto$ age$^{-0.5}$. Detailed surveys of solar-type
stars in open clusters \citep[beginning with the summary
in][]{Kraft67} have shown that the evolution in rotation period has a
mass-dependence.

Recently, \citet{Barnes07} used existing literature data to derive a
color-dependent version of the Skumanich law (``gyrochronology''). For
a given age, Barnes finds that the majority of solar-type stars in
clusters follow what he calls the {\it interface} or ``I'' rotational
sequence. The choice of nomenclature is theoretically motivated, as it
is believed that these stars are producing their magnetic flux near
the convective-radiative interface. Barnes dubs the population of
ultra-fast rotators the ``C'' or {\it convective} rotational sequence,
and posits that these stars lack large-scale dynamos, and hence break
their rotation very inefficiently \citep[see
also][]{Endal81,Stauffer84,Soderblom93}.  According to
\citet{Barnes07}, the rotation periods for I-sequence stars evolve
with age as:

\begin{eqnarray}
P(B-V, t) = f(B-V)\times g(t)   \label{gyro1}\\
f(B-V) = a((B - V)_{o} - c)^{b} \label{gyro2}\\
g(t) = t^{n}                    \label{gyro3}
\end{eqnarray}

With the age of the star $t$ given in Myr, Barnes finds $a$ =
0.7725\,$\pm$\,0.011, $b$ = 0.601\,$\pm$\,0.024, the ``color
singularity'' $c$ = 0.40 mag, and the time-dependence power law $n$ =
0.5189\,$\pm$\,0.0070.  In practice, Barnes segregates the I- and
C-sequence rotators at the 100 Myr {\it gyrochrone}, and does not
attempt to estimate ages for faster rotating stars. These coefficients
are claimed to satisfy the above gyrochronology relation for the Sun
and several young open clusters, and to match well a sample of
color-separated binaries with known rotation periods (e.g. $\alpha$
Cen, 61 Cyg, etc.).

\begin{figure}
\epsscale{1}
\plotone{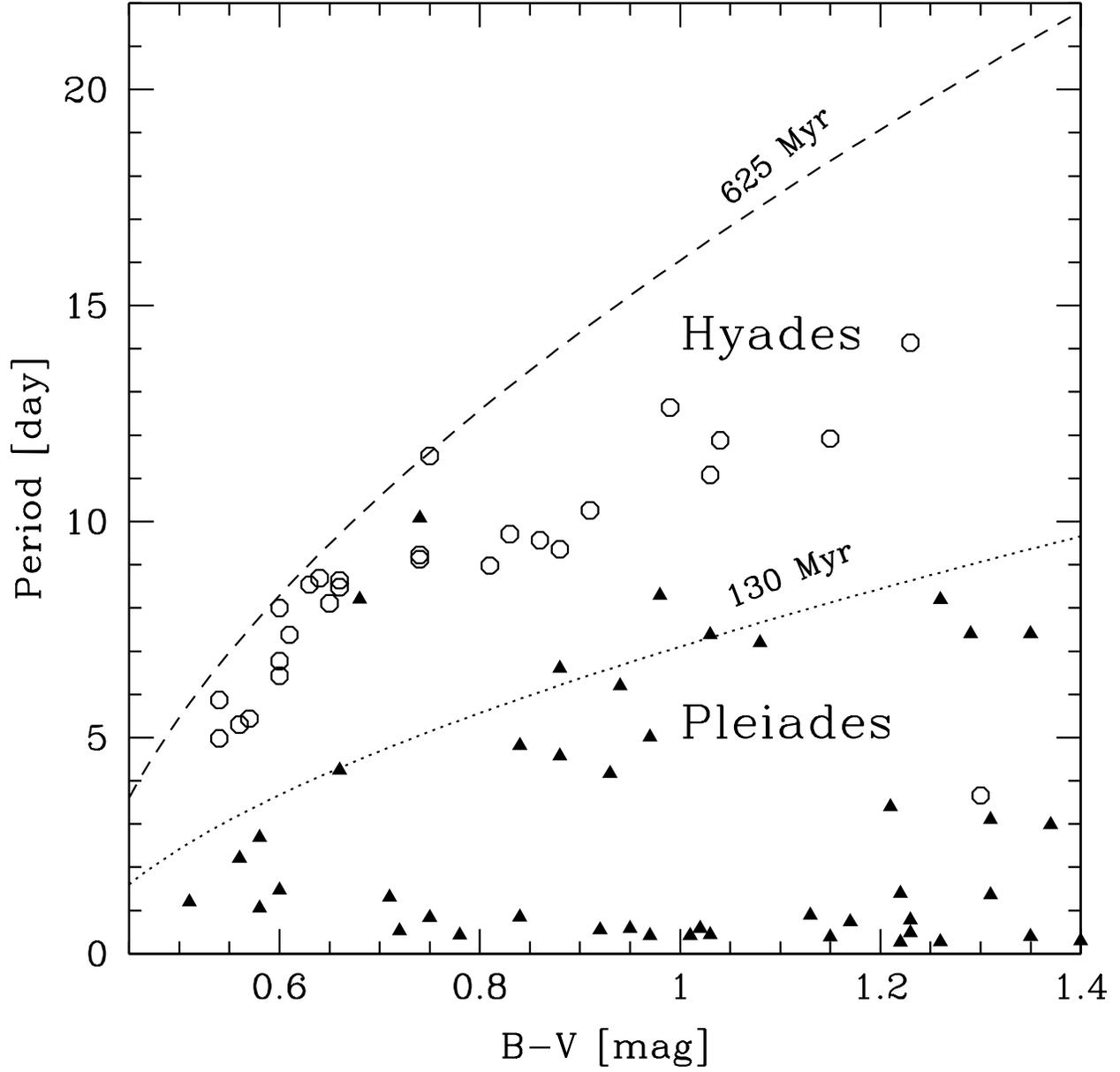}
\caption{Rotation period versus \bv\, for solar-type stars in
the Pleiades ({\it filled triangles}) and Hyades ({\it open circles})
compared to gyrochrones from \citet{Barnes07} for ages 130 Myr and 625
Myr. The offsets between the gyrochrones and the observed period
distributions for these benchmark clusters motivated us to rederive
the parameters in the gyro relations.\label{fig:per_gyro_barnes}}
\end{figure}

An independent assessment of data for the Sun, Hyades, and Pleiades
reveals discrepancies when using the gyrochronology relations from
\citet{Barnes07}. As illustrated in Fig. \ref{fig:per_gyro_barnes},
the Barnes ``gyrochrone'' for an age of 625 Myr over-predicts the
periods of Hyades members as a function of color by as much as 50\%,
suggesting the need for modification in $a$ and/or $b$. For the
Pleiades (130 Myr), the agreement is better overall, but disagreement
most prevalent for the bluer members, suggesting that the value of $c$
needs revision.  To produce suitable fits over a wide range of ages
within the Barnes formalism, we were forced to rederive the parameters
$a$, $b$, $c$, and $n$.

Considering the clusters of Tables \ref{tab:cluster} and
\ref{tab:cluster_other}, we find after a thorough literature search
that only a few have sufficient data on stellar rotation periods for
inclusion in this exercise.  They are the usual suspects: $\alpha$ Per
\citep{Prosser95}, Pleiades \citep{Prosser95,Krishnamurthi98}, M34
(Meibom et al., submitted), and Hyades
\citep{Radick87,Prosser95,Radick95,Paulson04}, and Henry
(priv. comm.).  Rotation data for benchmark clusters older than the
Hyades (such as Coma Ber, NGC 752, M 67, and NGC 188) are hard to come
by given the long mean periods of $>$10 days.  However, increased
interest in both planet searches and stellar oscillation studies may
soon rectify this situation.  We also include the Sun as an old anchor
datum, adopting a period of 26.09 days which is the latitudinal mean
observed by \citet{Donahue96} (the solar rotation ranges from $\sim$25
days near the equator to $\sim$32 days near the poles).

To rederive a gyro relation which more closely matches the cluster
sequences and the Sun, we include in the fit only the obvious
I-sequence rotators in the clusters, and omit the ultrafast C-sequence
rotators, as well as the two very slow rotators in the Pleiades (HII
2284 \& 2341).  For the four gyro parameters, we minimize the
residuals in period for the cluster data and solar datum, but
retaining only those fits that come within 0.1 day of the solar mean
rotation rate at its age. Our method forces perhaps undue statistical
significance upon this one data point (the Sun); however, as we are
lacking in cluster sequences or even single stars with accurate ages
$>$625 Myr, the solar datum is unique and thus extremely important to
reproduce. We also ignore the effects of metallicity on the cluster
sequences, working in color rather than mass.

\begin{figure}
\epsscale{1}
\plotone{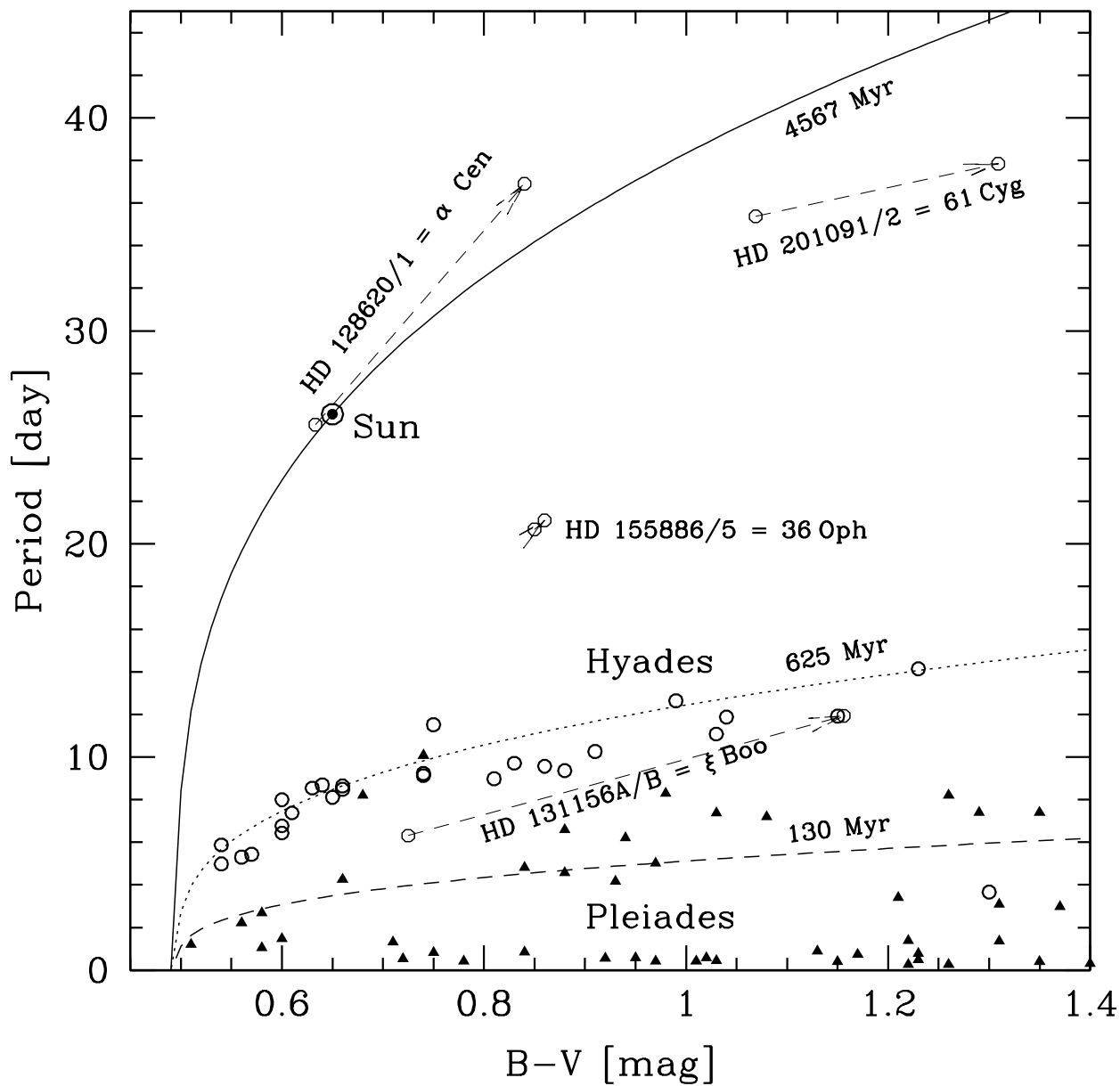}
\caption{Rotation period versus \bv\, for solar-type stars compared
to gyrochronology relations derived in this work.  Stars are
color-coded by anchor: Sun ({\it blue}), Hyades ({\it green}), M34
({\it magenta}), Pleiades ({\it red}).  In black are binary pairs,
which are presumed co-eval systems that follow the general sense of
the cluster data and the fitted gyrochrones.
\label{fig:bv_per}
}
\end{figure}

Our best estimate of the gyrochronology parameters are presented in
Table \ref{tab:gyronew}.  The errors reflect the uncertainties of the
parameters for $\Delta \chi^2$ = 1, where r.m.s = 1.23 day gives
$\chi^2_{\nu}$ = 1 for the best fit.  In Figure \ref{fig:bv_per} we
demonstrate the match of these coefficients to the data from which
they were established.

\begin{deluxetable}{cc}
\tablewidth{0pt}
\tablecaption{Revised Gyrochronology Parameters}
\tablehead{
{param.}&{value}
}
\startdata
a & 0.407\,$\pm$\,0.021\\
b & 0.325\,$\pm$\,0.024\\
c & 0.495\,$\pm$\,0.010\\
n & 0.566\,$\pm$\,0.008
\label{tab:gyronew}
\enddata
\end{deluxetable}

How well does our improved gyrochronology fit perform for the sample
four solar-type dwarf binaries with known periods
(\S\ref{Data_binaries})?  In Fig. \ref{fig:bv_per}, we also show that
the color-period lines connecting the binary components appear to
follow approximately the slopes of the curves predicted from our new
gyrochrone curve (\S\ref{Gyro}). In Table \ref{tab:bin_per_gyro} we
present revised estimates of the individual gyrochronological ages
based on our revised parameters for equations \ref{gyro1}-\ref{gyro3}.

Assuming the systems are coeval, our revised fit to the gyro equations
appears to yield stellar ages with precision of $\pm$0.05 dex
(1$\sigma$; $\pm$11\%) in log($\tau$/yr). This is comparable to the
precision claimed by \citet{Barnes07}; however the ages should be more
accurate as the Pleiades and Hyades color sequence is more accurately
modeled (c.f. Figures~\ref{fig:bv_per} vs ~\ref{fig:per_gyro_barnes}).
For the best studied system ($\alpha$ Cen), the inferred gyro age
(5.0\,$\pm$\,0.3 Gyr) compares well to recent estimates from modeling
asteroseismology data, which have been converging to a consensus age
of 6\,$\pm$\,1 Gyr in recent years: 4.85\,$\pm$\,0.5 Gyr
\citep[][]{Thevenin02}, $\sim$6.4 Gyr \citep[][]{Thoul03},
6.52\,$\pm$\,0.3 Gyr \citep[][]{Eggenberger04}, 5.2-7.1 Gyr
\citep[][]{Miglio05}.

We conclude that our improved gyrochronology fit is probably precise
to of order $\sim$0.05 dex in log($\tau$/yr) for I-sequence
rotators. This uncertainty does not include the absolute uncertainties
in the clusters age scale (which are probably of similar magnitude;
$\sim$15\%). Clearly, new samples of stars with well-constrained
rotation periods and ages at a range of colors are needed to constrain
the rotational evolution of solar-type stars at ages of $>$1 Gyr. Our
refined gyrochronology parameters represent our best attempt to
empirically parameterize the rotational evolution of solar-type stars
at present. However, we acknowledge that given the rapidly changing
data landscape for cluster rotation studies, superior rotation vs. age
relations may be soon available.

\begin{deluxetable}{llccc}
\setlength{\tabcolsep}{0.03in}
\tablewidth{0pt}
\tablecaption{Revised Gyro Ages for Field Binaries \label{tab:bin_per_gyro}}
\tablehead{
{(1)}     &{(2)} &{(3)}          &{(4)}           &{(5)}\\	  
{System}  &{HD}  &{log\,$\tau_A$}&{log\,$\tau_B$ }&{$\overline{\log \tau}$}\\
{}        &{}    &{(yr)}         &{(yr)}          &{(yr)}}        
\startdata 
$\xi$ Boo    & 131156AB & 8.47 & 8.70 & 8.59\\
$\alpha$ Cen & 128620/1 & 9.67 & 9.72 & 9.70\\
36 Oph       & 155886/6 & 9.28 & 9.28 & 9.28\\
61 Cyg       & 201091/2 & 9.57 & 9.53 & 9.55 
\enddata
\tablecomments{Columns: (1) common name, (2) HD name, (3) gyro age for
component A, (4) gyro age for component B, (5) mean gyro age for the
system. Gyro ages were estimated from the equation P = 
$a$((B - V)$_{o}$ - $c$)$^{b}$ $\times$ $t^{n}$, where
the coefficients are listed in Table \ref{tab:gyronew}.}
\end{deluxetable}

\subsection{Implications and Tests of New Gyro-Rossby Ages \label{Implications}}

\begin{figure}
\epsscale{1}
\plotone{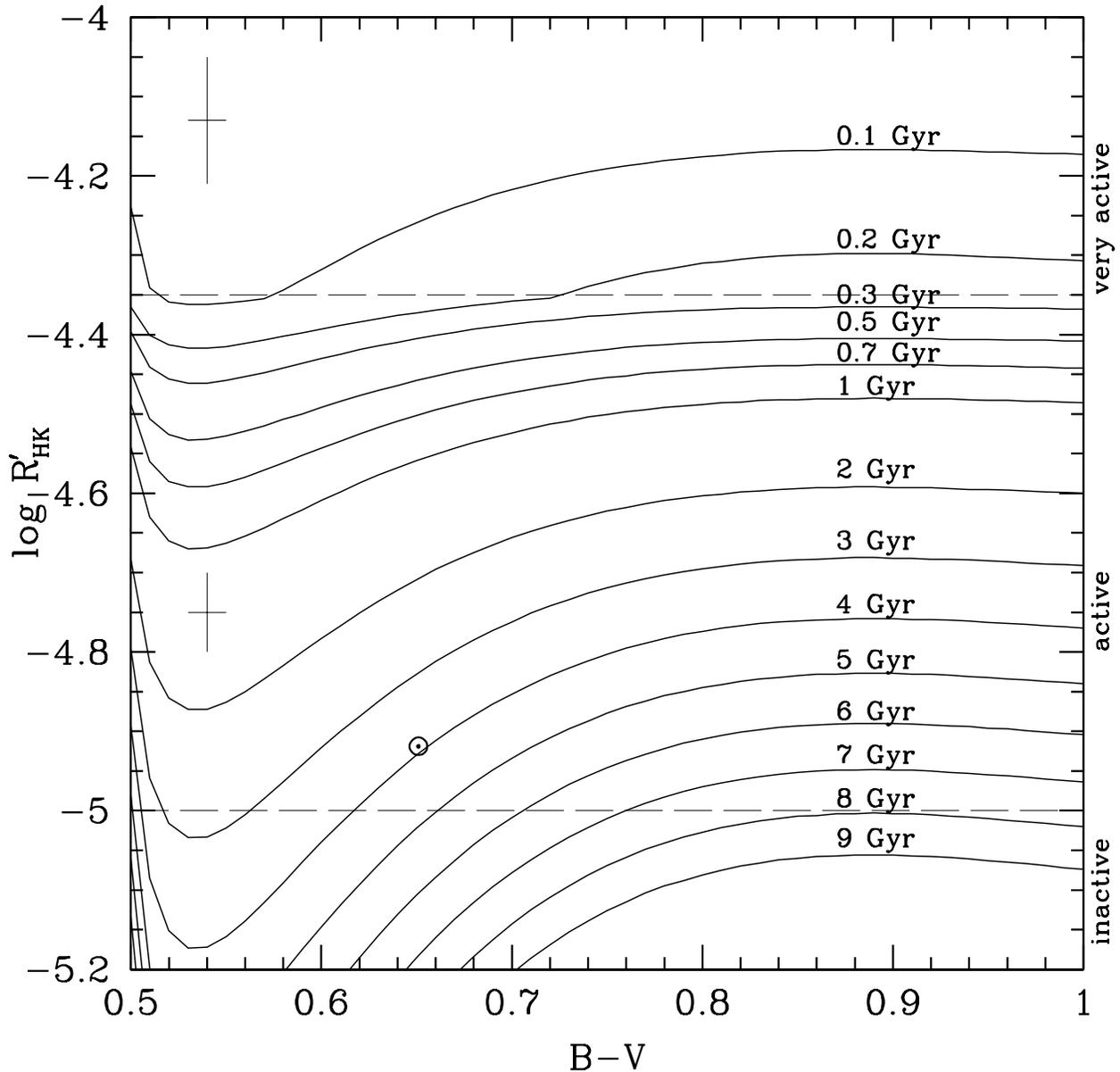}
\caption{Predicted chromospheric activity levels as a function of age
(``gyrochromochrones''), from combining the age-rotation relations in
\S\ref{Gyro} with the rotation-activity relations in \S\ref{Rossby}.
Typical uncertainty bars are shown in the very active and active regimes,
reflecting the r.m.s. in the Rossby number-activity fits, and typical
photometric errors. The behavior of the gyrochromochrones at the blue
end (i.e. the obvious upturn) is not well-constrained, and is
particularly sensitive to the $c$ parameter in the gyrochronology
fits.
\label{fig:gyrochrones}
}
\end{figure}

\begin{figure}
\epsscale{1}
\plotone{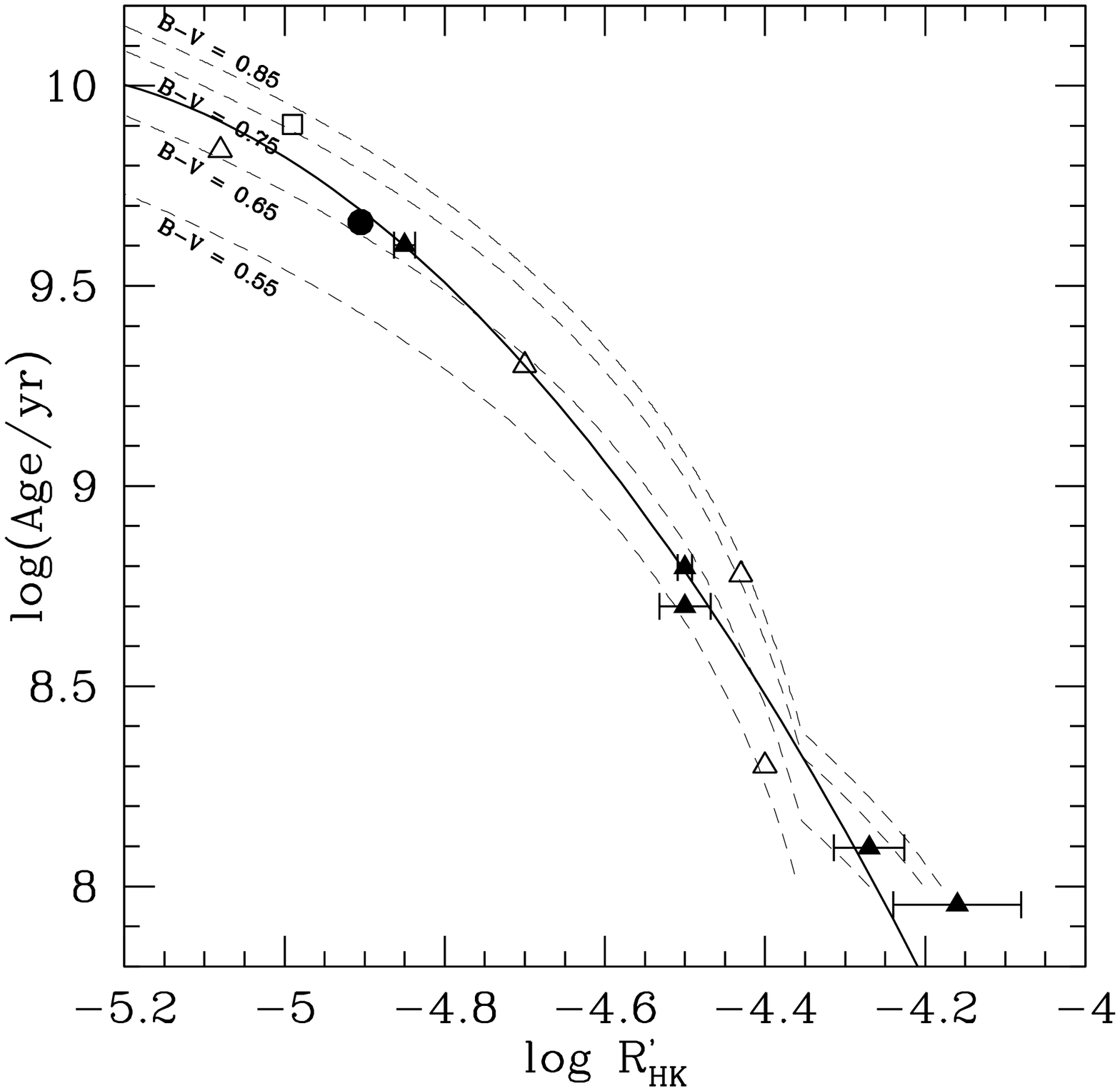}
\caption{Predicted \logrphk\, vs. age relation for solar-type dwarfs
of different colors ({\it dashed lines}). The cluster samples and mean
relation from Fig. \ref{fig:new_old_clusters} are plotted.  The {\it
dashed lines} represent the synthesis of the age-rotation
``gyrochronology'' relation (\S\ref{Gyro}) with the rotation-activity
relations (\S\ref{Rossby}). These ``gyrochromochrones'' show that the
assumption of an activity-age relation applicable to all solar-type
dwarfs in the color range (0.5 $<$ \bvo\, $<$ 0.9) is probably an
oversimplification. The kink in \logrphk\, corresponds to the
transition between the very active and active regimes.
\label{fig:cahk_predicted}
}
\end{figure}

Having calibrated the activity-rotation and rotation-age correlations
with the best available data, we can now use the results from
\S\ref{Rossby} and \S\ref{Gyro} to predict the evolution of \logrphk\,
as a function of age and color for solar-type stars. In
Fig. \ref{fig:gyrochrones}, we illustrate the predicted activity
tracks as a function of color and stellar age. In
Fig. \ref{fig:cahk_predicted}, we plot the predicted activity-age
relation for various colors of solar-type dwarfs.  Considering these
two plots leads us to a few conclusions. First, the subtle positive
mean slopes in $\Delta$\logrphk/$\Delta$\bv\, observed for the young
clusters in Fig. \ref{fig:cluster_bv_rhk} and Table \ref{tab:cluster}
can be understood in the context of mass-dependent rotation evolution
combined with an rotation-activity relation (a notable exception is
the old cluster M67). Second, the assumption of a single activity-age
relation applicable to the wide range of solar-type dwarf colors
($\sim$0.5 $<$ \bvo\, $<$ 0.9; Eqn. \ref{rhk_new_calib} and
\ref{rhk_new_calib_inverse}, and Fig. \ref{fig:new_old_clusters}) we
and others often considered is a poor assumption. The predicted
activity evolution curves in Fig. \ref{fig:gyrochrones} also warn that
the search for Maunder minimum candidates
\citep[e.g.][]{Donahue98,Wright04b} should take into account that
coeval stars may have different mean activity levels ($\sim$0.1-0.2
dex in \logrphk) as a function of \bv\, color.  The question remains,
{\it can we determine more accurate ages from a activity-rotation-age
algorithm compared to the standard activity-age relations?}

Similar to our analysis in \S\ref{Relation}, we wish to test the
consistency of our gyro-activity age predictions among two useful
types of samples: field binary stars and open cluster members.  In
each of these groups, the constituents are expected to be co-eval but
to display a range in mass, and to suffer from astrophysical scatter.
How well do the predicted ages agree among these presumably co-eval
stars?

\begin{figure}
\epsscale{1}
\plotone{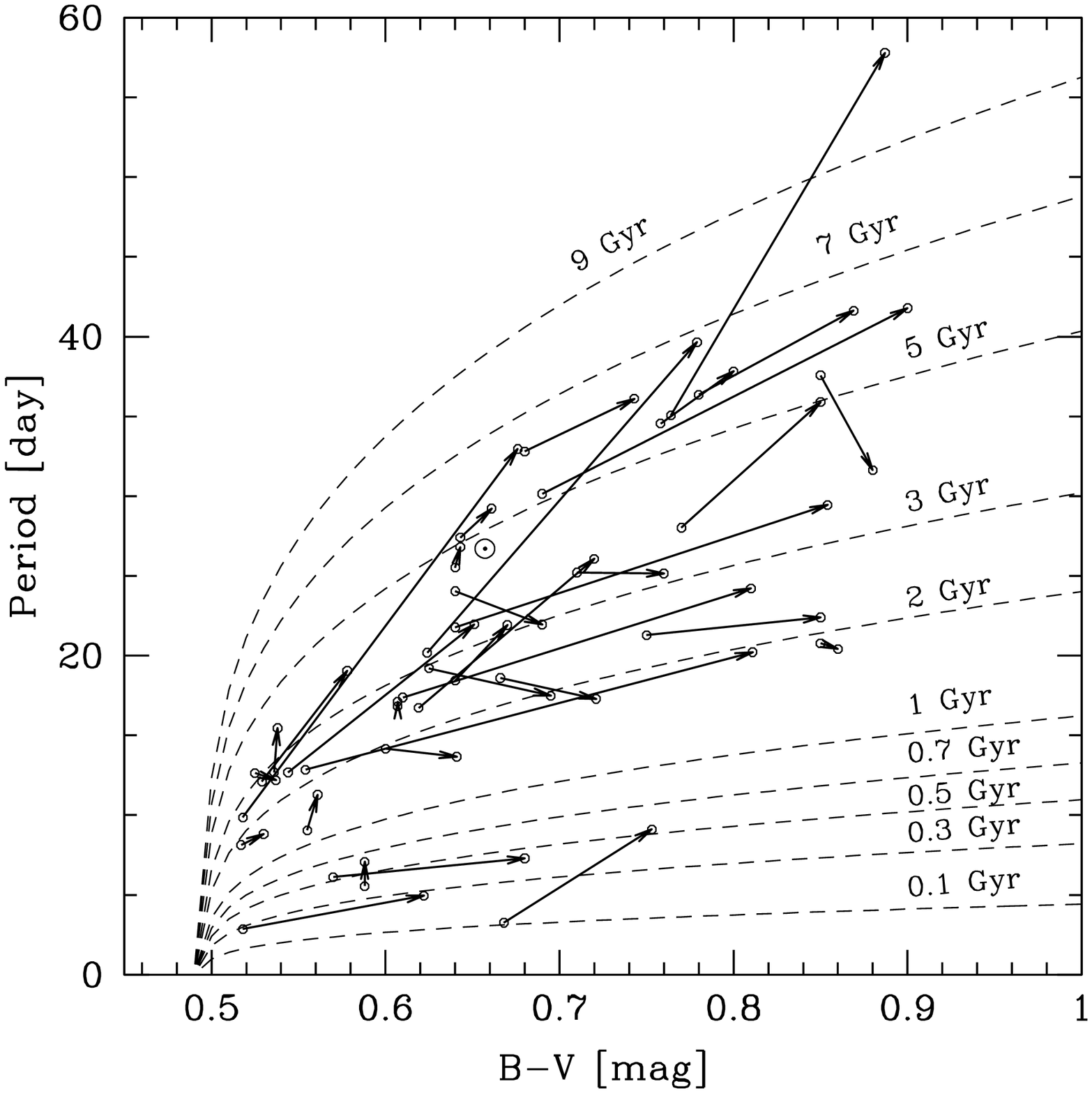}
\caption{Predicted rotation periods for field binary stars with
measured \logrphk. Periods were estimated from the activity-Rossby
relations (equations \ref{eqn:ro_inactive} and
\ref{eqn:ro_active}). Gyrochrone equations are from equations
\ref{gyro1}-\ref{gyro3} using the constants in Table
\ref{tab:gyronew}.
\label{fig:per_rhk}
}
\end{figure}

Our first test uses the 20 binary pairs of Table \ref{tab:pairs}.  We
convert the individual \rphk\ values to period via the \rphk\,
vs. Rossby number correlation, and use the gyrochronology relations to
estimate ages. The ages for these binaries are listed in Table
\ref{tab:pairs_ages}. The distribution of the periods (inferred from
the \rphk\, values) versus colors for the binaries are plotted in
Figure \ref{fig:per_rhk}, with the revised gyrochrones overlaid.
Excluding the known pathological system HD 137763 (footnote
\ref{HD_137763}), the remaining systems appear to give consistent ages
with a statistical r.m.s. of $\pm$0.07 dex ($\sim$15\%). Recall that
using the simple activity-age relation (Equation \ref{rhk_new_calib})
produced consistent ages with r.m.s. of $\sim$0.15 dex
($\sim$35\%). So for the sample of non-identical binaries, taking into
account the color-dependent rotational evolution appears to
significantly decrease the age uncertainties.

\begin{deluxetable}{llccc}
\setlength{\tabcolsep}{0.03in}
\tablewidth{0pt}
\tablecaption{Activity-Gyro Ages for Solar-type Binaries\label{tab:pairs_ages}}
\tablehead{
{(1)}      &{(2)}      &{(3)}          &{(4)}          &{(5)}    \\
{Primary}  &{Secondary}&{log\,$\tau_1$}&{log\,$\tau_2$}&{$\overline{\log \tau}$}\\
{}         &{}         &{(yr)}         &{(yr)}         &{(yr)}}        
\startdata
HD 531B    & HD 531A    & 8.01 &  8.73 & 8.37\,$\pm$\,0.36\\
HD 5190    & HD 5208    & 9.59 &  9.86 & 9.73\,$\pm$\,0.14\\
HD 13357A  & HD 13357B  & 9.42 &  9.28 & 9.35\,$\pm$\,0.07\\
HD 14082A  & HD 14082B  & 8.59 &  8.43 & 8.51\,$\pm$\,0.08\\
HD 23439A  & HD 23439B  & 9.80 & 10.11 & 9.96\,$\pm$\,0.15\\
HD 26923   & HD 26913   & 8.76 &  8.64 & 8.70\,$\pm$\,0.06\\
HD 53705   & HD 53706   & 9.56 &  9.89 & 9.72\,$\pm$\,0.16\\
HD 73668A  & HD 73668B  & 9.47 &  9.46 & 9.47\,$\pm$\,0.01\\
HD 103432  & HD 103431  & 9.60 &  9.54 & 9.57\,$\pm$\,0.03\\
HD 116442  & HD 116443  & 9.82 &  9.85 & 9.84\,$\pm$\,0.02\\
HD 134331  & HD 134330  & 9.42 &  9.61 & 9.52\,$\pm$\,0.10\\
HD 134439  & HD 134440  & 9.62 &  9.75 & 9.68\,$\pm$\,0.07\\
HD 135101A & HD 135101B & 9.85 &  9.85 & 9.85\,$\pm$\,0.00\\
HD 137763  & HD 137778  & 9.86 &  8.72 & 9.29\,$\pm$\,0.56*\\
HD 142661  & HD 142661B & 9.43 &  9.32 & 9.37\,$\pm$\,0.06\\
HD 144087  & HD 144088  & 9.42 &  9.37 & 9.39\,$\pm$\,0.02\\
HD 219175A & HD 219175B & 9.48 &  9.58 & 9.53\,$\pm$\,0.05
\enddata
\tablecomments{
Columns: (1) name of primary, (2) name of secondary, (3) activity-gyro age for
component A, (4) activity-gyro age for component B, (5) mean gyro age for the
system. (*) HD 137763 is a pathological case discussed in footnote \ref{HD_137763}.
}
\end{deluxetable}

Our second test involves the cluster stars from Table
\ref{tab:cluster_ca}.  Rather than, as illustrated in Figure
\ref{fig:cluster_rhk_hist}, adopting the mean activity level for a
cluster and turning it into a mean age which can be compared to
individually predicted ages, we convert the individual \rphk\, values
via the Rossby number correlation to period and use the gyrochronology
relations.  This method assumes that the stars are participating in
the so-called I-sequence identified by Barnes and are not ultra-fast
rotators of the so-called C-sequence.  If this is not true in reality,
some rapid rotators will have their ages underestimated via
gyrochronology/activity.  That the Rossby number vs \rphk\,
correlation of Figure \ref{fig:rorphk} breaks down or saturates at
high activity levels helps isolate us from this effect since those
stars will not have reliable conversions to period.  The resulting
dispersions (68\% CLs) in the ages inferred for the cluster members
are listed in Table \ref{tab:rhk_test}, along with the dispersions
observed for the two binary samples and the Sun. Also listed in Table
\ref{tab:rhk_test} is the inferred age dispersion for the same
samples when the simple activity-age relation (equation
\ref{rhk_new_calib}) is used to estimate ages.

From Table \ref{tab:rhk_test} we conclude the following regarding
adopting a simple activity-age relation (\S\ref{Relation}) versus an
activity-rotation-age prescription (\S\ref{Rossby} and \S\ref{Gyro}).
First, among the six stellar samples (four open clusters and two
binary samples), the activity-rotation-age technique resulted in
smaller age dispersions for 5 of the 6 samples (the exception being
M67). Quantifying the improvement is not so straightforward. The
improvement among 3 of the clusters (Pleiades, UMa, Hyades) was
typically a $\sim$10\% reduction in the age dispersion, equivalent to
removing a $\sim$0.1 dex source of systematic error. The two binary
samples show marked improvements in their age dispersions -- most
notably the dispersion in age estimates among the color-separated
binaries was reduced significantly by using the activity-rotation-age
technique rather than a simple activity-age relation.  The results for
M67 are somewhat perplexing, and hint that our activity-rotation-age
technique is not adequately modeling this $\sim$4-Gyr-old group.  This is
not surprising given that half of the M67 sample is hotter/bluer than
the Sun, and as Figure \ref{fig:bv_per} suggests, {\it the gyro
relations are not well-constrained for late-F/early-G stars for ages
older than the Hyades.} We conclude by stating that the
activity-rotation-age technique appears to give slightly more
consistent ages among the older samples tested than by using a simple
activity-age relation.

\subsection{Inferred Ages for the Nearest Solar-Type Dwarfs \label{near_stars}}

While a rigorous utilization of the revised age-deriving methods for
studying the star-formation history of the solar neighborhood is
beyond the focus of this study, we briefly discuss some implications
of our results for a small volume-limited sample of solar-type dwarfs.

We use our new and improved age-deriving methods to estimate the ages
for the 100 nearest solar-type dwarfs (Table \ref{tab:nearstars}).
The sample consists of the nearest known dwarfs with 0.5 $<$ \bv\, $<$
0.9 mag (the color region where both the \rhk\, calculations and
revised gyrochronology relations are constrained).  A few of the
entries are unresolved multiples, sometimes containing two or even
three solar-type stars (e.g. i Boo). Six evolved stars lying more than
one magnitude above the main sequence defined by \citet{Wright04} have
been omitted (i.e. $\Delta$M$_V$ $<$ -1: $\alpha$ Aur, $\eta$ Boo A,
$\mu$ Her, $\zeta$ Her, and $\beta$ Hyi).  When multiple \rhk\,
measurements were found in the literature, we gave highest priority to
those estimates that included the most observations. When multiple
single observations were published by different authors, we
preferentially adopted those from the largest surveys
\citep[e.g.][]{Duncan91,Henry96,Wright04}.  All parallaxes and V
magnitudes are from the Hipparcos catalog \citep{Perryman97}. MK
spectral types are preferentially taken from compilations by Keenan
and Gray and collaborators. Given the stated color, parallax, and
absolute magnitude constraints, this catalog is likely to be complete
for distances of $<$15 pc.

Estimated ages using our methods are listed in the final two columns
of Table \ref{tab:nearstars}. The first column of ages ($\tau_1$) are
from using the revised activity-age relation (\S3.2.2, Eqn. 3).  The
second column of ages ($\tau_2$) are those inferred from converting
the chromospheric activity levels to a rotation period via the Rossby
number, then converting the rotation period to an age using the
revised gyro relation (\S4, Eqns. 5-8, 10-12). {\it The final column
of ages $\tau_2$ are the preferred age estimates}.  The inferred
activity age for the extraordinarily active ZAMS star AB Dor is
$\sim$1 Myr, and clearly in error \citep[apparently by 2 orders of
magnitude;][]{Luhman05}. As AB Dor painfully illustrates, the
uncertainties in the inferred ages for the very active stars
(\logrphk\, $>$ -4.3) are large ($\sim$1 dex; c.f. Table 9). A
conservative estimate of the typical age uncertainty is $\sim$50\%\,
for the preferred ages $\tau_2$ of the lower activity stars.

\begin{figure}
\epsscale{1}
\plotone{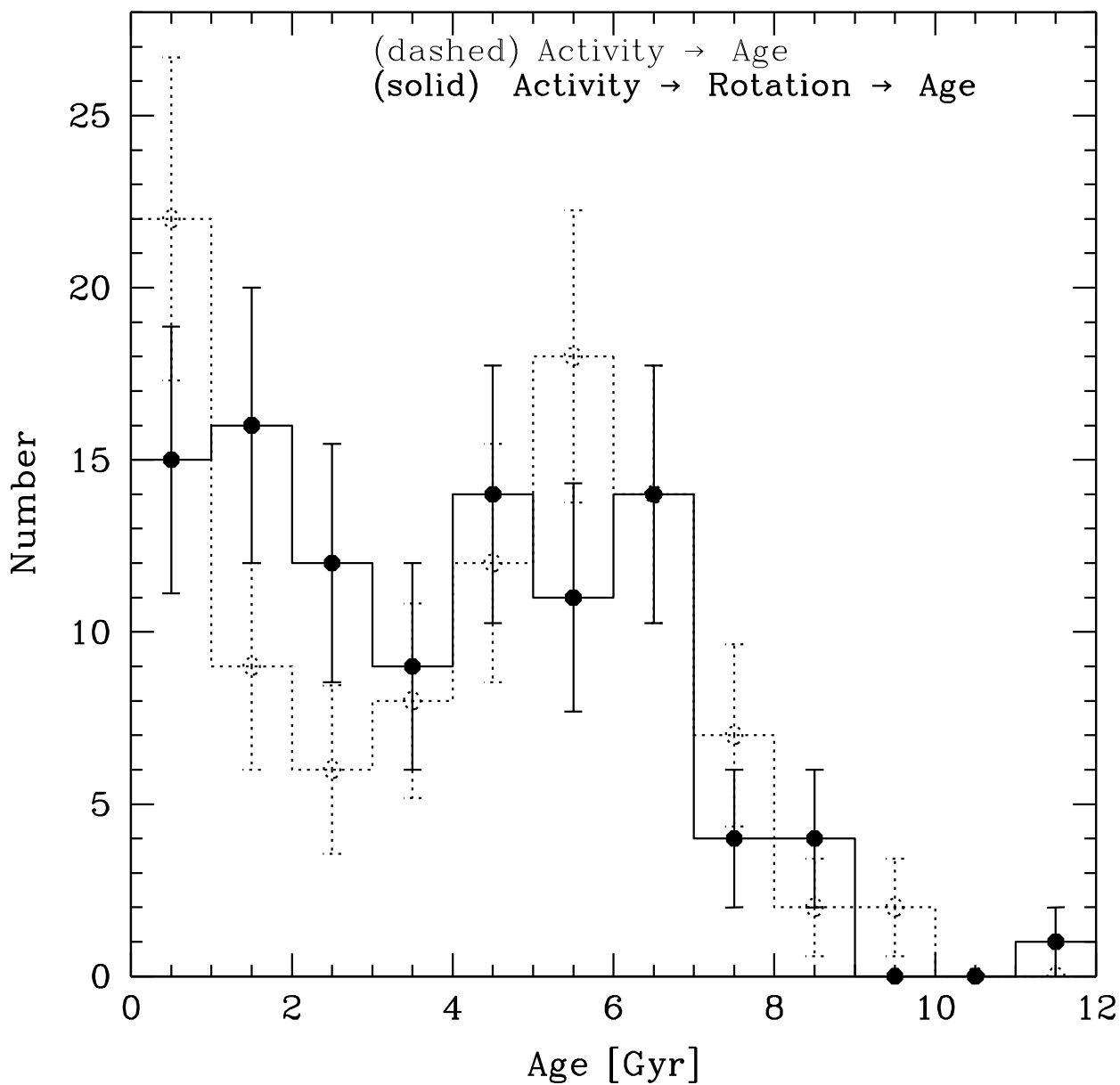}
\caption{Histogram of inferred ages for the nearest 100 solar-type
dwarfs (F7-K2V). {\it Dashed histogram} is for ages inferred directly
from activity using equation 3. {\it Solid histogram} is for ages
derived from converting activity to rotation period (\S4.1), then
converting rotation period and color to age using the revised gyro
relation (\S4.2). The ages inferred directly from activity show the
familiar lull near $\sim$3 Gyr noted in some studies
\citep[e.g.][]{Barry88}. Using the improved ages (from activity
$\rightarrow$ rotation $\rightarrow$ age), the inferred star-formation
rate appears to be smoother between 0-6 Gyr. \label{fig:hist_near} }
\end{figure}

In Fig. \ref{fig:hist_near}, we plot a histogram of the inferred ages
$\tau_1$ and $\tau_2$ for the sample of the 100 nearest solar-type
dwarfs. The histogram can not be strictly interpreted as a true
star-formation history as we have not accounted for disk heating
\citep[e.g.][]{Soderblom91,West08}. The effect preferentially removes
older, higher velocity stars from the local sample, but is subtle and
small for the youngest age bins. The ages inferred from the simple
activity-age (Eqn. 3; {\it dashed histogram}) shows a minimum at
$\sim$2-3 Gyr seen in previous studies which corresponds to the
``Vaughan-Preston gap'' \citep[][see also Fig. 7 and 8 of Henry et al.
1996]{Vaughan80,Barry88}. However, when we examine the histogram of
ages inferred from activity $\rightarrow$ rotation ({\it solid
histogram}), the minimum at $\sim$2-3 Gyr is not as obvious, revealing
a more or less smooth distribution of ages between 0-6 Gyr (with a
precipitous decrease at older ages, presumably due to disk heating and
loss of evolved higher-mass stars from the sample). Similarly, the
stellar birthrate during the past Gyr appears unremarkable compared to
the past $\sim$6 Gyr. These results also call into doubt previous
claims that the star-formation rate during the past Gyr has been
significantly enhanced \citep{Barry88}.

\clearpage
\begin{deluxetable}{llllccclrccrlrcc}
\tabletypesize{\scriptsize}
\setlength{\tabcolsep}{0.03in}
\tablewidth{0pt}
\tablecaption{Activity Ages for the 100 Nearest Solar-type Dwarfs \label{tab:nearstars}}
\tablehead{
{(1) }&{(2)} &{(3)} &{(4)}  &{(5)}     &{(6)}  &{(7)} &{(8)}     &{(9)} &{(10)} &{(11)} &{(12)}         &{(13)}&{(14)}&{(15)}&{(16)}\\
{HD}  &{HIP} &{GJ}  &{Alias}&{$\varpi$}&{\bv}  &{Ref.}&{\logrphk}&{Ref.}&{V}    &{M$_V$}&{$\Delta$M$_V$}&{SpT} &{Ref.}&{$\tau_1$}&{$\tau_2$}\\
{$\ldots$}&{$\ldots$}&{$\ldots$}&{$\ldots$} &{(mas)}   &{(mag)}&{$\ldots$}&{(dex)}   &{$\ldots$}&{(mag)}&{(mag)}&{(mag)}&{$\ldots$}&{$\ldots$}&{(Gyr)}&{(Gyr)}}
\startdata
166 & 544 & 5A & V439 And & 72.98$\pm$0.75 & 0.752 & 18 & -4.328 & 6 & 6.07 & 5.39 & -0.02 & G8V   & 9 & 0.2 & 0.2\\
1581 & 1599 & 17 & $\zeta$ Tuc & 116.38$\pm$0.64 & 0.572 & 17 & -4.839 & 12 & 4.23 & 4.56 & 0.25 & F9.5V & 10 & 3.8 & 2.1\\
3443 & 2941 & 25AB & HR 159 & 64.38$\pm$1.40 & 0.715 & 18 & -4.903 & 1 & 4.61 & 5.19 & -0.58 & G8V+G9V & 3 & 4.8 & 4.9\\
3651 & 3093 & 27A & 54 Psc & 90.03$\pm$0.72 & 0.850 & 18 & -4.991 & 1 & 5.88 & 5.65 & -0.28 & K0V   & 9 & 6.4 & 7.7\\
4391 & 3583 & 1021 & HR 209 & 66.92$\pm$0.73 & 0.640 & 17 & -4.55 & 12 & 5.80 & 4.93 & 0.23 & G5V Fe-0.8 & 10 & 0.8 & 0.9\\
4614 & 3821 & 34A & $\eta$ Cas & 167.99$\pm$0.62 & 0.574 & 17 & -4.958 & 6 & 3.46 & 4.59 & 0.20 & F9V   & 8 & 5.8 & 2.9\\
4628 & 3765 & 33 & HR 222 & 134.04$\pm$0.86 & 0.890 & 18 & -4.852 & 1 & 5.74 & 6.38 & 0.25 & K2.5V & 10 & 4.0 & 5.4\\
4813 & 3909 & 37 & 19 Cet & 64.69$\pm$1.03 & 0.514 & 18 & -4.78 & 9 & 5.17 & 4.22 & 0.32 & F7V   & 9 & 2.9 & 1.7\\
6582 & 5336 & 53A & $\mu$ Cas A & 132.40$\pm$0.60 & 0.695 & 17 & -4.964 & 6 & 5.17 & 5.78 & 0.65 & K1V Fe-2 & 9 & 5.9 & 5.3\\
7570 & 5862 & 55 & $\nu$ Phe & 66.43$\pm$0.64 & 0.571 & 18 & -4.95 & 15 & 4.97 & 4.28 & -0.20 & F9V Fe+0.4 & 10 & 5.7 & 2.8\\
10307 & 7918 & 67 & HR 483 & 79.09$\pm$0.83 & 0.618 & 18 & -5.02 & 11 & 4.96 & 4.45 & -0.14 & G1V   & 9 & 7.0 & 4.2\\
10360 & 7751 & 66A & HR 487 & 122.75$\pm$1.41 & 0.880 & 17 & -4.899 & 10 & 5.96 & 6.26 & 0.36 & K2V   & 10 & 4.8 & 6.2\\
10361 & 7751 & 66B & HR 486 & 122.75$\pm$1.41 & 0.850 & 17 & -4.839 & 10 & 5.81 & 6.26 & 0.33 & K2V   & 10 & 3.8 & 5.2\\
10476 & 7981 & 68 & 107 Psc & 133.91$\pm$0.91 & 0.836 & 18 & -4.912 & 1 & 5.24 & 5.87 & 0.02 & K1V   & 10 & 5.0 & 6.3\\
10700 & 8102 & 71 & $\tau$ Cet & 274.17$\pm$0.80 & 0.727 & 18 & -4.958 & 1 & 3.49 & 5.68 & 0.42 & G8.5V & 10 & 5.8 & 5.8\\
10780 & 8362 & 75 & V987 Cas & 100.24$\pm$0.68 & 0.804 & 18 & -4.681 & 1 & 5.63 & 5.64 & -0.06 & G9V   & 9 & 1.8 & 2.9\\
13445 & 10138 & 86A & HR 637 & 91.63$\pm$0.61 & 0.820 & 17 & -4.74 & 12 & 6.12 & 5.93 & 0.20 & K1V   & 10 & 2.4 & 3.7\\
13974 & 10644 & 92 & $\delta$ Tri A & 92.20$\pm$0.84 & 0.607 & 18 & -4.69 & 11 & 4.84 & 4.66 & 0.15 & G0V   & 8 & 1.9 & 1.5\\
14412 & 10798 & 95 & HR 683 & 78.88$\pm$0.72 & 0.724 & 18 & -4.85 & 21 & 6.33 & 5.81 & 0.57 & G8V   & 10 & 3.9 & 4.3\\
17925 & 13402 & 117 & EP Eri & 96.33$\pm$0.77 & 0.867 & 17 & -4.311 & 1 & 6.05 & 5.97 & -0.02 & K1.5V(k) & 10 & 0.1 & 0.2\\
19373 & 14632 & 124 & $\iota$ Per & 94.93$\pm$0.67 & 0.595 & 18 & -5.02 & 11 & 4.05 & 3.94 & -0.50 & F9.5V & 9 & 7.0 & 3.7\\
20630 & 15457 & 137 & 96 Cet & 109.18$\pm$0.78 & 0.681 & 18 & -4.420 & 1 & 4.84 & 5.03 & 0.05 & G5V   & 16 & 0.3 & 0.4\\
20766 & 15330 & 136 & $\zeta^1$ Ret & 82.51$\pm$0.54 & 0.641 & 18 & -4.646 & 12 & 5.53 & 5.11 & 0.38 & G2V   & 10 & 1.5 & 1.5\\
20794 & 15510 & 139 & 82 Eri & 165.02$\pm$0.55 & 0.708 & 17 & -4.998 & 12 & 4.26 & 5.35 & 0.18 & G8V   & 10 & 6.6 & 6.1\\
20807 & 15371 & 138 & $\zeta^2$ Ret & 82.79$\pm$0.53 & 0.600 & 18 & -4.787 & 12 & 5.24 & 4.83 & 0.36 & G0V   & 10 & 3.0 & 2.0\\
22049 & 16537 & 144 & $\epsilon$ Eri & 310.75$\pm$0.85 & 0.881 & 18 & -4.455 & 1 & 3.72 & 6.18 & 0.10 & K2V(k) & 10 & 0.4 & 0.8\\
22484 & 16852 & 147 & 10 Tau & 72.89$\pm$0.78 & 0.575 & 18 & -5.12 & 21 & 4.29 & 3.60 & -0.70 & F9IV-V & 8 & 8.8 & 4.2\\
26965 & 19849 & 166A & 40 Eri & 198.24$\pm$0.84 & 0.820 & 18 & -4.872 & 1 & 4.43 & 5.92 & 0.14 & K0.5V & 10 & 4.3 & 5.6\\
30495 & 22263 & 177 & 58 Eri & 75.10$\pm$0.80 & 0.632 & 18 & -4.49 & 11 & 5.49 & 4.87 & 0.19 & G1.5V CH-0.5 & 10 & 0.6 & 0.6\\
34411 & 24813 & 197 & $\lambda$ Aur & 79.08$\pm$0.90 & 0.630 & 18 & -5.067 & 6 & 4.69 & 4.18 & -0.48 & G1V   & 9 & 7.9 & 5.0\\
36705 & 25647 & $\ldots$ & AB Dor & 66.92$\pm$0.54 & 0.830 & 18 & -3.88 & 10 & 6.88 & 6.01 & 0.18 & K2Vk  & 10 & $<$0.1 & $<$0.1\\
37394 & 26779 & 211 & V538 Aur & 81.69$\pm$0.83 & 0.840 & 18 & -4.454 & 1 & 6.21 & 5.77 & -0.11 & K0V   & 9 & 0.4 & 0.8\\
38858 & 27435 & 1085 & HR 2007 & 64.25$\pm$1.19 & 0.639 & 18 & -4.87 & 11 & 5.97 & 5.01 & 0.29 & G2V   & 9 & 4.3 & 3.2\\
39587 & 27913 & 222 & 54 Ori & 115.43$\pm$1.08 & 0.594 & 18 & -4.426 & 1 & 4.39 & 4.70 & 0.27 & G0V CH-0.3 & 10 & 0.4 & 0.3\\
41593 & 28954 & 227 & V1386 Ori & 64.71$\pm$0.91 & 0.814 & 18 & -4.42 & 6 & 6.76 & 5.82 & 0.07 & G9V   & 9 & 0.3 & 0.6\\
43834 & 29271 & 231 & $\alpha$ Men & 98.54$\pm$0.45 & 0.720 & 17 & -4.94 & 12 & 5.08 & 5.05 & -0.14 & G7V   & 10 & 5.5 & 5.5\\
52698 & 33817 & 259 & NLTT 17311 & 68.42$\pm$0.72 & 0.894 & 17 & -4.64 & 12 & 6.71 & 5.89 & -0.20 & K1V(k) & 10 & 1.4 & 2.5\\
63077 & 37853 & 288A & 171 Pup & 65.79$\pm$0.56 & 0.589 & 18 & -4.97 & 21 & 5.36 & 4.45 & 0.05 & F9V   & 10 & 6.0 & 3.2\\
69830 & 40693 & 302 & HR 3259 & 79.48$\pm$0.77 & 0.754 & 18 & -4.95 & 21 & 5.95 & 5.45 & 0.03 & G8+V  & 10 & 5.7 & 6.1\\
72673 & 41926 & 309 & HR 3384 & 82.15$\pm$0.66 & 0.784 & 18 & -4.95 & 21 & 6.38 & 5.95 & 0.39 & G9V   & 10 & 5.7 & 6.5\\
72905 & 42438 & 311 & 3 UMa & 70.07$\pm$0.71 & 0.618 & 18 & -4.375 & 1 & 5.63 & 4.86 & 0.27 & G0.5V & 8 & 0.2 & 0.2\\
75732 & 43587 & 324A & 55 Cnc A & 79.80$\pm$0.84 & 0.860 & 17 & -5.04 & 21 & 5.96 & 5.47 & -0.55 & K0IV-V & 9 & 7.4 & 8.7\\
82885 & 47080 & 356A & 11 LMi & 89.45$\pm$0.78 & 0.770 & 18 & -4.638 & 1 & 5.40 & 5.16 & -0.35 & G8+V  & 9 & 1.4 & 2.3\\
86728 & 49081 & 376A & 20 LMi & 67.14$\pm$0.83 & 0.676 & 18 & -5.06 & 21 & 5.37 & 4.50 & -0.45 & G4V   & 9 & 7.7 & 6.2\\
95128 & 53721 & 407 & 47 UMa & 71.04$\pm$0.66 & 0.624 & 18 & -5.02 & 11 & 5.03 & 4.29 & -0.34 & G1V   & 8 & 7.0 & 4.4\\
100623 & 56452 & 432A & HR 4458 & 104.84$\pm$0.81 & 0.811 & 18 & -4.89 & 21 & 5.96 & 6.06 & 0.33 & K0-V  & 10 & 4.6 & 5.8\\
101501 & 56997 & 434 & 61 UMa & 104.81$\pm$0.72 & 0.723 & 18 & -4.546 & 1 & 5.31 & 5.41 & 0.17 & G8V   & 9 & 0.8 & 1.2\\
102365 & 57443 & 442A & HR 4523 & 108.23$\pm$0.70 & 0.664 & 18 & -4.95 & 12 & 4.89 & 5.06 & 0.18 & G2V   & 10 & 5.7 & 4.5\\
103095 & 57939 & 451A & CF UMa & 109.21$\pm$0.78 & 0.751 & 18 & -4.896 & 1 & 6.42 & 6.61 & 1.19 & K1V Fe-1.5 & 9 & 4.7 & 5.3\\
104304 & 58576 & 454 & HR 4587 & 77.48$\pm$0.80 & 0.770 & 17 & -4.92 & 21 & 5.54 & 4.99 & -0.47 & G8IV  & 10 & 5.1 & 5.9\\
109358 & 61317 & 475 & $\beta$ CVn & 119.46$\pm$0.83 & 0.585 & 17 & -4.99 & 11 & 4.26 & 4.64 & 0.23 & G0V   & 9 & 6.4 & 3.3\\
114710 & 64394 & 502 & $\beta$ Com & 109.23$\pm$0.72 & 0.572 & 18 & -4.745 & 1 & 4.23 & 4.42 & 0.13 & G0V   & 8 & 2.5 & 1.5\\
115617 & 64924 & 506 & 61 Vir & 117.30$\pm$0.71 & 0.709 & 18 & -5.001 & 1 & 4.74 & 5.09 & -0.07 & G7V   & 10 & 6.6 & 6.1\\
118972 & 66765 & 1175 & NLTT 34858 & 64.08$\pm$0.81 & 0.855 & 18 & -4.39 & 12 & 6.92 & 5.95 & 0.00 & K0V(k) & 10 & 0.3 & 0.4\\
120136 & 67275 & 527A & $\tau$ Boo & 64.12$\pm$0.70 & 0.508 & 18 & -4.731 & 1 & 4.50 & 3.54 & -0.33 & F7IV-V & 8 & 2.3 & 1.6\\
128620 & 71683 & 559A & $\alpha$ Cen A & 742.12$\pm$1.40 & 0.633 & 2 & -5.002 & 12 & -0.01 & 4.34 & -0.82 & G2V   & 10 & 6.6 & 4.4\\
128621 & 71681 & 559B & $\alpha$ Cen B & 742.12$\pm$1.40 & 0.840 & 2 & -4.923 & 12 & 1.35 & 5.70 & -0.47 & K2IV  & 10 & 5.2 & 6.5\\
131156 & 72659 & 566A & $\xi$ Boo A & 149.26$\pm$0.76 & 0.720 & 17 & -4.344 & 6 & 4.72 & 5.59 & 0.37 & G7V   & 9 & 0.2 & 0.2\\
131511 & 72848 & 567 & DE Boo & 86.69$\pm$0.81 & 0.833 & 18 & -4.52 & 11 & 6.00 & 5.69 & -0.19 & K0V   & 9 & 0.7 & 1.3\\
133640 & 73695 & 575 & i Boo ABC & 78.39$\pm$1.03 & 0.647 & 18 & -4.637 & 6 & 4.83 & 4.30 & -0.47 & G1V+G8V+K0V & 13 & 1.4 & 1.5\\
135599 & 74702 & $\ldots$ & V739 Ser & 64.19$\pm$0.97 & 0.830 & 18 & -4.52 & 21 & 6.92 & 5.96 & 0.13 & K0V   & 9 & 0.7 & 1.3\\
136352 & 75181 & 582 & $\nu^2$ Lup & 68.70$\pm$0.79 & 0.639 & 18 & -4.91 & 12 & 5.65 & 4.83 & 0.11 & G2-V  & 10 & 5.0 & 3.6\\
140538 & 77052 & 596.1A & $\psi$ Ser & 68.16$\pm$0.87 & 0.684 & 18 & -4.80 & 11 & 5.86 & 5.03 & 0.02 & G5V   & 8 & 3.2 & 3.2\\
140901 & 77358 & 599A & HR 5864 & 65.60$\pm$0.77 & 0.715 & 18 & -4.72 & 12 & 6.01 & 5.10 & -0.10 & G7IV-V & 10 & 2.2 & 2.7\\
141004 & 77257 & 598 & $\lambda$ Ser & 85.08$\pm$0.80 & 0.603 & 17 & -5.004 & 1 & 4.42 & 4.07 & -0.43 & G0IV-V & 9 & 6.7 & 3.8\\
142373 & 77760 & 602 & $\chi$ Her & 63.08$\pm$0.54 & 0.563 & 18 & -5.18 & 1 & 4.60 & 3.60 & -0.63 & G0V Fe-0.8 & 9 & 9.7 & 4.4\\
144579 & 78775 & 611A & LHS 3152 & 69.61$\pm$0.57 & 0.734 & 18 & -4.97 & 21 & 6.66 & 5.87 & 0.57 & K0V Fe-1.2 & 9 & 6.0 & 6.1\\
144628 & 79190 & 613 & NLTT 42064 & 69.66$\pm$0.90 & 0.856 & 18 & -4.94 & 12 & 7.11 & 6.32 & 0.37 & K1V   & 10 & 5.5 & 6.8\\
145417 & 79537 & 615 & LHS 413 & 72.75$\pm$0.82 & 0.815 & 18 & -5.06 & 12 & 7.53 & 6.84 & 1.09 & K3V Fe-1.7 & 10 & 7.7 & 8.8\\
146233 & 79672 & 616 & 18 Sco & 71.30$\pm$0.89 & 0.652 & 18 & -4.93 & 11 & 5.49 & 4.76 & -0.05 & G2V   & 9 & 5.3 & 4.1\\
147513 & 80337 & 620.1A & HR 6094 & 77.69$\pm$0.86 & 0.625 & 18 & -4.45 & 20 & 5.37 & 4.82 & 0.19 & G1V CH-0.4 & 10 & 0.4 & 0.4\\
147584 & 80686 & 624 & $\zeta$ TrA & 82.61$\pm$0.57 & 0.550 & 17 & -4.56 & 12 & 4.90 & 4.49 & 0.31 & F9V   & 10 & 0.9 & 0.6\\
149661 & 81300 & 631 & 12 Oph & 102.27$\pm$0.85 & 0.827 & 18 & -4.583 & 1 & 5.77 & 5.82 & 0.01 & K0V(k) & 10 & 1.0 & 1.9\\
154577 & 83990 & 656 & NLTT 44221 & 73.07$\pm$0.91 & 0.893 & 17 & -4.815 & 15 & 7.38 & 6.70 & 0.58 & K2.5V(k) & 10 & 3.4 & 4.8\\
155885 & 84405 & 663B & 36 Oph B & 167.08$\pm$1.07 & 0.860 & 14 & -4.559 & 1 & 5.11 & 6.23 & 0.25 & K0V   & 4 & 0.9 & 1.7\\
155886 & 84405 & 663A & 36 Oph A & 167.08$\pm$1.07 & 0.850 & 14 & -4.570 & 1 & 5.07 & 6.19 & 0.26 & K0V   & 4 & 1.0 & 1.8\\
156274 & 84720 & 666A & 41 Ara & 113.81$\pm$1.36 & 0.777 & 18 & -4.941 & 12 & 5.47 & 5.75 & 0.28 & G9V   & 4 & 5.5 & 6.3\\
157214 & 84862 & 672 & 72 Her & 69.48$\pm$0.56 & 0.619 & 18 & -5.00 & 11 & 5.38 & 4.59 & -0.01 & G0V   & 8 & 6.6 & 4.1\\
158633 & 85235 & 675 & HR 6518 & 78.14$\pm$0.51 & 0.759 & 18 & -4.93 & 21 & 6.44 & 5.90 & 0.46 & K0V   & 5 & 5.3 & 5.9\\
160269 & 86036 & 684AB & 26 Dra AB & 70.98$\pm$0.55 & 0.602 & 18 & -4.62 & 22 & 5.23 & 4.49 & 0.00 & F9V+K3V & 7 & 1.3 & 1.1\\
160691 & 86796 & 691 & $\mu$ Ara & 65.46$\pm$0.80 & 0.700 & 17 & -5.04 & 20 & 5.12 & 4.20 & -0.90 & G3IV-V & 10 & 7.4 & 6.5\\
165341 & 88601 & 702A & 70 Oph A & 196.62$\pm$1.38 & 0.860 & 18 & -4.586 & 6 & 4.25 & 5.50 & -0.48 & K0-V  & 9 & 1.1 & 1.9\\
165908 & 88745 & 704A & 99 Her A & 63.88$\pm$0.55 & 0.528 & 18 & -5.02 & 21 & 5.08 & 4.11 & 0.11 & F9V mw & 8 & 7.0 & 2.9\\
166620 & 88972 & 706 & HR 6806 & 90.11$\pm$0.54 & 0.876 & 18 & -4.955 & 1 & 6.38 & 6.15 & 0.10 & K2V   & 9 & 5.8 & 7.1\\
170657 & 90790 & 716 & NLTT 46596 & 75.71$\pm$0.89 & 0.861 & 18 & -4.65 & 21 & 6.81 & 6.21 & 0.22 & K2V   & 10 & 1.5 & 2.6\\
172051 & 91438 & 722 & HR 6998 & 77.02$\pm$0.85 & 0.673 & 18 & -4.90 & 21 & 5.85 & 5.28 & 0.35 & G6V   & 10 & 4.8 & 4.1\\
176051 & 93017 & 738AB & HR 7162 & 66.76$\pm$0.54 & 0.594 & 18 & -4.874 & 1 & 5.20 & 4.32 & -0.11 & F9V+K1V & 7 & 4.3 & 2.6\\
182488 & 95319 & 758 & HR 7368 & 64.54$\pm$0.60 & 0.804 & 18 & -5.06 & 6 & 6.37 & 5.42 & -0.27 & K0V   & 16 & 7.7 & 8.7\\
185144 & 96100 & 764 & $\sigma$ Dra & 173.41$\pm$0.46 & 0.786 & 18 & -4.832 & 1 & 4.67 & 5.87 & 0.27 & G9V   & 9 & 3.7 & 4.7\\
188512 & 98036 & 771A & $\beta$ Aql & 72.95$\pm$0.83 & 0.855 & 18 & -5.173 & 1 & 3.71 & 3.03 & -2.93 & G9.5IV & 10 & 9.6 & 11.4\\
190248 & 99240 & 780 & $\delta$ Pav & 163.73$\pm$0.65 & 0.751 & 18 & -4.999 & 12 & 3.55 & 4.62 & -0.78 & G8IV  & 10 & 6.6 & 6.9\\
190404 & 98792 & 778 & LHS 481 & 64.17$\pm$0.85 & 0.815 & 18 & -4.98 & 21 & 7.28 & 5.75 & 0.57 & K1V   & 9 & 6.2 & 7.3\\
191408 & 99461 & 783A & HR 7703 & 165.24$\pm$0.90 & 0.868 & 18 & -4.988 & 12 & 5.32 & 6.41 & 0.39 & K2.5V & 10 & 6.4 & 7.7\\
192310 & 99825 & 785 & HR 7722 & 113.33$\pm$0.89 & 0.878 & 18 & -5.048 & 10 & 5.73 & 6.00 & -0.06 & K2+V  & 10 & 7.5 & 8.9\\
196761 & 101997 & 796 & HR 7898 & 68.28$\pm$0.82 & 0.722 & 18 & -4.92 & 21 & 6.36 & 5.53 & 0.32 & G8V   & 10 & 5.1 & 5.2\\
205390 & 106696 & 833 & NLTT 51629 & 67.85$\pm$0.92 & 0.884 & 17 & -4.53 & 15 & 7.14 & 6.30 & 0.23 & K1.5V & 10 & 0.7 & 1.4\\
207129 & 107649 & 838 & HR 8323 & 63.95$\pm$0.78 & 0.601 & 18 & -4.80 & 12 & 5.57 & 4.60 & 0.12 & G0V Fe+0.4 & 10 & 3.2 & 2.1\\
211415 & 110109 & 853A & HR 8501 & 73.47$\pm$0.70 & 0.605 & 17 & -4.86 & 12 & 5.36 & 4.69 & 0.13 & G0V   & 10 & 4.1 & 2.6\\
217014 & 113357 & 882 & 51 Peg & 65.10$\pm$0.76 & 0.666 & 18 & -5.08 & 6 & 5.45 & 4.52 & -0.37 & G2V+  & 10 & 8.1 & 6.1\\
224930 & 171 & 914A & 85 Peg A & 80.63$\pm$3.03 & 0.673 & 17 & -4.875 & 1 & 5.80 & 5.33 & 0.29 & G5V Fe-1 & 9 & 4.4 & 3.8\\
\enddata
\tablecomments{References:
(1) \citet{Baliunas96},
(2) \citet{Bessell81},
(3) \citet{Christy69},
(4) \citet{Corbally84},
(5) \citet{Cowley67},
(6) \citet{Duncan91}, calculated using equations in \citet{Noyes84},
(7) \citet{Edwards76},
(8) \citet{Gray01},
(9) \citet{Gray03},
(10) \citet{Gray06},
(11) \citet{Hall07}, 
(12) \citet{Henry96},
(13) \citet{Hill89},
(14) \citet{Hoffleit91},
(15) \citet{Jenkins06},
(16) \citet{Keenan89},
(17) \citet{Mermilliod91},
(18) \citet{Perryman97},
(19) \citet{Roman50},
(20) \citet{Saffe05},
(21) \citet{Wright04},
(22) estimated from {\it ROSAT} All-Sky Survey X-ray emission
\citep{Voges99,Voges00} via equation A1 (see also \S2.3).  
}
\end{deluxetable}

\clearpage
\section{Summary}

The primary goal of this study was derive a well-calibrated conversion
between activity and age for stars younger than the Sun. To achieve
this, we compiled from the literature \rphk, \rx, and rotation period
data for members of stellar associations and clusters; in particular,
we have populated for the first time the young end of the
chromospheric activity-age relation.  We also used updated/modern ages
for many young associations and clusters.  We then fit the following
relations critical to assessing stellar ages of solar-type dwarfs: a
chromospheric activity-age relation, a chromospheric activity-rotation
relation, a coronal activity-rotation relation, and a rotation-age
``gyrochronology'' relation. Our main results drawn from study of the
rotation and activity observed among binary stars and star cluster
members with 0.5 $<$ \bv\, $<$ 0.9 can be summarized as:

$\bullet$ We provide an improved \logrphk\, vs. age relation for
solar-type stars which constrains especially the young, high-activity
end relative to the relations of \citet{Soderblom91, Donahue93,
Lachaume99}. The activity-age relation for solar-color stars appears
to be absolutely calibrated to the modern cluster age scale to
$\sim$$\pm$0.07 dex in log($\tau$/yr) for stars older than the
Pleiades, and perhaps to only $\sim$$\pm$0.23 dex accuracy in
log($\tau$/yr) for stars younger than the Pleiades.  For young stars
recently arriving on the MS (e.g. the Pleiades), \logrphk\, is not
very useful as a quantitative age estimator as the inferred
r.m.s. spread in ages derived from chromospheric activity is an order
of magnitude. For older samples ($>$0.5 Gyr) and typical \logrphk\,
measurements, it appears that our calibration can estimate the ages of
solar-type dwarfs to $\sim$$\pm$0.25 dex ($\sim$60\%; 1$\sigma$)
accuracy, accounting for systematic errors in the calibration, random
errors due to astrophysical scatter, variability of \logrphk, and
measurement errors. This activity-age relation, however, does not
account for color-dependent evolution effects which appear to be
present.

$\bullet$ We corroborate previous studies which find a tight relation
between chromospheric activity and rotation for stars with -5.0 $<$
\logrphk\, $<$ -4.35, as well as coronal X-ray activity and rotation
for stars with -7 $<$ \logrx\, $<$ -4 (both via the Rossby number).
In their respective saturated regimes (\logrphk\, $>$ -4.35, \logrx\,
$>$ -4), the correlation between chromospheric and coronal activity is
poor. For stars with long-term \logrphk\, averages and well-determined
periods, we find that rotation period can predict mean \logrphk\, to
$\pm$0.05 dex (1$\sigma$) accuracy. For stars with multi-decadal
average \logrphk\, measurements (e.g. Mt. Wilson HK sample),
\logrphk\, can be used to predict Rossby number (period divided by
convective turnover time) to $\pm$0.1 (1$\sigma$) accuracy.  For
shorter baseline \logrphk\, measurements this uncertainty in Rossby
number is larger, with the limit of a single \logrphk\, measurement
probably capable of predicting the Rossby number to $\sim$0.2-0.3
1$\sigma$ accuracy.  Similarly, fractional X-ray luminosity \rx\, for
non-saturated X-ray emitters can be used to infer Rossby number to
$\sim$0.25 1$\sigma$ accuracy.

$\bullet$ We provide an improved gyrochronology relation (period as a
function of color and age), which fits the young cluster data better
than the coefficients provided by \citet{Barnes07}.  For so-called
I-sequence rotators, the new fit is statistically accurate to $\pm$1.2
days in rotation between the age of the Pleiades and Sun. Our revised
gyro relation predicts self-consistent ages with statistical accuracy
of $\pm$0.06 dex (14\%; 1$\sigma$) for solar-type stars with
well-determined periods.

$\bullet$ Combining our activity-rotation relation (via the Rossby
number; \S\ref{Rossby}) and our improved gyrochronology relations
(rotation-color-age; \S\ref{Gyro}), we predict the evolution of
activity as a function of color for solar-type dwarf stars.  Our
activity-rotation-age calibration appears to yield slightly better
ages than using an activity-age relation alone.  Statistical analysis
of binary samples suggest that the activity-rotation-age technique can
estimates ages of $\sim$$\pm$0.1 dex accuracy, whereas analysis of the
cluster samples suggests an accuracy of more like $\sim$$\pm$0.2 dex.

\acknowledgments

We thank Mark Giampapa, David Soderblom, John Stauffer, Jason Wright,
Debra Fischer, Sallie Baliunas, S\o ren Meibom, and Sydney Barnes for
discussions and input.  We acknowledge Greg Henry for allowing us
access to his rotation period data for young main sequence stars in
advance of publication.  EM is supported through a Clay Postdoctoral
Fellowship from the Smithsonian Astrophysical Observatory.

\appendix

\section{X-ray vs. Chromospheric Activity}

\citet{Sterzik97} demonstrated that fractional X-ray luminosity
(\loglxlbol\, or \logrx, hereafter) and \logrphk\, are well-correlated
over a wide range of masses and ages for solar-type dwarfs, and
studies of the Sun and other solar-type dwarfs show that enhanced
coronal activity traces enhanced chromospheric activity temporally as
well \citep[e.g.][]{Hempelmann03}. Whereas \rphk\, appears to drop by
$\sim$1 dex (see Figure \ref{fig:new_old_clusters}) between the T
Tauri epoch ($\sim$1-10 Myr) and the age of the Sun ($\sim$5 Gyr),
\logrx\, declines by $\sim$3 dex \citep{Preibisch05}.  Further, the
saturation of \logrx\ \citep{Preibisch05} appears to occur at earlier
ages than the saturation of \logrphk\ \citep{White07}.  We conclude
that at the high activity end, \logrx\ may be a better diagnostic of
age than \logrphk.

The \logrphk\, vs.  \logrx\, relation of \citet{Sterzik97} could be
improved in two ways. First, their sample is X-ray-biased, as it only
includes stars with \logrphk\, measurements that were detected in the
RASS. Secondly, the relation is poorly constrained at the high
activity end due to the relative rarity of extremely young solar-type
stars within 25 pc. To ameliorate this situation, we fit a \logrphk\,
vs.  \logrx\, relation to an unbiased sample of solar-type dwarfs, and
check that it fits the high-activity regime for solar-type stars.  A
convenient X-ray-unbiased sample of solar-type stars is the
Baliunas-Donahue sample of 28 solar-type dwarfs from the Mt. Wilson HK
survey. This sample has well-determined rotation periods measured over
$>$5 seasons by \citet{Donahue96} and well-determined mean \logrphk\,
values from the Mt. Wilson survey \citep{Baliunas96}. Fortunately,
{\it all} of these stars were detected in X-rays with ROSAT, and X-ray
luminosities and $R_X$ values were calculated by the authors (\S2.2).
An auxiliary sample of X-ray-{\it biased} solar-type stars was also
constructed, so that the \logrx\, vs. \logrphk\, relation fit to the
X-ray-unbiased sample could be verified in the high activity
regime. This auxiliary sample is comprised of 199 solar-type dwarfs
from the literature with \logrphk, \logrx, and rotation period
measurements. This sample was based on the compilation of
\citet{Pizzolato03}, but added to, quality checked, and brought up to
date.
 
\begin{figure}
\epsscale{1}
\plotone{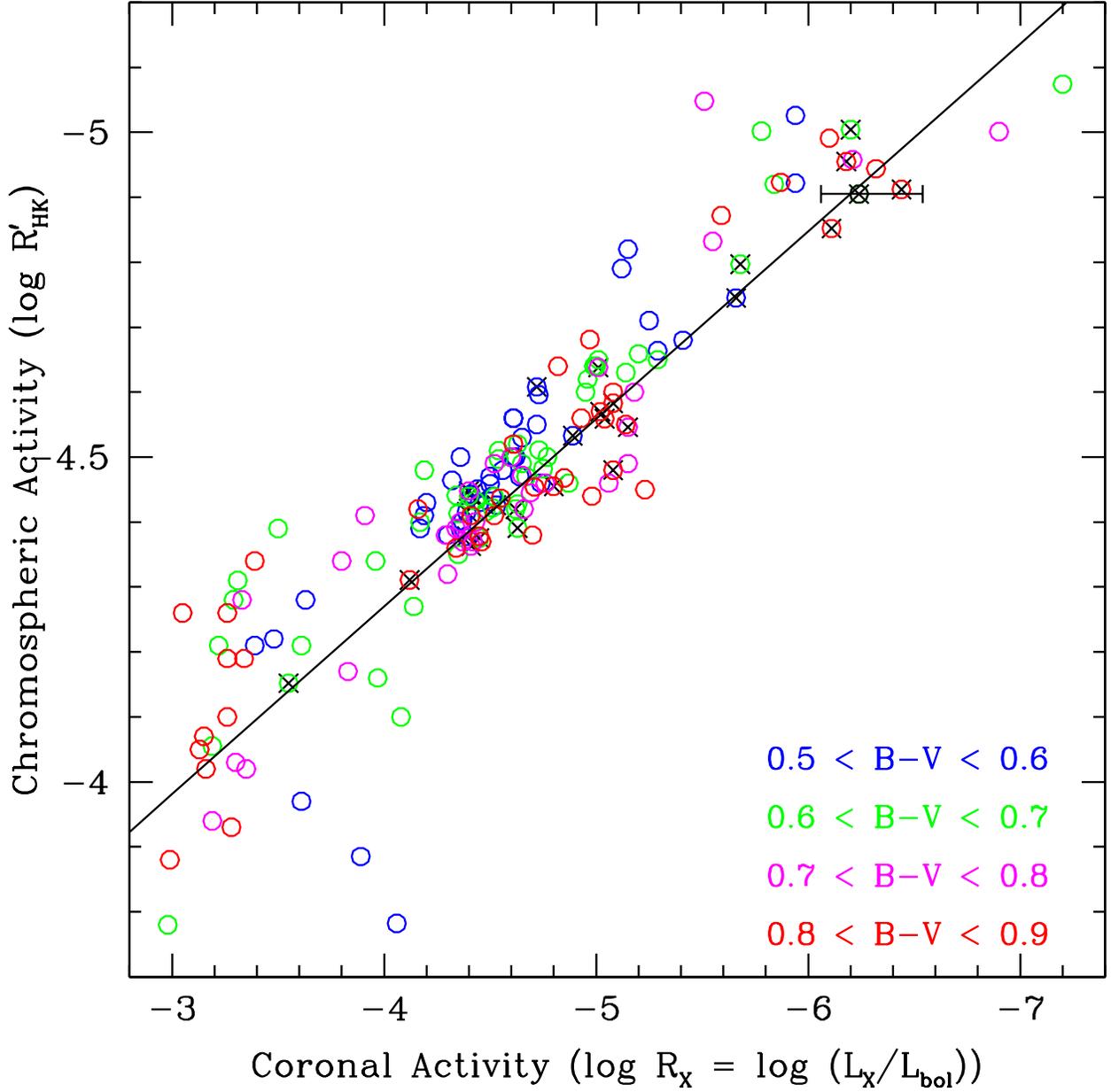}
\caption{\logrx\, vs. \logrphk\, for stars in our sample
of solar-type stars with known rotation periods and chromospheric and
X-ray activity levels.  Donahue-Baliunas stars with well-determined
periods also have dark Xs. Color bins are illustrated in the legend.
The Solar datum uses the mean \logrphk\, calculated in \S1 and the
mean \logrx\, calculated from \citet{Judge03} (with systematic
uncertainty of 50\% in \logrx\, plotted).
\label{fig:logrx}}
\end{figure}

We show in Figure \ref{fig:logrx} the correlation between the coronal
and chromospheric activity indices for both the Baliunas-Donahue
(X-ray unbiased) and auxiliary (X-ray biased) samples.  For the
X-ray-unbiased sample, the X-ray and chromospheric indices are
remarkably well correlated (Pearson $r$ = 0.96). We calculate the OLS
bisector linear regression following \citet{Isobe90}. We find

\begin{equation}
{\rm log}\,R^{'}_{HK}\, = (-4.54 \pm 0.01) + (0.289 \pm 0.015)\,({\rm log}\,R_X + 4.92)
\label{eqn:rhk_rx}
\end{equation} 

\noindent with an r.m.s. scatter of 0.06 in \logrphk. The inverse
relation is:

\begin{equation}
{\rm log}\,R_X\,  = (-4.90 \pm 0.04) + (3.46 \pm 0.18)\,({\rm log}\,R^{'}_{HK} + 4.53)
\label{eqn:rx_rhk}
\end{equation} 

\noindent with an r.m.s. of 0.19 dex ($\sim$55\%) in \logrx.  Equation
\ref{eqn:rx_rhk} is statistically consistent with the relation found
by \citet{Sterzik97}, but our uncertainties are $\sim$2$\times$
smaller. Linear fits were also made for \logrphk\, vs. \logrx, and its
inverse, for the X-ray-based auxiliary sample.  The result fits gave
slopes statistically consistent with that estimated for the
Baliunas-Donahue X-ray-unbiased sample, but with y-intercepts favored
towards giving larger \logrx\, values (e.g. the X-ray-biased fit would
predict \logrx\, for the solar \logrphk\, value higher by $\sim$0.2
dex compared to the X-ray-unbiased fit). We find that equations A1 and
A2 are satisfactory for the high-activity stars also, so the fits are
appropriate for the full range of \logrx\, and \logrphk\, values seen
for solar-type field dwarfs and pre-MS stars.  The scatter in both
relations increases substantially as the transition from the "active"
regimes in both sequences to the "very active" regime above about
-4.35 in \logrphk\, and the "saturated" regime above about -4 in
\logrx\, is approached.

If one combines equations \ref{rhk_new_calib} and \ref{eqn:rhk_rx},
one can derive an X-ray activity vs. age relation for solar-type
dwarfs:

\begin{equation}
\rm{log}\,\tau\, = 1.20 - 2.307\, \rm{log}\,R_X\, - 0.1512\, \rm{log}\,R_X^2
\label{eqn:age_rx}
\end{equation}

From the cluster X-ray data compiled in \citet{Pizzolato03}, it
appears that the spread in \logrx\, among solar-type dwarfs in young
clusters is $\sim$$\pm$0.2-0.6 dex (68\% CL). If the chromospheric
activity levels for the 4 Gyr-old members of M 67 \citep{Giampapa06}
are converted to \logrx\, via Equation \ref{eqn:rx_rhk}, one would
predict a $\pm$0.4 dex (68\% CL) spread in \logrx\, values among its
solar-type members. Based on this, a $\sim$$\pm$0.4 dex (68\%CL)
spread in \logrx\, values for a coeval population can be adopted, and
should be factored into any age uncertainty inferred from Equation
\ref{eqn:age_rx}. 

The \citet{Baliunas96} \logrphk\, values are long-term averages from
$\sim$20 years of Mt. Wilson HK observations, whereas the \logrx\,
values typically represent only a few-hundred second snapshot in the
star's life. The correlation suggests that one can predict a
multi-decadal average of \logrphk\, to within $\pm$0.1 1$\sigma$
accuracy for a solar-type star from a few hundred seconds of X-ray
data.  Given the current state of X-ray and chromospheric activity
data in the literature, we believe that these r.m.s values are
representative of how accurately these variables can be used to
predict one another.

\end{document}